\documentclass[preprint]{aastex}

\newcommand{\kms}{\, {\rm km\, s}^{-1}}
\newcommand{\ikms}{(\kms)^{-1}}

\newcommand{\lya}{Ly$\alpha$}
\newcommand{\lyaf}{Ly$\alpha$ forest}
\newcommand{\bF}{\bar{F}}
\newcommand{\bC}{\bar{C}}
\newcommand{\bS}{\bar{S}}
\newcommand{\xrei}{x_{rei}}
\newcommand{\lr}{\lambda_{{\rm rest}}}

\newcommand{\vdF}{{\mathbf \delta_F}}
\newcommand{\vdS}{{\mathbf \delta_S}}
\newcommand{\vdf}{{\mathbf \delta_f}}
\newcommand{\vdn}{{\mathbf \delta_n}}
\newcommand{\vdC}{{\mathbf \delta_C}}
\newcommand{\vdX}{{\mathbf \delta_X}}
\newcommand{\PF}{$P_F(k,z)$}
\newcommand{\PX}{$P_X(k,z)$}
\newcommand{\lcdm}{$\Lambda$CDM}

\begin{document}

\title{The Lyman-$\alpha$ Forest Power Spectrum from the Sloan Digital Sky 
Survey}

\author{Patrick McDonald\altaffilmark{1}, Uro\v s Seljak\altaffilmark{1}, 
Scott Burles\altaffilmark{2},
David~J.~Schlegel\altaffilmark{3},
David~H.~Weinberg\altaffilmark{4},
David Shih\altaffilmark{1},
Joop~Schaye\altaffilmark{5},
Donald~P.~Schneider\altaffilmark{6}, 
J. Brinkmann\altaffilmark{7},
Robert J. Brunner\altaffilmark{8},
Masataka Fukugita\altaffilmark{9}
}

\altaffiltext{1}
{Physics Department, Princeton University, Princeton, NJ 08544, USA; 
pm,useljak@princeton.edu}

\altaffiltext{2}
{Physics Department, MIT, 77 Massachusetts Av., Cambridge MA 02139, USA}

\altaffiltext{3}
{Princeton University Observatory, Princeton, NJ 08544, USA}

\altaffiltext{4}
{Department of Astronomy, Ohio State University, Columbus, OH 43210, USA}

\altaffiltext{5}
{School of Natural Sciences, Institute for Advanced Study, Einstein Drive, 
Princeton NJ 08540, USA}

\altaffiltext{6}
{Department of Astronomy and Astrophysics,
The Pennsylvania State University, University Park, PA 16802, USA}

\altaffiltext{7}
{Apache Point Observatory, 2001 Apache Point Rd, Sunspot, NM 88349-0059, USA}

\altaffiltext{8}
{NCSA and Department of Astronomy, University of Illinois, Urbana, IL 61801, 
USA}

\altaffiltext{9}
{Inst. for Cosmic Ray Research, Univ. of Tokyo, Kashiwa 277-8582, Japan} 

\begin{abstract}

We measure the power spectrum, \PF, of the transmitted flux in the \lyaf\
using 3035 high redshift quasar spectra from the
Sloan Digital 
Sky Survey. This sample is almost two orders of magnitude larger than 
any previously available data set, yielding statistical errors of
$\sim 0.6$\% and $\sim 0.005$ on, respectively, the overall 
amplitude and logarithmic slope of \PF.
This unprecedented statistical power 
requires a correspondingly careful analysis of the data and 
of possible systematic contaminations in it. 
For this purpose we reanalyze the raw spectra to make use of information
not preserved by the standard pipeline.
We investigate the details of the noise in the data, resolution of 
the spectrograph, sky subtraction, quasar continuum, and metal absorption.
We find that background sources such as metals contribute 
significantly to the total power and have to be subtracted properly. 
We also find clear evidence for SiIII correlations 
with the \lyaf\ and suggest a simple model to account for this 
contribution to the power. 
While it is likely that our newly developed analysis technique 
does not eliminate 
all systematic errors in the \PF\ measurement below the 
level of the statistical errors, 
our tests indicate that any residual systematics in the analysis 
are unlikely to affect the inference of cosmological parameters from \PF. 
These results should provide an essential ingredient for all future
attempts to constrain modeling of structure formation, cosmological
parameters, and theories for the origin of primordial fluctuations.

\end{abstract}


\section{Introduction}

Although the \lyaf\ was discovered many decades ago \citep{1971ApJ...164L..73L},
it has only recently emerged as one of 
the prime tracers of the large scale structure in the Universe. 
The high resolution measurements using the Keck HIRES spectrograph 
\citep{1994SPIE.2198..362V} have been largely reproduced using 
hydrodynamical simulations
\citep{1994ApJ...437L...9C,1995ApJ...453L..57Z,1996ApJ...457L..51H,
1998MNRAS.301..478T}
and semi-analytical models \citep{1998MNRAS.296...44G}.
The picture that has emerged from these studies
is one in which the neutral gas responsible for the absorption is in a 
relatively
low density, smooth environment, which implies a simple connection 
between the gas and the underlying dark matter.
The neutral fraction of the gas is determined
by the interplay between the recombination rate (which depends on the
temperature of the gas) and ionization caused by
ultraviolet photons. Photoionization heating
and expansion cooling cause the gas density and temperature to be tightly
related, except where mild shocks heat up the gas. This
leads to a tight relation between the absorption
and the gas density.
Finally, the gas density is
closely related to the dark matter density on large scales, while on small
scales the effects of thermal broadening and Jeans smoothing must 
be included.
In the simplest picture described here, all of the physics ingredients are
known and can be modeled. 
The fact that one can trace the fluctuations over a range 
of redshifts ($2 \lesssim z \lesssim 6$ using ground based spectrographs) 
and over a range 
of scales, which are typically smaller than the scales of other tracers, 
is the main strength of this method. It becomes particularly powerful 
when combined with cosmic microwave background (CMB)  anisotropies or other 
tracers that are sensitive to larger scales. Such a combination is 
sensitive to the shape of the primordial spectrum of fluctuations, 
which is one of the few observationally accessible 
probes of the early universe. These observations are therefore 
directly testing the models of the early universe such as inflation. 

\lyaf\ observations and constraints on cosmology
have been explored by several groups in the past.
Most of the analyses focused on the 
power spectrum, $P_F(k)$, of the fluctuations in the \lyaf\ flux,
\begin{equation}
\delta_F(\lambda)=
\exp[-\tau(\lambda)]/\left<\exp(-\tau)\right>-1~,
\end{equation}
where $\tau$ is the optical depth to \lya\ absorption. 
The first such work was by \cite{1998ApJ...495...44C}, followed 
by 
\cite{2000ApJ...543....1M}, 
\cite{2002ApJ...581...20C},
and \cite{2003astro.ph..8103K}.
These groups were limited 
to a few dozen spectra at most.
Recent theoretical analyses, in addition to above, have been performed by
\cite{2002MNRAS.334..107G}, \cite{2001astro.ph..11230Z}, and 
\cite{2003MNRAS.342L..79S}. 
In the latter two of these papers the degeneracy between the amplitude
and slope of the primordial power spectrum and the
normalization of the optical depth-density relation [most sensitive
to the intensity of UV background, and typically parametrized in 
terms of the mean transmitted flux fraction, 
$\bF\equiv \left<\exp(-\tau)\right>$]
was emphasized, which leads to a 
significant expansion of the allowed range of cosmological parameters 
relative to what one would have inferred from the errors on the flux 
power spectrum alone. 
\cite{2003MNRAS.342L..79S} have shown that the current \lyaf\
constraints 
are consistent with the \lcdm\ model favored by recent CMB data,
testing it in a regime of redshift and length scale not probed
by other measurements, but that within the \lcdm\ framework they
do not add much leverage on parameter values beyond that
afforded by the CMB data alone.

An important practical implication of the theoretical breakthroughs of 
the 1990s is that large scale structure in the \lyaf\ can be effectively
studied with moderate resolution spectra.  Once the spectrum is modeled
as a continuous phenomenon rather than a collection of discrete lines,
there is no need to resolve every feature.  Some of the studies cited
above use high resolution ($\sim 0.08$\AA) spectra, some use moderate
resolution ($\sim 1-3\AA$) spectra, and some use a mix of the two.

The goal of this paper is to present a new measurement of the \lyaf\ 
power spectrum, based on $\sim 3000$ Sloan Digital Sky Survey (SDSS; 
\cite{2000AJ....120.1579Y}) spectra that 
probe the \lyaf\ at a resolution $R \sim 2000$ ($\sim 2.5$\AA\ FWHM).
This sample is almost two orders of magnitude 
larger than anything that was available before. 
As such it greatly increases 
the statistical power of the \lyaf, making it comparable to the CMB
from WMAP.  
At the same time, the required tolerance of systematic
errors also increases by the same amount. This requires a careful 
investigation of all  
of the sources of systematic errors, and a large portion of this paper 
is devoted to the issue of possible systematics in the data 
and their influence on the parameters of interest. 
We also discuss how the analysis we perform and results we obtain 
differ from what can be done using the standard spectral pipeline
outputs in the public SDSS data.
In part because of the practicalities of work in a large,
multi-institutional collaboration, and in part because of the importance
of obtaining an accurate measurement with well understood
statistical and systematic errors, the \lya\ forest power spectrum
has been pursued by two independent groups within the SDSS, one
led by P. McDonald and U. Seljak, and the other by L. Hui and
A. Lidz.  The
methods employed are different and have been
developed independently.  Results of the alternative analysis
will be presented elsewhere (Hui et al., in preparation).

We only present the observational measurement of the SDSS
\lyaf\ power spectrum in the current paper. 
Independent of any theoretical interpretation, this basic result 
should be robust on the scales 
for which we give results, $0.0013~\ikms <k<0.02~ \ikms$, where
$k \equiv 2 \pi / \lambda$ if $\lambda$ is the wavelength
of a Fourier mode (not to be confused with spectral wavelength), 
here measured in $\kms$ (note that
throughout the paper we frequently use velocity in place of 
observed wavelength, with the understanding that all that enters 
into our calculations are velocity differences between pixels of 
measured spectra, 
defined by $\Delta v = c~ \Delta \ln(\lambda)$ -- we do not measure
power on scales large enough for the imperfections in this expression
to become relevant). 
The choice of $k$-range is determined by the continuum fluctuations 
on the low end and 
spectral resolution at the high end.
We note also that the useful range is limited not only by these uncertainties,
which are 
related to the data analysis, but also by the uncertainties in the 
theoretical modeling and/or additional astrophysical effects. 
We will address these latter issues in more detail in a separate publication. 
However, 
we  do not completely decouple the theory from the 
data analysis. For example, when discussing the importance of systematic 
errors it is useful to understand how they 
would affect cosmological results like the
slope and amplitude of the matter power spectrum, so much 
of our discussion of systematics is devoted to this issue. 

The common usage of the term \lyaf\ is to describe the \lya\ absorption 
by neutral hydrogen in the relatively low density bulk of the IGM. 
In this paper we include damped-\lya\ systems (DLAs) in the definition
of the ``forest'', so it includes all HI-\lya\ absorption.  
We could try to remove DLAs before measuring \PF, because they 
are more difficult to simulate than the lower optical depth absorption;
however, we believe the advantage of removal is illusory.  If the DLAs were
located randomly within the IGM (which they certainly are not completely), 
it would be simple to include them in the theory using their known column 
density distribution.  If they are 
not located randomly, the regions obscured by DLAs in the spectra are 
special, so the effect of removing the DLAs still must be understood 
using simulations.  We leave the handling of the effects of
DLAs as a problem for the theory, which we will address elsewhere.

Absorption by metals is also difficult to simulate accurately, so we would
like to remove this contribution to \PF.  This is
relatively easy to do for transitions with 
wavelength $\lambda \gtrsim 1300$ \AA, but it
is basically impossible for transitions with $\lambda \lesssim \lambda_\alpha$,
because the metal features always appear mixed with HI-\lya.  
We will subtract the power 
measured in the rest wavelength range $1268~{\rm \AA} < \lr < 1380$ \AA\ 
from our 
measurement of the power in the forest, which removes 
the effect of transitions with longer wavelength, but we leave shorter
wavelength transitions as part of the forest.  The only significant
contaminant of this kind that we can identify is SiIII absorption at 
$1206.50$\AA,
and we develop a simple and effective way 
to account for this in the theory. 
We refer to our final background-subtracted power spectrum as \PF, and
use $P_{\lambda_1,\lambda_2}(k,z)$ for the raw power measured 
in the interval $\lambda_1<\lr<\lambda_2$.  
We are using the range $1041~{\rm \AA}<\lr<1185$ \AA\ for the \lyaf. 

The outline of this paper is as follows. 
In \S \ref{secdataprep}, we describe the selection of our data set and the
preparation of spectra for the measurement of \PF.   In
\S \ref{secpowerextract} we describe the method used to measure the
power spectrum  
and estimate the error bars, test the procedure, and give the basic
results.  We perform consistency checks on the results and discuss
systematic errors in \S \ref{secconchecks}, which is followed by a 
brief recipe for using our results in \S \ref{directionsforuse}, and
conclusions in \S \ref{conclusions}.  

\section{Data Selection and Preparation \label{secdataprep}}

We describe the sample of quasar spectra that 
we use in \S \ref{secsampselect}.  
In \S \ref{secBAL} we explain how we remove broad absorption
line (BAL) quasars from the sample.
In \S \ref{secexpcomb} we explain how we combine spectra from
different exposures for the same quasar and use the differences 
between exposures to understand the noise in the data.
We discuss the resolution of the spectra in \S \ref{secrescal}.
Finally, in \S \ref{secmeandiv} we describe how we divide each 
spectrum by an estimate of the quasar continuum, the expected 
mean absorption level 
in the spectrum, and a spectral calibration vector (see below), 
to produce the 
vectors of transmission-fluctuation estimates, $\vdf$, 
for each quasar, from which we will measure \PF. 

\subsection{SDSS Observations and Sample Selection \label{secsampselect}}

The Sloan Digital Sky Survey \citep{2000AJ....120.1579Y} uses 
a drift-scanning 
imaging camera \citep{1998AJ....116.3040G} and a 640 fiber double 
spectrograph on a dedicated 2.5
m telescope. It is an ongoing survey to image 10,000 sq. deg. of 
the sky in the SDSS $ugriz$ AB magnitude system 
\citep{1996AJ....111.1748F,2002AJ....123..485S} and to obtain 
spectra for $\sim 10^6$ galaxies and $\sim 10^5$ quasars. The astrometric 
calibration is good to better than $0\farcs 1$ rms per 
coordinate \citep{2003AJ....125.1559P}, 
and the photometric calibration is accurate to 3\% or 
better \citep{2001AJ....122.2129H,2002AJ....123.2121S}. 
The data sample used in this paper 
was compiled in Summer 2002 and is a 
combination of data releases one \citep{2003AJ....126.2081A} and two 
\citep{abazajian04}.

About 13\% of the spectroscopic survey targets
are quasar candidates selected based on their colors 
\citep{2002AJ....123.2945R}. 
The magnitude limit for UV-excess objects is $i=19.1$, while
additional high-redshift candidates ($z>3$) 
are targeted to $i=20.2$.
Fibers are allocated according to a tiling algorithm
\citep{2003AJ....125.2276B}, with the galaxy sample and the 
quasar sample being the top priorities. 
The remaining 8\% of fibers serve for calibration purposes.

SDSS spectra are obtained using plates holding 640 fibers, each of which 
subtends 3$\arcsec$ on the sky; the spectra cover $3800-9200\mbox{\AA}$.
The pixel width is a slowly varying function of wavelength, 
but is typically $\sim 70 \kms$.  The resolution also varies, but is 
typically also $\sim 70 \kms$ rms (i.e., the resolution is $1800<R<2100$ 
and there are $\sim 2.4$ pixels per FWHM resolution element).  
All quasars have multiple spectra, usually taken one after the other 
(timescales of a fraction of an hour), so 
the quasar variability can be ignored (in the opposite case 
it would act as an additional source of noise).
The co-added spectra in the official SDSS release use 
local spline interpolation onto a uniform grid of pixels of width 
$\Delta \log_{10}(\lambda)=0.0001$,
and do not guarantee the noise to be uncorrelated.  
We therefore redo this step starting from the individual exposures. This 
is discussed in more detail below. 
Spectral flux errors per pixel in most cases are 
about $1\times 10^{-17}$ erg s$^{-1}$ cm$^{-2}$ \AA$^{-1}$.
Redshifts are automatically assigned by the SDSS spectral classification 
algorithm, which is based on $\chi^2$ fitting of templates to each spectrum 
(Schlegel et al., in preparation). 

\begin{figure}
\plotone{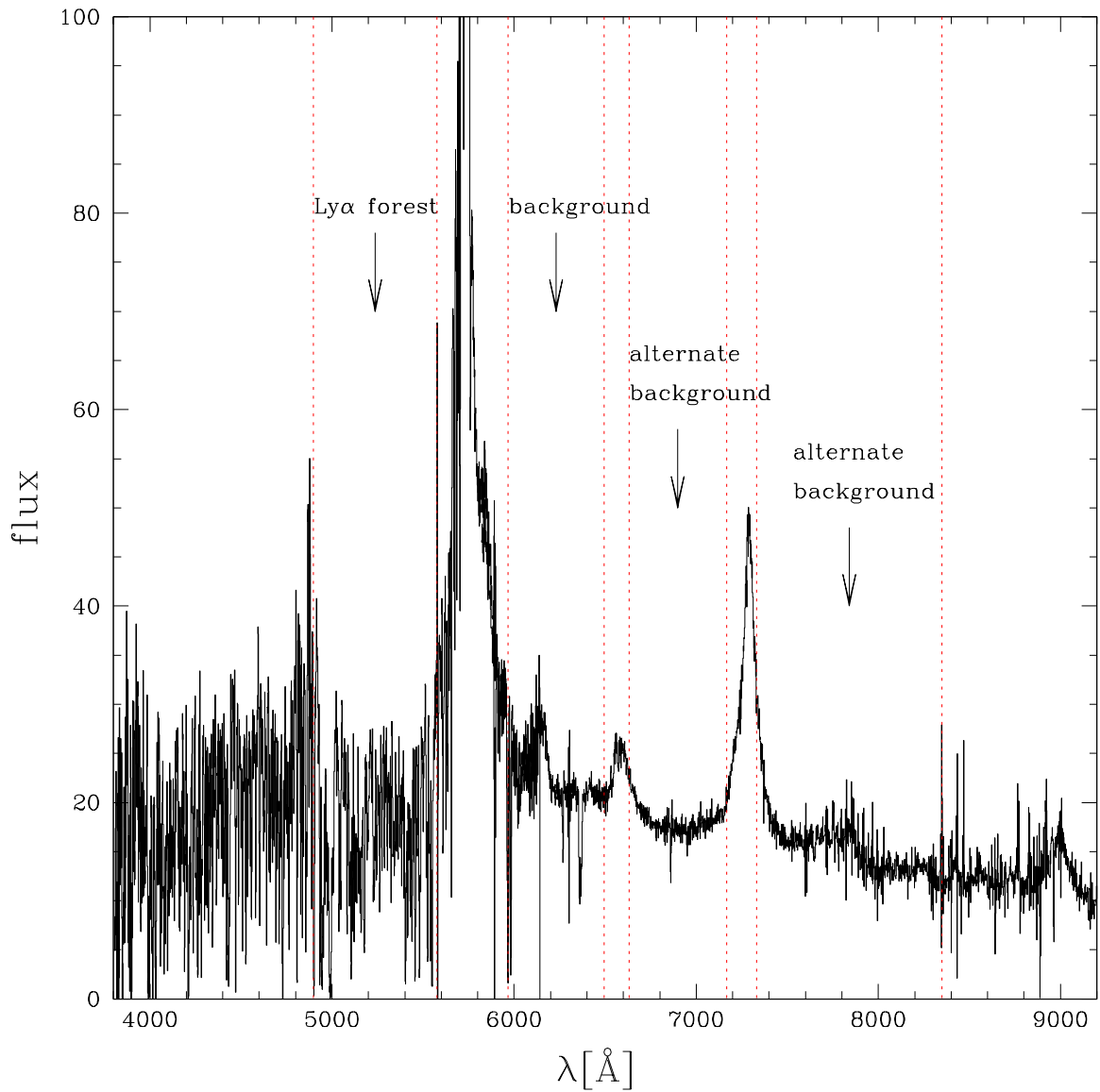}
\caption{
Example spectrum of a $z=3.7$ quasar with unusually high S/N. 
The regions we use to measure the \lyaf\ power and background
power are indicated by vertical dotted lines, along with a couple
of alternate regions that we will discuss (note that the
background and \lyaf\ observed in the same quasar spectrum correspond 
to different redshifts).
}
\label{fullspecexamp}
\end{figure}

We limit ourselves to quasars 
with redshift $z_q>2.3$ when measuring the power in the \lyaf\ region of
spectra, so that each spectrum contains a significant stretch of the 
\lyaf\ above the detector cutoff at 3800 \AA\ (which corresponds to
\lya\ absorption at $z=2.12$). 
We use the sample compiled in Summer 2002, 
cut down to 3035 spectra by 
eliminating some plates of questionable quality,
some spectra where two different redshift estimation codes disagree, and
some BAL quasars (see below).
Figure \ref{zhist} shows the redshift distribution of the data. 
\begin{figure}
\plotone{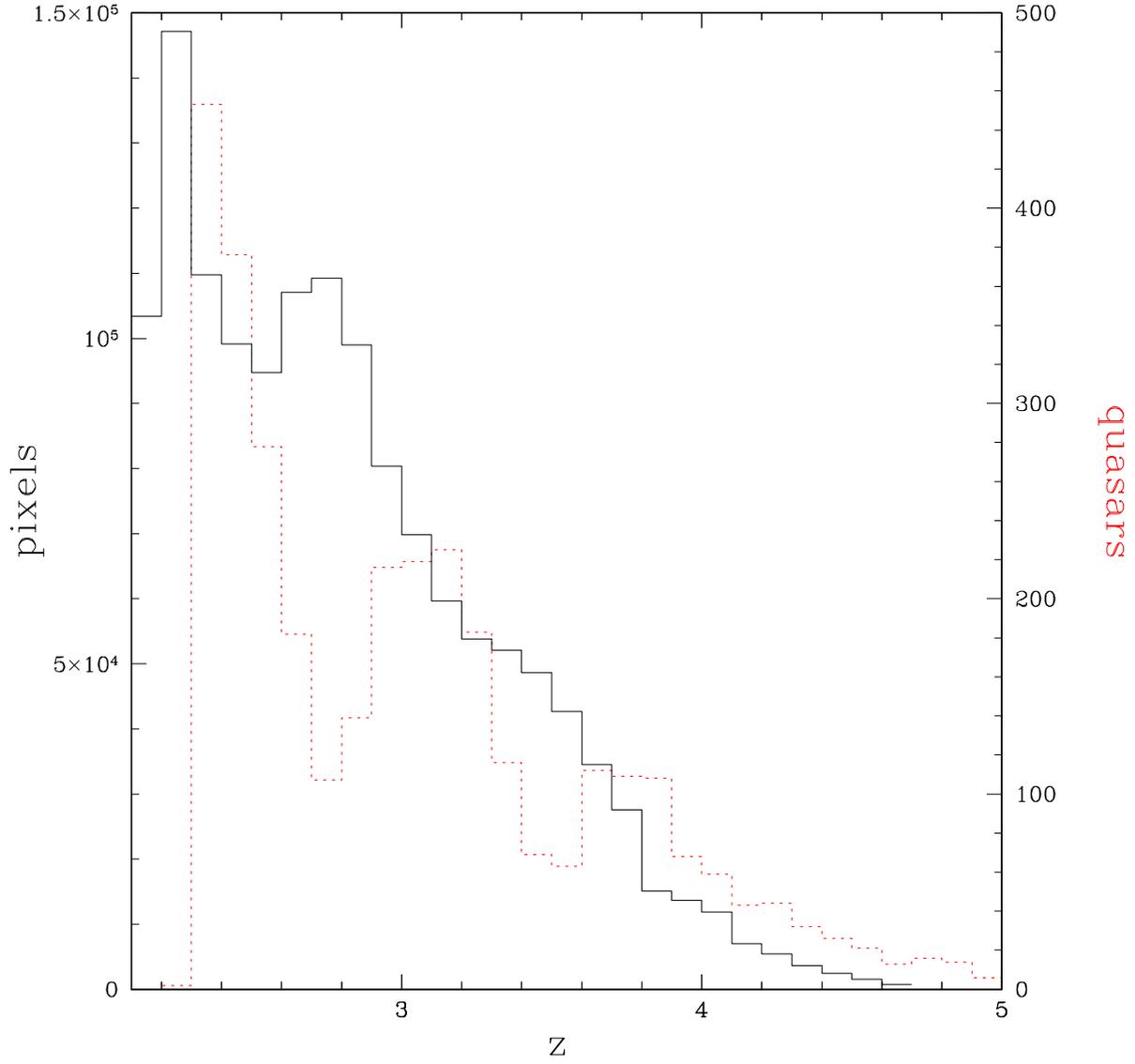}
\caption{
The distribution of the spectral pixels used to probe the \lyaf\ (black,
solid histogram; scale on left axis), and
the redshift distribution of our primary sample of 3035
quasars (red, dotted histogram; right axis).  
Note the gap at $z\sim 2.7$ in the quasar redshift 
distribution, 
caused by a class of stars being indistinguishable from quasars in the 
SDSS photometry \citep{2002AJ....123.2945R}. }
\label{zhist}
\end{figure}
The dashed, red histogram shows the distribution of quasar
redshifts.  The solid, black histogram shows the  
distribution of pixels in the range $1041<\lr<1185$\AA.
Note that there is a gap in the quasar redshifts 
around $z \sim 2.7$, which is due to the 
stellar locus crossing the quasar locus in the 5-color SDSS photometry
\citep{2002AJ....123.2945R}.
Figure \ref{fullspecexamp} shows an example SDSS spectrum of a $z=3.7$ 
quasar.  
This spectrum is unusual in that most have lower S/N, and most quasars
are at lower redshift.

We employ an additional sample of $\sim 8000$ spectra with $z_q<2.3$, 
so that we can study the full observed wavelength range,
$3800~{\rm \AA}\lesssim \lambda \lesssim 9200~{\rm \AA}$,
outside the confusion of the \lyaf.
As we discuss in \S \ref{secbacksubtract}, we compute a non-negligible 
background power term (probably mostly metal absorption), by 
measuring the power in the wavelength range $1268~{\rm \AA} < \lr <
1380$ \AA.  Using only the primary sample, we would not be able to 
compute this term for observed wavelengths below $\sim 4400$ \AA.

We remove several wavelength regions from our analysis because of
calibration problems: $\lambda<3800$ \AA, $5575~{\rm \AA}<\lambda<5583$ \AA, 
$5888~{\rm \AA} <\lambda<5894$ \AA, $6296~{\rm \AA}<\lambda<6308$ \AA, and 
$6862~{\rm \AA}<\lambda<6871$ \AA\
(the last two have no direct effect on the results we present). 
Most of these problems are due to strong sky lines.

\subsection{BAL Removal \label{secBAL}}

Our sample was initially examined by eye, and the most extreme broad 
absorption line (BAL) quasars were removed (see \cite{2002ApJS..141..267H} 
for a discussion of BALs).  When we first measured the background power in the
region $1409< \lr <1523$\AA, we found that the most extreme outliers in 
power were still obvious BALs (this was not true of the \lyaf\ region).  
To test the importance of these systems
to our \lyaf\ power measurement, we removed a further 147 quasars using the 
following automated method:  Each spectrum is smoothed by a Gaussian with
rms width $280\kms$.  The continuum within the region $1420 <\lr <1535$\AA\
is redefined by dividing by the mean flux-to-continuum ratio in the region.
A quasar is identified as a BAL quasar if 
the region $1420 <\lr <1535$\AA\ contains a $2000\kms$ long 
continuous set of pixels 
that all fall more than 20\% above or below our estimated continuum 
(we initially identified wide regions with flux above the continuum 
out of simple
curiosity, but found that these are in practice almost always 
obvious BAL quasars where the continuum has been biased low by the 
BAL feature).  We iterate the
continuum redefinition twice, computing the new mean after throwing out
pixels more than 20\% below the previous mean, but this makes almost
no difference to the results.  Note that the $280 \kms$ smoothing was
applied to allow easier identification of BALs in noisy spectra.
As we show below, this BAL cut makes essentially no difference to our 
\PF\ result, although it does have a noticeable effect on the 
power measurement in the region $1409< \lr <1523$\AA.

\subsection{Combining Exposures and Calibrating the Noise \label{secexpcomb}}

\begin{figure}
\plotone{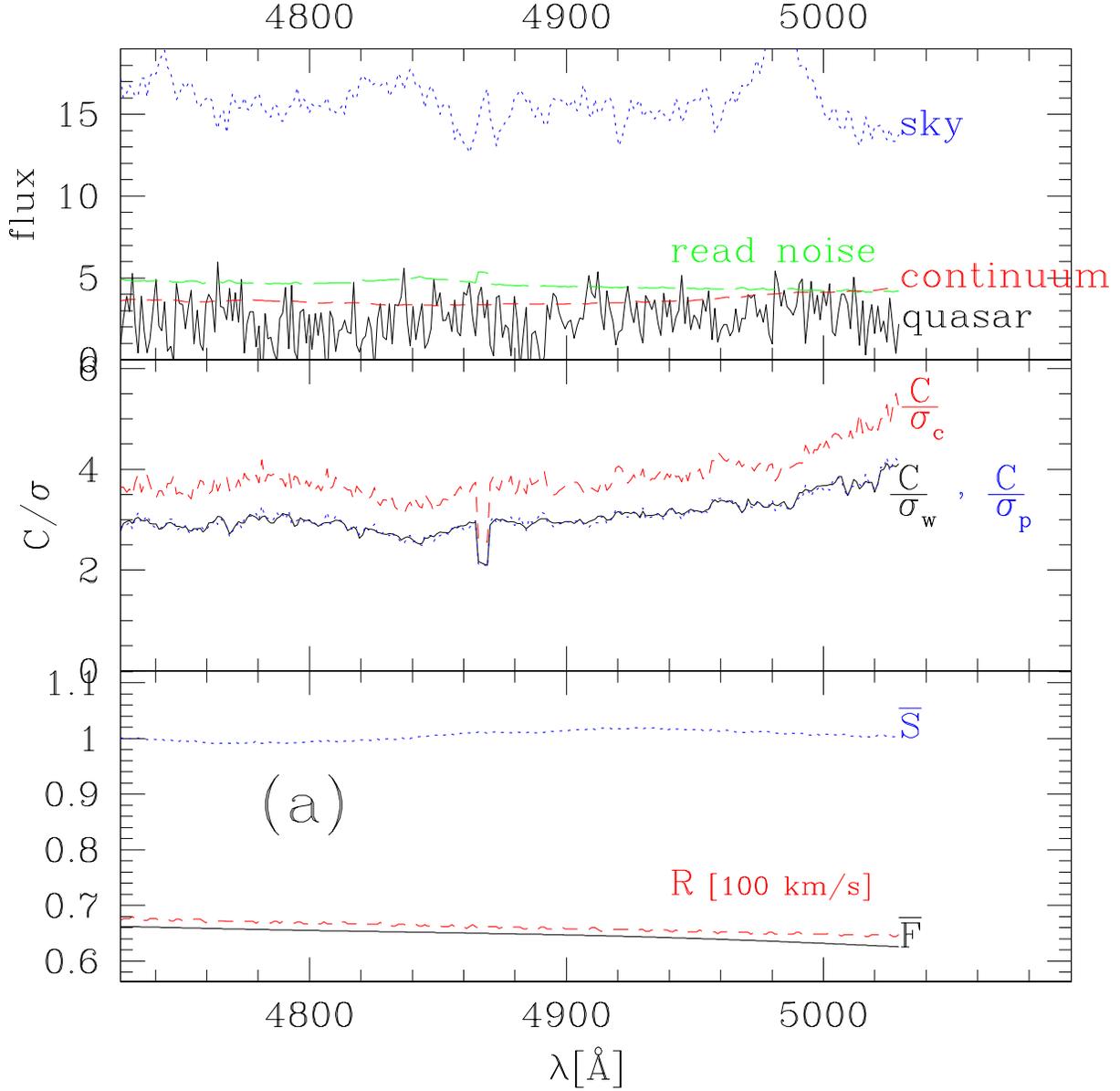}
\caption{
Examples of the chunks of spectra used to measure power, with
(a,b) showing quasars at $z_q=$(3.24,2.45) over the rest wavelength
range $1113~{\rm \AA}<\lr<1185$ \AA, and (c) showing a quasar at 
$z_q=3.30$ over the rest wavelength range $1041~{\rm \AA}<\lr< 1113$ \AA.
Top panel: quasar flux (solid black line), sky flux 
(dotted blue line), our continuum estimate (red short-dashed line),
and the read-noise as an equivalent photon flux (green long-dashed line).  
Middle panel:  S/N level shown as a ratio of our continuum to
the different rms noise levels (see text), 
$\sigma_w$ (black solid line), $\sigma_p$, (blue 
dotted line), and $\sigma_c$ (red dashed line).
Bottom panel:  Calibration correction vector, $\bS$ (blue dotted line),
rms resolution in units of 100 km/s (red dashed line), and evolution of
the mean transmission fraction, $\bF(z)$ (black solid line).
The perfect degeneracy in our analysis between the overall 
normalization of the
continuum and $\bF(z)$ has been broken arbitrarily, so only the evolution 
of $\bF(z)$ is meaningful (see text).
}
\label{deltaFvector}
\end{figure}

SDSS obtains multiple (at least 3) exposures for each quasar.
We combine the individual exposure spectra to produce a single
spectrum, using a nearest-grid-point method that produces 
uncorrelated noise and a reasonably well-defined sampling window.  
For each pixel we record estimates of wavelength, quasar flux,
resolution, sky flux, read-noise, and two different total noise
estimates.
The first noise estimate, which we will call simply $\sigma_p$ 
($p$ for pipeline), is computed
using the error array given for the exposure spectra 
by the spectral reduction pipeline.
The second noise estimate, which we call $\sigma_c$
($c$ for component), 
is computed by summing the read-noise and the noise 
implied by estimates of the number of photons corresponding 
to the quasar flux and sky flux.
The two noise estimates generally do not agree, but this is not
a problem for us because we ultimately recalibrate the noise
(next). 
Finally, we record $\chi^2/\nu$ for each pixel, computed by treating the
determination of the combined flux value for each pixel as a one parameter
fit to the measurements given by the different exposures.
Examples of the more important of these quantities in Ly$\alpha$ forest 
regions are shown in Figure \ref{deltaFvector}.
\begin{figure}
\plotone{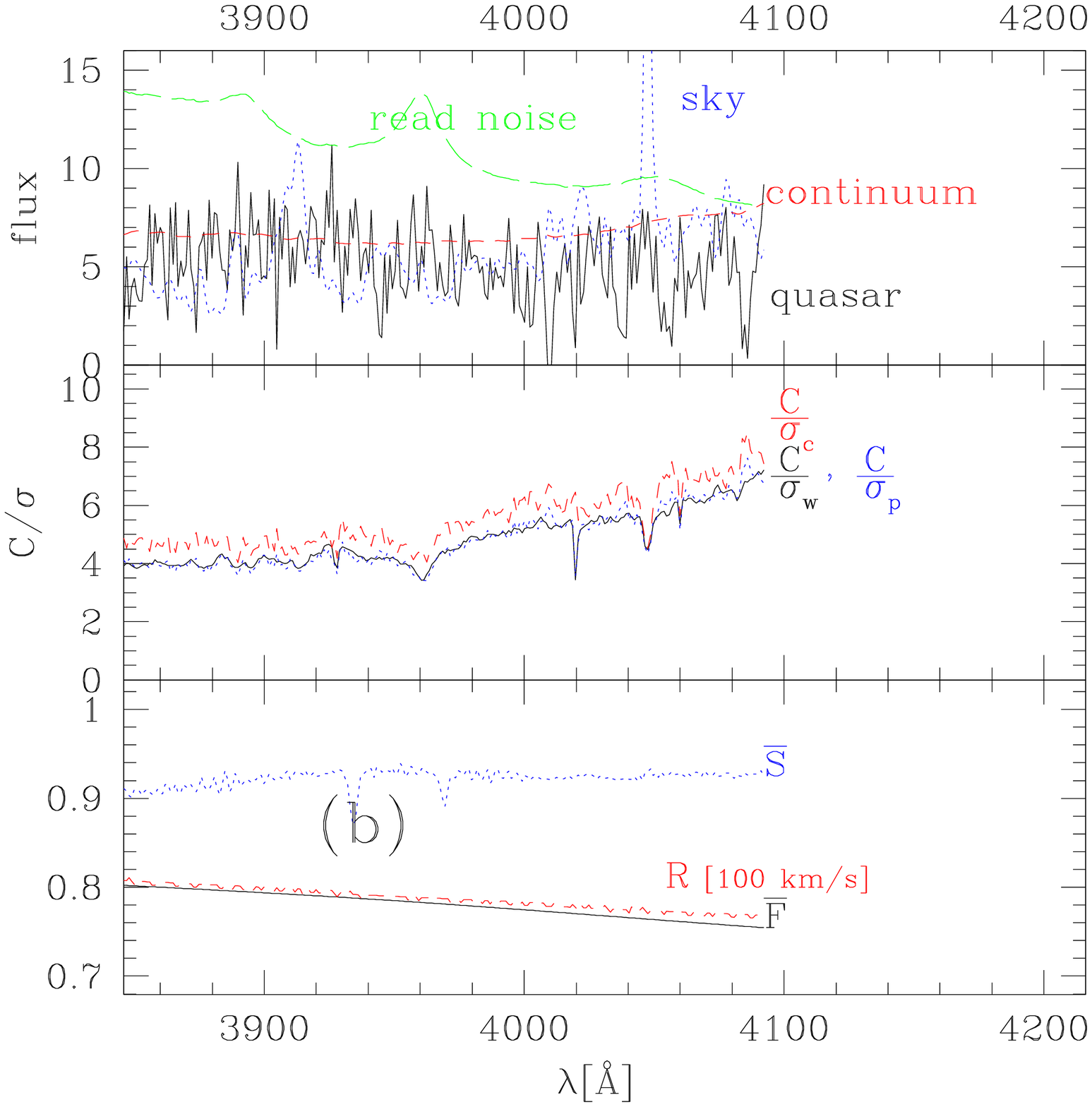}
\end{figure}
\begin{figure}
\plotone{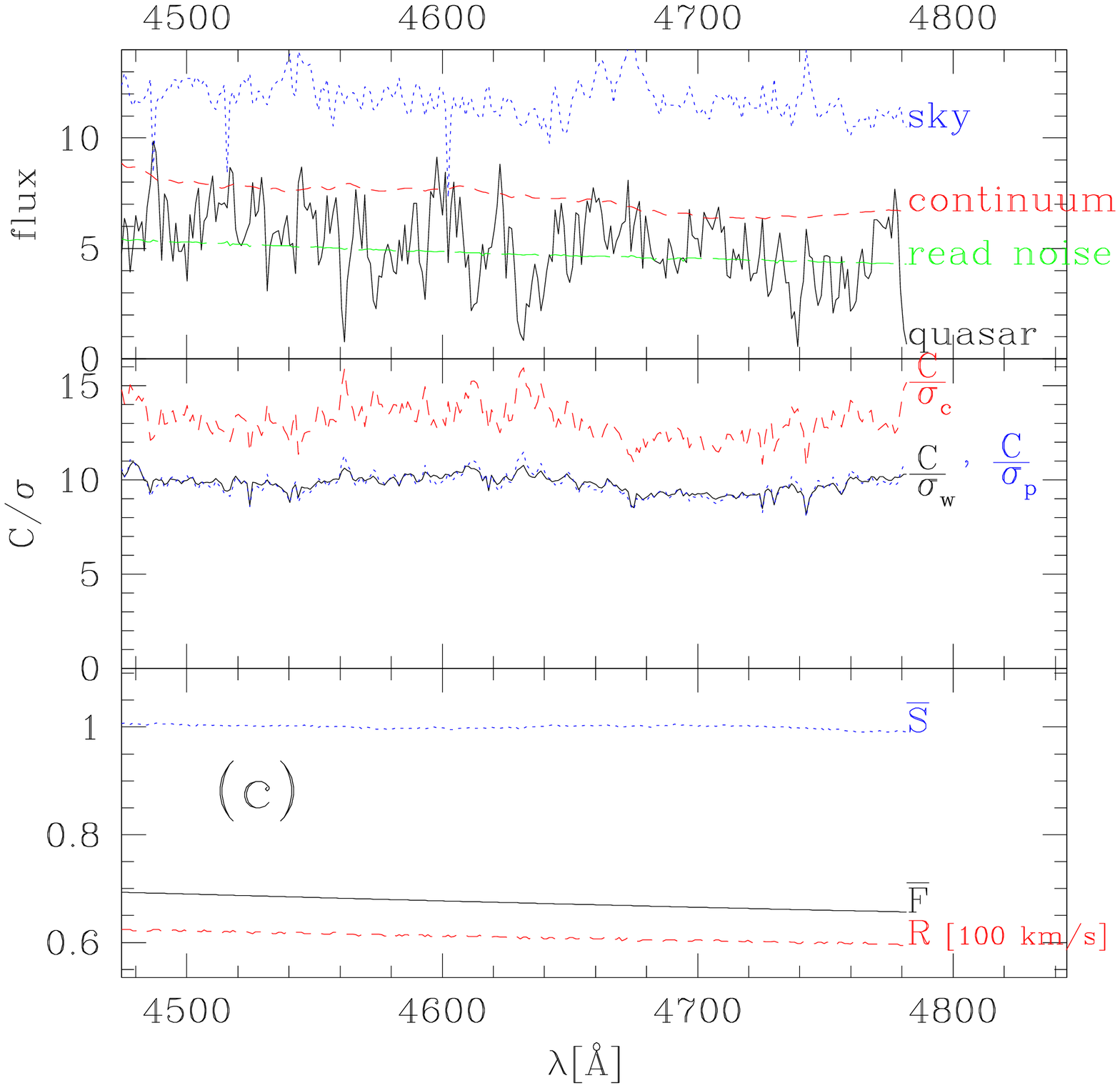}
\end{figure}
For comparison 
to the sky and quasar flux levels, we have converted the
Gaussian read-noise into the flux of photons that would contribute
the same noise variance.  Several elements of Figure \ref{deltaFvector}
(e.g., the estimation of the quasar continuum and $\sigma_w$) 
will be described later in this paper. 

\begin{figure}
\plotone{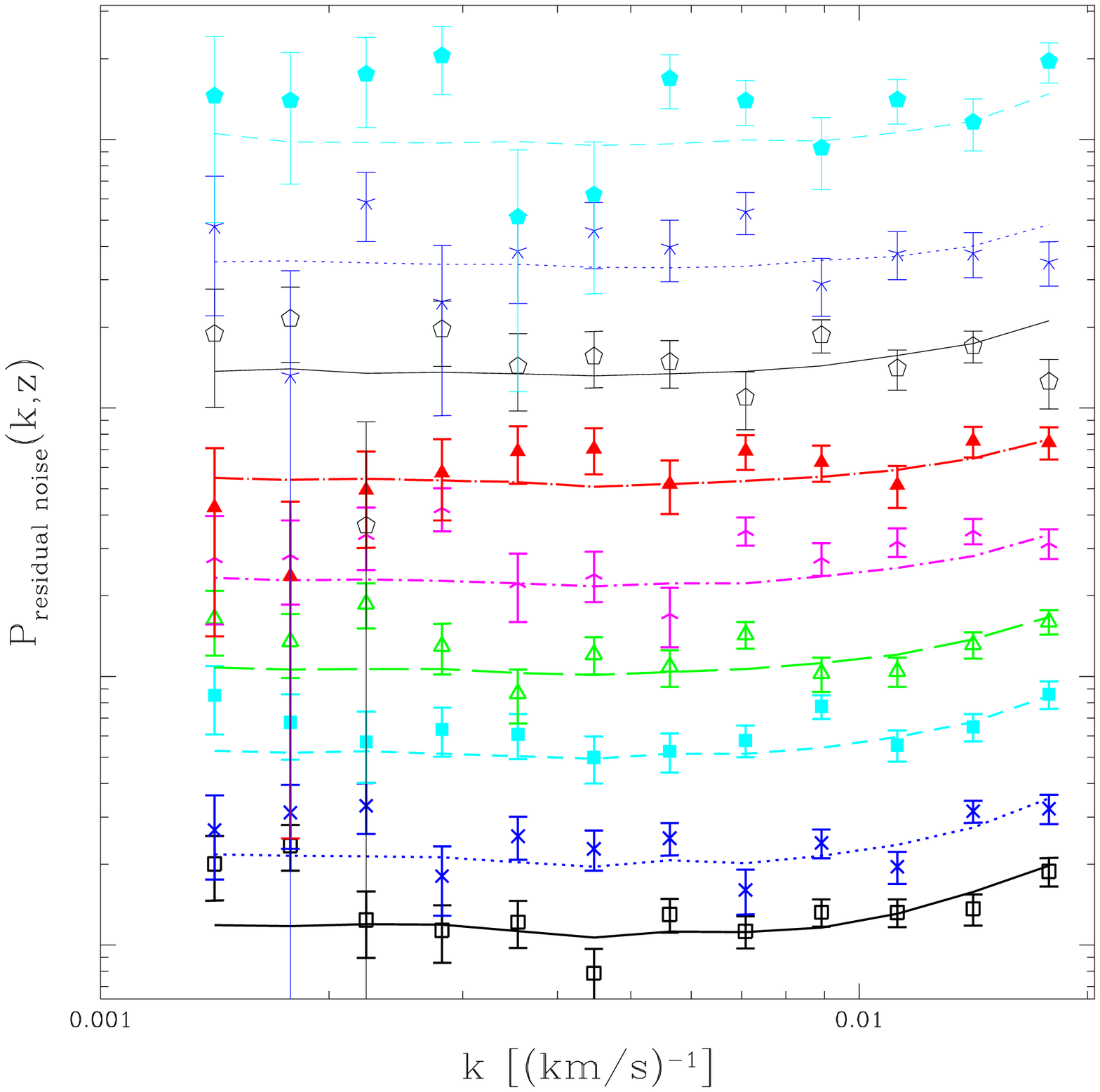}
\caption{The points 
show the measured power in difference spectra,
created by subtracting separate exposures for the same quasar.  
Noise power has been subtracted based on the standard pipeline 
noise estimates for each exposure.  
The lines show 16\% of the subtracted noise term.  
The different colors, lines, and symbols identify redshift bins, 
in a pattern that we will use repeatedly throughout the paper.
From bottom to top ---
z=2.2:  black, solid line, open square; 
z=2.4:  blue, dotted line, 4-point star (cross); 
z=2.6:  cyan, dashed line, filled square; 
z=2.8:  green, long-dashed line, open triangle; 
z=3.0:  magenta, dot-dashed line, 3-point star; 
z=3.2:  red, dot-long-dashed line, filled triangle; 
z=3.4:  black, thin solid line, open pentagon; 
z=3.6:  blue, thin dotted line, 5-point star; 
z=3.8:  cyan, thin dashed line, filled pentagon.
The different redshifts have
been shifted vertically by arbitrary amounts on this 
logarithmic plot. 
}
\label{diffspecpow}
\end{figure}

The noise estimate from the standard SDSS 
pipeline is only approximate.  The accuracy of the noise estimate 
required for our purpose is much higher than anticipated 
when the pipeline was developed.  For this reason 
we use the differences between single-exposure spectra for
the same quasar to determine the noise properties of the
data.  We construct difference spectra by combining the flux-calibrated 
exposures with alternating sign for each exposure, i.e., we use exactly
the same procedure that we normally use to produce combined spectra
from the exposures,
except half of the exposures are subtracted instead of added, so the
mean result is zero
(we drop the last exposure when there are an odd number -- this is
not the most efficient method possible, but we do not need it to be).  
The result is a direct measure of the
exposure-to-exposure changes.  We measure the power spectrum of 
these difference spectra using the method described in \S 3.1, 
including noise subtraction based on the pipeline noise estimates for 
the pixels.  The result is shown in Figure \ref{diffspecpow} (points
with error bars).  
We obtain a clear detection of power, where there should be none
if the spectra differ only by the noise estimate from the pipeline 
which is being subtracted.
If we assume that the noise has been underestimated by a constant
factor, and fit for that factor using the error covariance matrix
estimated by bootstrap resampling, we find a decent
fit: $\chi^2=141.6$ for 107 degrees of freedom (formally, this
fit is not good because $\chi^2$ is unlikely to be this high by 
chance). 
This fit uses our usual points in $0.0013~\ikms<k<0.02~\ikms$.  
The best fit value of the excess noise contribution
is $16.1\pm0.4$\% of the original noise estimate, indicating that the 
rms noise was underestimated by 8\%.  The best fit and goodness
of fit do not change if we add points on larger scales.  The quality
of the fit begins to degrade as we add points with larger $k$, but the
best fit value changes by only 1\% (in power) out to the Nyquist 
frequency of the data.  
Of course, we have no reason to expect a single
redshift-independent factor to describe the relation between the 
true and pipeline noise, so the formally bad $\chi^2$ is not 
a fundamental problem.  We check for systematic change with 
redshift by allowing a power law dependence, 
$P_{\rm residual~noise} \propto [3.75/(1+z)]^d$, but
find no significant detection ($d=0.07\pm 0.20$).  Our final method
will effectively account for evolution anyway, as described below.

A $k$ dependence different than expected for white noise could be 
a problem for us, so we check for this by fitting for a power law
dependence, $P_{\rm residual~noise} \propto (k/k_p)^b$ [with
$k_p=0.0074~\ikms$], 
finding $b=-0.111\pm 0.025$, a significant detection ($\chi^2$ is now
a reasonable 123.3 for 106 dof).  Allowing a
running of the power law, 
$P_{\rm residual~noise} \propto (k/k_p)^{b+1/2~c~\ln(k/k_p)}$,
does not improve the fit ($c=-0.046\pm 0.066$).  
The slope we find corresponds to a $\sim 20$\% change in 16\% of
the noise power at the extremes of our $k$ range, i.e., only $\sim 3$\%
of the total noise power, which is a relatively small fraction of the
\lyaf\ power except at the highest $k$ (see Figure \ref{noisefraction}
below).  We henceforth assume that the extra noise is proportional
to $k^{-0.111}$ rather than white (this makes $<1$\% difference in
the final results except for the one highest $k$, lowest $z$ point
where the difference is 2\%).

How accurate is this noise estimate based on differences between 
exposures?
Our difference spectra will 
contain a component of the \lyaf\ power if the calibration between
exposures is not perfect.  
The power in this term would be suppressed relative
to the \lyaf\ power by the fractional calibration error squared,
so it would be very small unless the exposure-to-exposure calibration
errors were quite large.
The fact that a simple one parameter
extra-noise model fits reasonably well, in the face of
variation in redshift, 
noise amplitude, and $k$, argues against calibration errors being a
big problem.  
More convincingly, we measure nearly the same excess noise contribution 
($14.2\pm0.5$\%) and slope ($b=-0.135\pm 0.028$)
in the region $1268~{\rm \AA}<\lr<1380$ \AA\ as we do in the \lyaf.
This argues against any connection to leaking \lyaf\ power.
Note that the effective absolute level of noise in the 
$1268~{\rm \AA}<\lr<1380$ \AA\ region is about half that in the 
\lyaf\ region, so this test shows that the fraction of extra noise 
does not depend strongly on the noise level itself.

Pixels in different exposures are not perfectly aligned, 
and misalignment can allow \lyaf\ power to leak into our difference 
spectra.
To test this alternative explanation for the apparent 
excess noise in the spectra, we split the spectra into two groups
with approximately equal weight, based on the rms misalignment in
the forest region (the alignment is known from the wavelength calibration 
of the exposures, which is thought to be practically perfect).
We find the same excess noise power in both the poorly 
aligned group ($16.1\pm0.6$\%, $b=-0.086\pm0.036$) and the better aligned group
($15.3\pm 0.6$\%, $b=-0.123\pm0.036$), suggesting that the
excess power is not due to misalignment. 
Furthermore, the presence of a similar level of excess noise power 
outside of the forest region again argues against leakage.
We therefore believe that our noise estimate is considerably more 
accurate than the noise estimate from the SDSS pipeline.

In our initial power spectrum analysis we multiplied the noise-power
estimated from the pipeline errors by the factor 1.16 for all spectra;
however, when we split the data based on the mean value of $\chi^2/\nu$ 
for the exposure combination (see \S \ref{secsubsamples}) we found that 
the large and small $\chi^2/\nu$ subsamples disagreed significantly 
on the \PF\ results.  We  
eliminated this problem by estimating the noise-correction factor 
individually for each spectrum, by fitting to the power in the difference
spectrum for that quasar.
The mean extra power from these fits is still close to 16\%, but there is 
considerable scatter.  When we use these individual estimates, the
correlation between measured \PF\ and $\chi^2/\nu$ disappears, i.e,
the mean value of $\chi^2/\nu$ for a spectrum's exposure 
combination
was a good indicator of the amount by which the noise in each spectrum
was misestimated.  
Note that there are statistical errors in these noise estimates for each 
spectrum, of the same order as the error for which we are trying to 
correct; however,
there is no systematic bias associated with these errors, and the random
error they contribute is automatically included in our final bootstrap errors.
In fact, including the spectrum-by-spectrum noise estimate reduces the 
bootstrap errors slightly on small scales, verifying that these estimates 
are on average more accurate than the original noise estimates.
It is not known why the noise is misestimated
by the standard pipeline.  Tests at this level have not been done 
before.

Our final data product will be a measurement of \PF binned
in $k$ and $z$, i.e., a matrix $P_{F,ij}$ where $i$ labels bins
with $z_i$ and $j$ labels bins with $k_j$.  We will also 
give the noise power that was subtracted, $P_{N,ij}$, 
in the same bins.
We suggest allowing a 5\% rms freedom in the noise amplitude in each
$z$ bin when performing model fits, i.e., for each bin subtract
$f_i P_{N,ij}$ from $P_{F,ij}$, and add
$(f_i/0.05)^2$ to $\chi^2$.  This is probably overly conservative
for any one bin, but implies a combined freedom $\pm0.05/3$ (for 9 bins) 
on an overall noise misestimation. This seems prudent, even though it 
is not really required by any test we have performed. 

\subsection{Accuracy of the Resolution \label{secrescal}}

The resolution of the SDSS spectra is estimated using lines from 
calibration lamps mounted on the telescope structure.
Shifts of the detector pixel grid relative to a fixed observed wavelength
frame during an exposure, which we will call flexure, are expected to be 
the dominant source of error in this spectral resolution estimate.  
We tried estimating the rate of 
shifting for each pixel by differencing the wavelength calibrations of 
adjacent exposures (this calibration is
determined very precisely for
each exposure using the positions of sky lines).  
The implied extra smoothing only changes the power by $\sim 2$\% at our
highest $k$ bin.   

\begin{figure}
\plotone{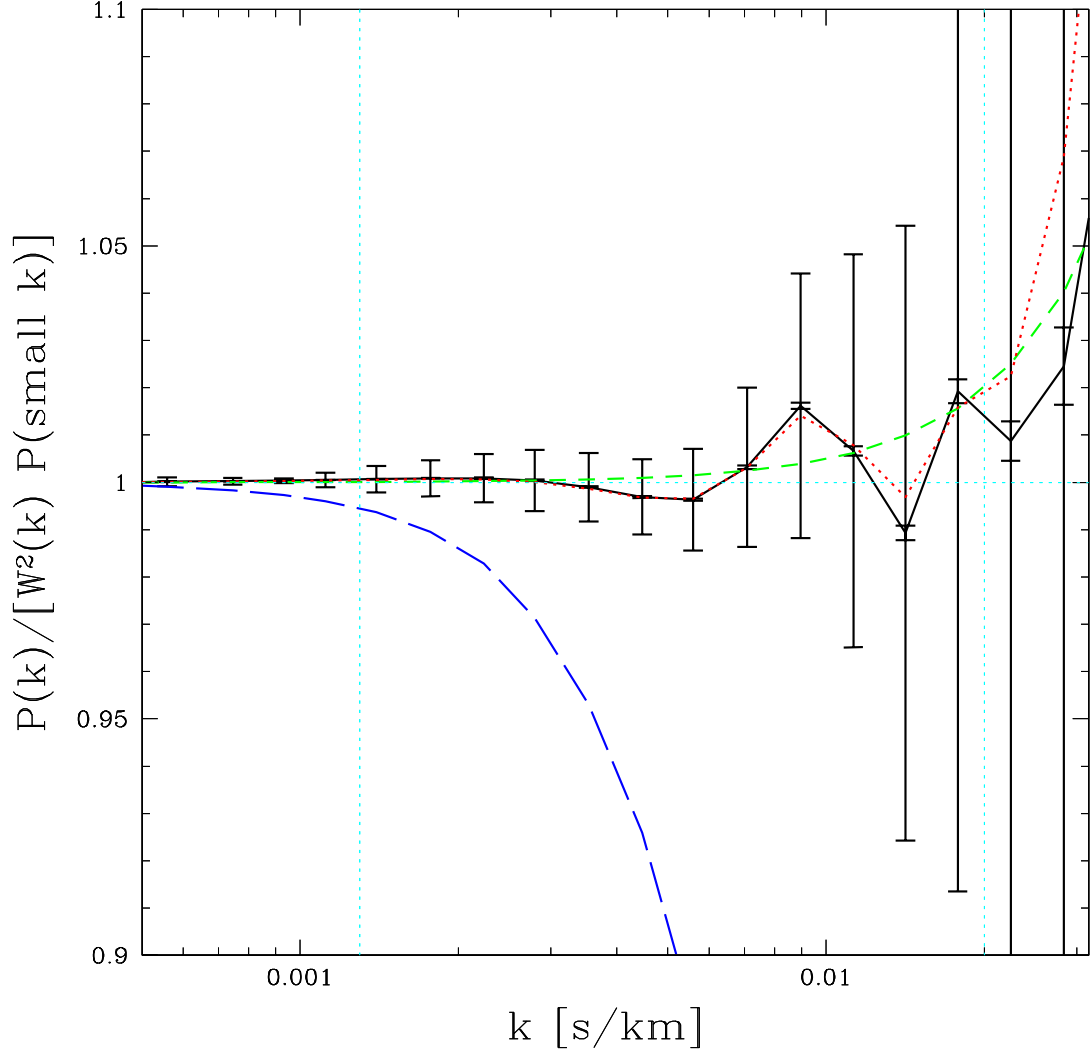}
\caption{
Resolution test.  The solid, black line with error bars shows the power 
measured
in $\sim 3000$ sky spectra in the range $5560{\rm \AA}< \lambda<5598$\AA\ 
(dominated by the
strong sky line at 5577\AA) divided by the asymptotic small $k$ power 
and by the estimated resolution/pixelization kernel $W^2(k)$ for each
spectrum.  
If the resolution estimate 
was perfect, and the sky line was narrow and the only flux present, this 
division
would give exactly 1.  The large error bars are the spectrum-to-spectrum 
variation,
the small ones are the error on the mean.  The blue, long-dashed line shows 
the power
not divided by $W^2(k)$, i.e., basically an averaged version of $W^2(k)$, 
which 
drops to $\sim 0.25$ by $k=0.02\ikms$.  The red, dotted line shows the result
of our test 
for mock spectra constructed with a Gaussian at 5579\AA\ and two more 
at 5566 and 5591 \AA\ with 0.003 times its amplitude, representing OH lines.
The green, short-dashed line shows $\exp[(k~7\kms)^2]$.  
The vertical, cyan, dotted lines 
bound the $k$ region in which we will present \lyaf\ results, while the 
horizontal, cyan, dotted line just guides the eye to 1.
}
\label{skypower5579}
\end{figure}

\begin{figure}
\plotone{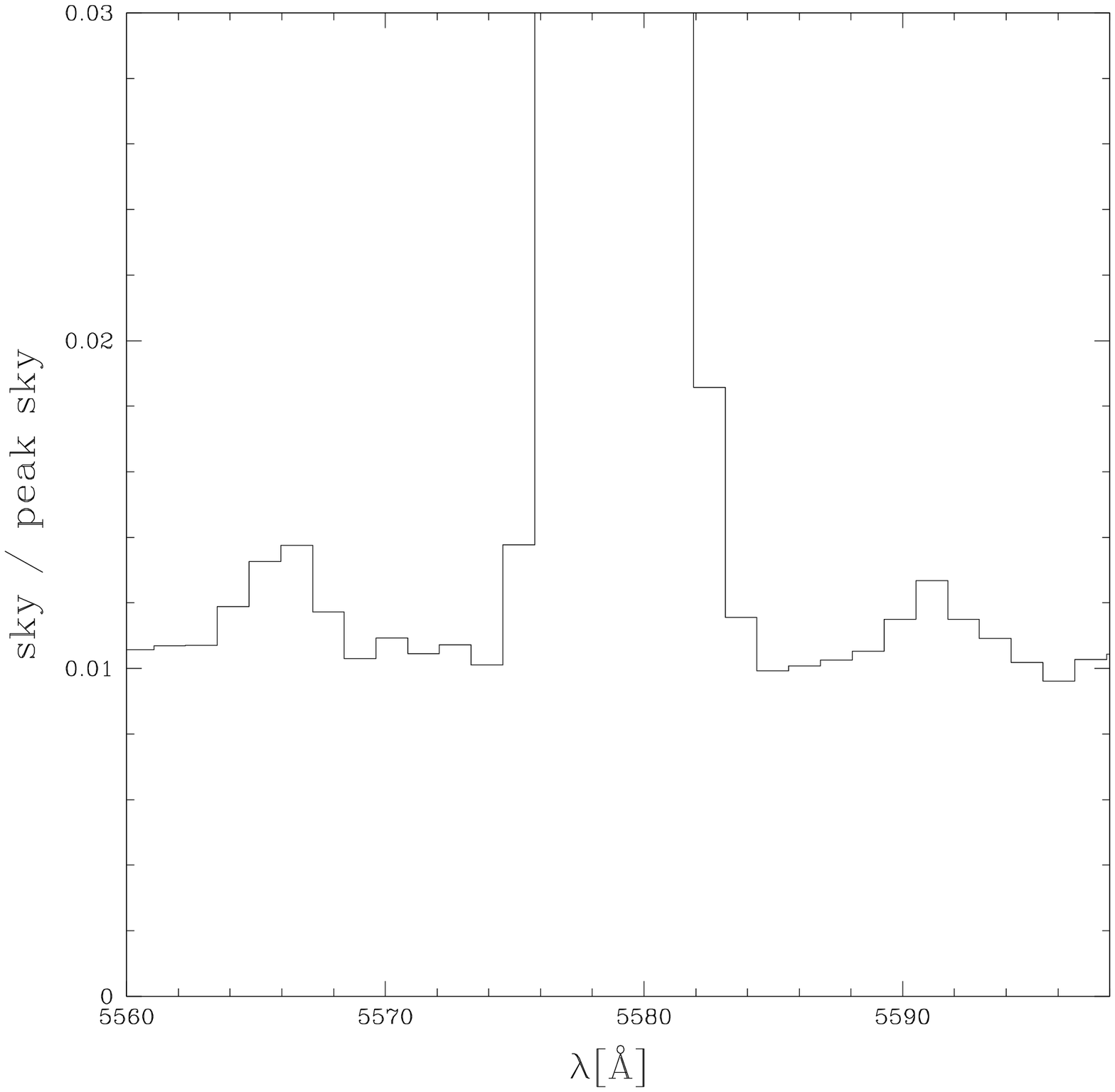}
\caption{
Example of the sky flux near the sky line at vacuum wavelength 5579\AA, 
relative to the peak of the line (one spectrum only). 
}
\label{zoom5579}
\end{figure}

The strong sky line at 5577 \AA\ 
provides a good opportunity
to measure the resolution more directly
(note that the spectral wavelengths 
are in vacuum, and heliocentric, so this and other sky lines 
generally appear shifted from their standard wavelength).  
We measure the power spectrum in $\sim 3000$ sky spectra 
in the range $5560{\rm \AA}< \lambda<5598$\AA.
If the sky line has negligible width
and the smoothing has a Gaussian shape with rms width $R$, the power spectrum
should be proportional to $W^2(k,R,l)=\exp[-(k R)^2] [\sin(k~l/2)/(k~l/2)]^2$,
where $l$ is the pixel width
(the pixelization effect is subdominant but not negligible).  
In Figure \ref{skypower5579}
we show the measured power averaged over all the sky spectra after dividing
each individual measurement by $W^2(k,R,l)$, where $R$ and $l$ are the local
values (they are to a good approximation constant over the 
range we are looking at), and also dividing each measurement by the value
at a low $k$ where the resolution should not have any effect.
The result is remarkably close to unity, indicating that the estimated 
resolution 
is an accurate representation of the true resolution.  
What are the small wiggles? 
Figure \ref{zoom5579}
shows an example of the region we Fourier transform to measure the power.
We believe the small features to the sides of the main line are 
OH lines at 5564 and 5589 \AA\ \citep{2003PASP..115..869S}.
We test this explanation for the wiggles by constructing mock 
sky spectra that simply have a 
delta function at 5579\AA\ and two more with 0.003 times the
main line's amplitude at 5566 and 5591 \AA, convolved with the 
resolution and pixelization (0.003 was chosen to give the
best fit to the wiggles).
The red, dotted line in Figure \ref{skypower5579} shows the 
same resolution test using the mocks.  We see that the wiggles
are essentially perfectly reproduced.
In conclusion:  the resolution profile appears to be perfectly
Gaussian, with 
exactly the width expected from the given resolution.  
There is apparently no room for even
a 2\% level effect from flexure.  
We are prevented from performing the same kind of
measurement using other sky lines by similar features which are always much
larger relative to the central line.

We suggest that fits to \PF\ include a multiplicative uncertainty on the
overall power, of the form $\exp(\alpha k^2)$, where $\alpha$ is a 
single parameter in the fit subject to the rms constraint 
$\sigma_\alpha = (7\kms)^2$.  This allows for a $\sim 2$\% change in 
the smoothing kernel at our highest $k$, similar to our estimate
of the error from flexure.  This error estimate is somewhat
arbitrary, but the evidence we have presented suggests that it should
be smaller, so our estimate is conservative.

Note that this resolution test, and the noise calibration, cannot be used
directly with the standard pipeline spectra, where the exposures are 
combined in a different way.
The reader may be confused at this point about how our spectra differ from
the standard publicly available set, so we give the following summary:
\begin{itemize}
\item Our nearest-grid-point combination of the exposures produces 
uncorrelated noise in pixels (to the extent that the noise in the 
exposures was uncorrelated, which is expected from the way they 
are extracted), while the standard pipeline uses a local splining
procedure which does a good but not perfect job of preventing 
noise correlation. 
\item When combining exposures we record the effect of different pixel sizes,
misalignment of the pixels, and flexure of the detector during 
exposures, which can influence the effective resolution.
\item We record the contribution of quasar flux, sky flux, and read-noise
to the total noise in each pixel.  Knowing the contribution from quasar 
flux is important if pixel-by-pixel noise weighting is to be used, because
the correlation between flux level and noise amplitude can lead to biases
(see the end of \S\ref{secmeandiv}).
\item We correct for the bias in the exposure combination associated 
with cross-correlation between the noise variance level in exposure pixels 
and the quasar flux in the pixel (a different incarnation of the problem
alluded to in the previous point). 
\item The noise is recalibrated for each spectrum by differencing the 
exposures.  The noise variance in the standard pipeline exposure
spectra is underestimated by
on average 16\%, on top of any error related to the exposure 
combination, and the power measured
in the difference spectra is slightly tilted relative to white noise.  
\end{itemize}
The last point is the most important.  

\subsection{Determination of the Continuum and Mean Absorption Level 
\label{secmeandiv}}

Our goal is to measure the power spectrum of the fluctuations in 
the transmitted flux fraction through the IGM, 
$\delta_F(\lambda)=F(\lambda)/\bar{F}-1$, where 
$F(\lambda)=\exp[-\tau(\lambda)]$ 
and $\tau(\lambda)$ is the \lyaf\ optical depth (as defined in \S 1). 
However, the spectrum of 
each quasar is the product of $F$ and the quasar continuum (note that
we use ``continuum'' to refer to all the flux emitted by the quasar, 
including emission lines), further
complicated by errors in the detector calibration and absorption by
longer wavelength transitions.  
The details of the procedure we use to separate these contributions 
will be presented elsewhere, here we give the basic idea and key 
results that are relevant for the flux power spectrum determination. 

We use an iterative procedure to determine 
the components of the data model 
\begin{equation}
f^i=A_q~ \bC(\lr^i)~(1+\delta_C^i)~ 
\bS(\lambda^i)~(1+\delta_S^i)~ \bF(z^i)~ (1+\delta_F^i)+n^i~,
\label{datamodel}
\end{equation}
where $f^i$ is the raw flux in pixel $i$, $n^i$ is the noise, $A_q$ 
is the overall normalization of the quasar spectrum,
$\bC(\lr)$ is the mean quasar continuum shape, $\delta_C$ are fluctuations
around the mean continuum, $\bS(\lambda)$ is a mean generalized
calibration vector (this includes wavelength dependent
calibration errors in the SDSS spectra, but also the mean absorption 
by metal
lines with resonance wavelength $\lambda\gtrsim 1300$\AA), $\delta_S$ are
fluctuations around $\bS$, such as individual metal lines or
variable calibration errors, $\bF(z)$ is the mean \lyaf\ absorption 
at a given redshift,
and $\delta_F$ are the fluctuations in \lyaf\ absorption.
Note that here, as most places in this paper, 
$z^i=\lambda^i/\lambda_\alpha-1$ is the redshift of
gas that would produce \lya\ absorption in the pixel, not the redshift
of the quasar.
Briefly, we determine $A_q$ for each spectrum, the global functions 
$\bC(\lr)$, $\bS(\lambda)$, and $\bF(z)$, and a set of 
principal component analysis (PCA) eigenvectors
that describe $\delta_C$ by, in turn, computing each component of
the model from
all the spectra while holding all the others fixed. 
E.g., $\bF(z)$ is estimated in bins of $z$ by averaging 
$f^i / A_q~\bC(\lr^i)~\bS(\lambda^i)~(1+\delta_C^i)$ over all the 
pixels in the \lyaf\ that fall in each bin.  We separate 
$\bS(\lambda)$ from $\bF(z)$ by measuring $\bS(\lambda)$ 
in the rest 
wavelength range $1268~{\rm \AA}<\lr<1380$ \AA, i.e., outside the
\lyaf.  A few iterations suffice to determine
all the components of the model independently.  
Three examples of the results are shown in Figure \ref{deltaFvector}.
The full details of this 
procedure will be presented in a separate paper focused on a 
precise determination of $\bF(z)$.

In preparation for measuring the power spectrum, we divide each 
quasar spectrum by our estimate of 
$A_q~ \bC(\lr)~ (1+\delta_C)~ \bS(\lambda)~ \bF(z)$.
The power in the $\bS(\lambda)$ and $\bF(z)$ terms is completely negligible
(always $<0.5$\% of \PF\ and usually much less). 
$\delta_C$ are
represented for each quasar by N PCA eigenvectors.  We have tried several 
different values for N ranging from 0 to 13, and find that the power 
spectrum results
depend slightly (but not critically) on the value, as we discuss below. 
For our final results we use $N=0$, i.e., we only divide by a mean 
continuum, although we will also show how using $N=13$ affects the 
cosmological fit results.  We do 
not use the PCA continua because we are not satisfied with their 
robustness, and division by them frequently actually
increases the resulting power slightly.  
This may indicate that the error introduced
by allowing additional freedom in the continuum is larger than the 
continuum fluctuations that we are trying to remove.  
Our study of 
PCA continua was primarily aimed at determining $\bF(z)$ rather than 
the power spectrum, so we cannot be certain that the PCA method could 
not be used productively in a power spectrum measurement if it was 
more carefully optimized for that application.
Because we know that our continuum estimate (which involves
an extrapolation from outside the \lyaf\ region) 
is not perfect 
within the \lyaf, we further divide each chunk of spectrum that will 
be used to make a power spectrum estimate by its mean
(optimally computed considering both observational noise and 
absorption variance).  
We call our resulting observed data vector $\vdf=\vdF+\vdS+\vdn$, where
$\vdn$ is the normalized noise fluctuation  and we are ignoring the cross-terms 
between $\vdF$, $\vdC$, and $\vdS$.
As we describe in detail 
below, $\delta_S$ is treated as a 
random noise background and its statistical properties are determined 
by measuring the power spectrum
in the rest wavelength range $1268<\lr<1380$\AA, where $\delta_F\equiv 0$
(and $\bF\equiv 1$). 

A small but non-negligible detail of our procedure is hidden within
our description of the normalization of the spectra.  When we estimate the
mean to divide by, we weight the computation optimally using the covariance 
matrix, $C_{ij}$, of the pixels ($C$ is discussed in more detail in our 
explanation of the 
power spectrum measurement below).  $C$ includes \lyaf\ 
fluctuations and measurement noise.  We do not use our best estimate of 
the measurement noise directly for the weighting, because the noise variance 
estimate is correlated with the measured flux in the pixel, which leads
to a bias: the mean is underestimated because lower flux pixels have lower 
noise.  The original noise estimate is 
$\sigma_p^2 = \gamma~(f_{quasar}+f_{sky})+ \sigma^2_{read noise}$,
where $f_{quasar}$ is the flux from the quasar, $f_{sky}$ is the flux from
the sky, and $\gamma$ accounts for the conversion between the units of 
flux and photons (this description is slightly idealized since the 
reduction of 
two-dimensional CCD data to a spectrum introduces some complications).  
To remove the correlation between flux and noise, we
subtract $\gamma f_{quasar}$ from $\sigma_p^2$ and add 
$\gamma \left<f_{quasar}\right>$, where 
$\left<f_{quasar}\right>=
A_q~ \bC(\lr)~ (1+\delta_C)~ \bS(\lambda)~ \bF(z)$. 
We call the final result $\sigma_w$ ($w$ for weight; see Figure \ref{deltaFvector}
for some comparisons of the noise estimates). 
The estimate of $\gamma$ we have from the spectral reduction
pipeline is not perfect, so our replacement of the correlated part of the
noise amplitude is imperfect.  We make a final, very small, correction to the
mean estimation based on a direct computation of the cross-correlation 
between the flux and noise amplitude.  We use the same decorrelated noise 
amplitudes for weighting the power spectrum extraction (discussed below); 
however, the bias is completely insignificant in that case (i.e., \PF\ 
computed using the original noise estimates for weighting is practically 
identical).

\section{Power Spectrum Determination \label{secpowerextract}}

The high precision of the \PF\ measurement obtainable 
using the SDSS data sample requires unprecedented (in  this field) 
care in the design
and testing of the procedure used to produce it.  
We describe the basic method that we use to extract a power spectrum and 
estimate the errors in \S \ref{secextractmethod}. 
In \S \ref{sectestfake} we present
the test of the full method as implemented in our code, 
using mock data sets.  
In \S \ref{secrawpower} we give the raw result
for the measurement of power in the \lyaf\ region.
 
We aim to measure the power spectrum of $\delta_F$, representing
the correlation of fluctuations in the \lyaf\ absorption only; however,
the covariance matrix of the data vector $\vdf$ is
\begin{equation}
\left<\vdf \vdf^T\right>=\left<\vdF \vdF^T\right>+
\left<\vdS \vdS^T\right>+
\left<\vdn \vdn^T\right>
\label{expectdeltaf}
\end{equation}
(the three components of $\vdf$ as we have defined it should
be uncorrelated).  
The noise term in equation (\ref{expectdeltaf}) is relatively 
easy to compute and subtract.  We estimate and
subtract most of $\left<\vdS \vdS^T\right>$
by defining 
\begin{equation}
P_F(k,z) = P_{1041,1185}(k,z)-P_{1268,1380}(k,z)~,
\label{backsubeq}
\end{equation}
where $z$ is always defined by $z=\lambda/\lambda_\alpha-1$, 
so that we are subtracting power measured in the same observed
wavelength ranges, not the same quasar spectrum 
(we remind the
reader that we have defined $P_{\lambda_1,\lambda_2}$ to 
mean the power measured in the range $\lambda_1 <\lr <\lambda_2$).
As it appears in $P_{1268,1380}(k,z)$, $z$ is the redshift of gas 
that would produce \lya\ absorption in this part of the quasar
spectrum, if it was not at a 
higher redshift than the quasar, i.e., $z$ is really just an 
indicator of observed wavelength.  
The subtraction in equation (\ref{backsubeq})
will completely remove the power due to transitions with
$\lambda>1380$ \AA, including SiIV (a doublet absorbing
at rest wavelengths 1393.75 and 1402.77 \AA) 
and CIV (another doublet at 1548.20 and 1550.78 \AA).
Note that this subtraction of metal power is exact, not an approximation
[except for the approximation that 
$(1+\delta_F)(1+\delta_S) \simeq 1+\delta_F+\delta_S$], 
because we are determining the metal power in 
exactly the same observed wavelength range as the \lyaf\ power from
which it is being subtracted, i.e., the same gas, at the same redshift,
is doing the absorbing both inside and outside the forest, so the 
absorption will have identical statistical properties.
This background subtraction will also remove any 
strictly observed-wavelength-dependent power introduced by the detector, 
such as spectrum to spectrum variations in the calibration of the detector.
We implement it in \S \ref{secbacksubtract}.

\subsection{Core Method \label{secextractmethod}}

In this subsection we describe our method for extracting the 
power spectrum, \PX, from any selected rest wavelength range $X$
(\S \ref{secbandpowerest}), and estimating its statistical uncertainty
(\S \ref{secbootstrap}).   

\subsubsection{Band-Power Estimation \label{secbandpowerest}}

We estimate $P_X(k)$ using the 
quadratic estimation method, which is essentially 
a fast iterative implementation 
of the maximum likelihood estimator (we follow the expressions 
as given in \cite{1998ApJ...506...64S}).
This method is optimal for a Gaussian probability distribution. 
While the power spectrum estimates are not Gaussian distributed, 
the deviations are small, as shown below. 
We measure the power in flat bands with edges given
by $\log_{10}(k_i)=-4.2+0.1~i$ where $i$ ranges from 0 to 30 (to 
produce 29 bands), although we will not give results for some of 
the large and small-scale bands when we think they are unreliable.
Defining $\vdf=\vdX+\vdn$, where $\vdX$ are the 
fluctuations we are measuring (e.g., $\vdX=\vdF+\vdS$ within the forest) 
and $\vdn$ are the normalized noise fluctuations, 
a band-power estimate, $\hat{P}_k$, for each chunk of spectrum is 
given by 
\begin{equation}
\hat{P}_k = \frac{1}{2} \sum_{k'} F^{-1}_{k k'} (\vdf^T
C^{-1} Q_{k'} C^{-1} \vdf -b_{k'})~,
\label{Pestimator}
\end{equation}
where $C=\left<\vdf \vdf^T\right>=S+N$,
$S=\left<\vdX \vdX^T\right>$, $N=\left<\vdn \vdn^T\right>$,
$Q_k=\partial S/\partial P_k$,
\begin{equation}
F_{k k'}=\frac{1}{2}{\rm tr}(C^{-1} Q_k C^{-1} Q_{k'}) ~ 
\label{fishmat}
\end{equation} 
is the Fisher matrix and the noise bias is
\begin{equation}
b_k={\rm tr}(C^{-1}Q_k C^{-1} N)~.
\end{equation}
Note that we could include the background power explicitly in these 
equations as a noise source when measuring the power in the \lyaf\ region,
but we ignore this because its contribution is too small to 
change the weighting significantly.  We will subtract it from the 
estimates later.  The noise subtraction term, $b_k$, is computed using
the pipeline noise estimates, $\sigma_p$ (not $\sigma_w$), with the
amplitude corrected as discussed above based on the differences between
exposures.
In principle, $S$ in these equations should be the true covariance 
matrix; however, as we discuss below, we use the measured covariance
from a previous iteration of the power spectrum determination instead. 

Except in a few cases that we will identify as they arise, when we 
set out to measure the power in a defined 
rest-wavelength region (e.g., $1041 < \lr < 1185$ \AA\ for the \lyaf\ 
region) we first use equation (\ref{Pestimator}) to estimate the 
power separately in halves of the region in 
each spectrum (e.g., $1041<\lr<1113$ \AA\ and $1113<\lr<1185$ \AA).  
Our choice of half-spectra is a compromise between
competing desires for resolution in redshift and wavenumber.
The full length of the forest in a spectrum corresponds to a redshift 
interval, $\Delta z \simeq 0.4$, that is unnecessarily large.  While
the precision of the measured power spectrum would support smaller than 
half-spectrum chunks to give finer redshift resolution 
than $\Delta z \simeq 0.2$, the shorter chunks would limit the $k$-space 
resolution.  Note that we could have used full chunks and still achieved 
the same $z$-resolution by more carefully applying the estimator equation,
as we discuss below, but this would increase the computational time 
without much improvement in the final errors on the scales of relevance.

After computing estimates $\hat{P}_k$ for each half-spectrum, we perform
a weighted average to determine \PX\ in 
redshift bins centered at $z_i=2+0.2~i$ where $i=1..13$ (in this paper we
only present results up to $i=9$).
Each bin is the average of
the power in all the half-spectra for which the redshift of the 
central pixel falls within $\pm 0.1$ of the bin center (we discuss
below how we correct for an asymmetric distribution of data within
a bin).   
We combine sets of estimates using the Fisher matrices (equation
\ref{fishmat}) for the weighting.  In practice this means that we sum the 
quantity in parentheses in equation (\ref{Pestimator}) over all estimates 
and multiply
the result by the inverse of the sum of the Fisher matrices for each 
individual estimate.
Our procedure would be optimal for Gaussian data, which the
\lyaf\ is not; however, when we use the Gaussian approximation   
to compute the errors on the measured power 
the results are not much different
from the more accurate bootstrap errors (see \S \ref{secbootstrap}), 
so we conclude that our method is not far from optimal.  

Whenever we have a finite length of spectrum, there will be mixing 
between the power in different bins. 
Variable noise or gaps in the data will produce more mixing.
This mixing is described in terms of a window matrix, which is given by the 
Fisher matrix in equation \ref{fishmat}.
In our standard procedure,
the power spectrum estimates in equation \ref{Pestimator} are multiplied with 
the inverse of Fisher matrix and are thus deconvolved with the 
window, which removes the mixing of other modes into the bin one 
is estimating (however, the bins are still correlated). 
This method thus produces a diagonal window matrix, so each combined
estimate of \PX\ represents
exactly the range of $k$ corresponding to its bin.  
Our tests below show
that there is no practical problem with instability in the Fisher 
matrix inversion (the window matrices are close to diagonal to begin 
with).  A diagonal window matrix is desirable from a theoretical standpoint 
because our ability to compute the power spectrum from simulations is 
limited at both low $k$ (by limited box size) and high $k$ (by simulation
resolution and complexity of physics).  
In the few cases where we use the power without deconvolution, we are using
the estimator ${\mathcal N}_k~({\mathbf F}~{\mathbf P}_X)_k$, where 
${\mathcal N}_k\equiv\left(\sum_{k^\prime}F_{k k^\prime}\right)^{-1}$ 
\citep{1998ApJ...506...64S}.
 
To compute the weight matrix $C$, we need an estimate of $S$, i.e., the
power spectrum we are trying to measure.  We solve this problem by computing 
\PX\ iteratively.  The first estimate is made assuming $S=0$.  In 
subsequent iterations we compute $S$ from the previous estimate of \PX.
This procedure converges quickly (the difference between
$S=0$ and a reasonable estimate of $S$ is significant, but once $S$ is in 
the right ballpark it does not matter what it is exactly). 
We add a large constant (10.0) to all elements of the weight matrix, to 
remove all direct sensitivity of our power measurement to the mean of 
the chunk. This makes very little difference to the results on the 
scales we present. We are however still sensitive to the mean 
estimate from when we divided the spectrum by it.  Even if the mean 
estimates
are correct on average, the statistical error on the mean for each 
spectrum can still lead to a bias.  If the errors on the mean estimate
were small and uncorrelated with the fluctuations in the flux field, 
the bias would be $1+3~\sigma_m^2$,
where $\sigma_m$ is the error on the mean
[to lowest order in $\sigma_m$, i.e.,
the bias is $\left<1/(1+\delta_m)^2\right>\simeq 
1+3\left<\delta_m^2\right>$,
where $\delta_m$ is the fractional error in the mean, and 
$\left<\delta_m^2\right>=\sigma^2_m$].
We divide each estimate by 
this factor as part of our standard procedure; however, as we discuss 
below when we test our code on mock 
spectra, this approximation is not sufficient and we will need to 
include another small, $k$-dependent, factor determined numerically 
using the mock spectra (this is the only use of the mock spectra
other than for testing). 

The reader may at this point be wondering what redshift 
the resulting $P_X(k,z_i)$ should be taken to represent, i.e.,  
$z_i$ is not necessarily the center of
weight of the data, and neither is the mean redshift of the pixels in the bin, 
considering the rather complicated weighting in equation (\ref{Pestimator}).
In fact, the effective redshift is not even the same for each $k$-bin in the
same $z$-bin.
We resolve this question  -- $P_X(k,z_i)$ represents the 
power spectrum at precisely $z_i$ (to first order) --
in our construction of 
$S^{a b}=\left<\delta^a_X \delta^b_X\right>$ and 
$Q^{a b}_k=\partial S^{a b}/\partial P_{k,i}$, where $a$ and $b$ label pixels
at redshifts $z_a$ and $z_b$, and $i$ labels the redshift bin
in which this chunk of spectrum falls.  
To account for the evolution from $z_a$ and $z_b$ to $z_i$,  
we define a power spectrum growth factor, 
$D_{k,i}(z)=[(1+z)/(1+z_i)]^{\alpha_{k,i}}$, where
\begin{equation}
\alpha_{k,i}=\left. \frac{d\ln[P_X(k,z)]}{d\ln(1+z)}\right|_{z_i} \simeq
\frac{\ln(P_{k,i+1}/P_{k,i-1})}{\ln[(1+z_{i+1})/(1+z_{i-1})]}
\label{DFalpha}
\end{equation}
(we use a one-sided derivative estimate instead of 
equation \ref{DFalpha} for the first and last redshift bins).
Now $Q_k^{a b}=D_{k,i}(z_{a b}) \left. Q_k^{a b}\right|_{z_i}$, where 
$z_{a b}=(z_a+z_b)/2$ and 
$\left. Q_k^{a b}\right|_{z_i}$ is computed as if the pixels were 
located at the center of the bin.  
Finally, $S^{a b}=\sum_k Q_k^{a b} P_k(z_i)$.
This correction may be difficult to understand intuitively at 
first, but it is really quite simple.  
The modification of $Q$ just corrects the power spectrum 
estimate for the excess (dearth) of power that we expect 
for pixels in the high (low) redshift ends of the bin. 
The correction to $S$ affects the weighting, simply
producing a more accurate $S$ at the redshift of the 
pixels in question.  

Note that an alternative method would be to treat the points $P_X(k,z_i)$
as simply parameters of a continuous power spectrum defined by some form
of interpolation.  This would mean $S^{a b}$ would have non-zero derivative
with respect to more than one of the power spectrum bins (e.g., usually 
two for 
linear interpolation).  This method would be elegant, and probably produce 
narrower
effective window functions in the $z$ direction; however, it will increase
the correlation in the $z$ direction between measurement errors, because the 
same pixels would contribute to more than one power spectrum point.
Since this more sophisticated method would allow long chunks of spectra 
to be used
without degrading our $z$ resolution, it would be most useful if we were 
trying to measure the power on even larger scales.

\begin{figure}
\plotone{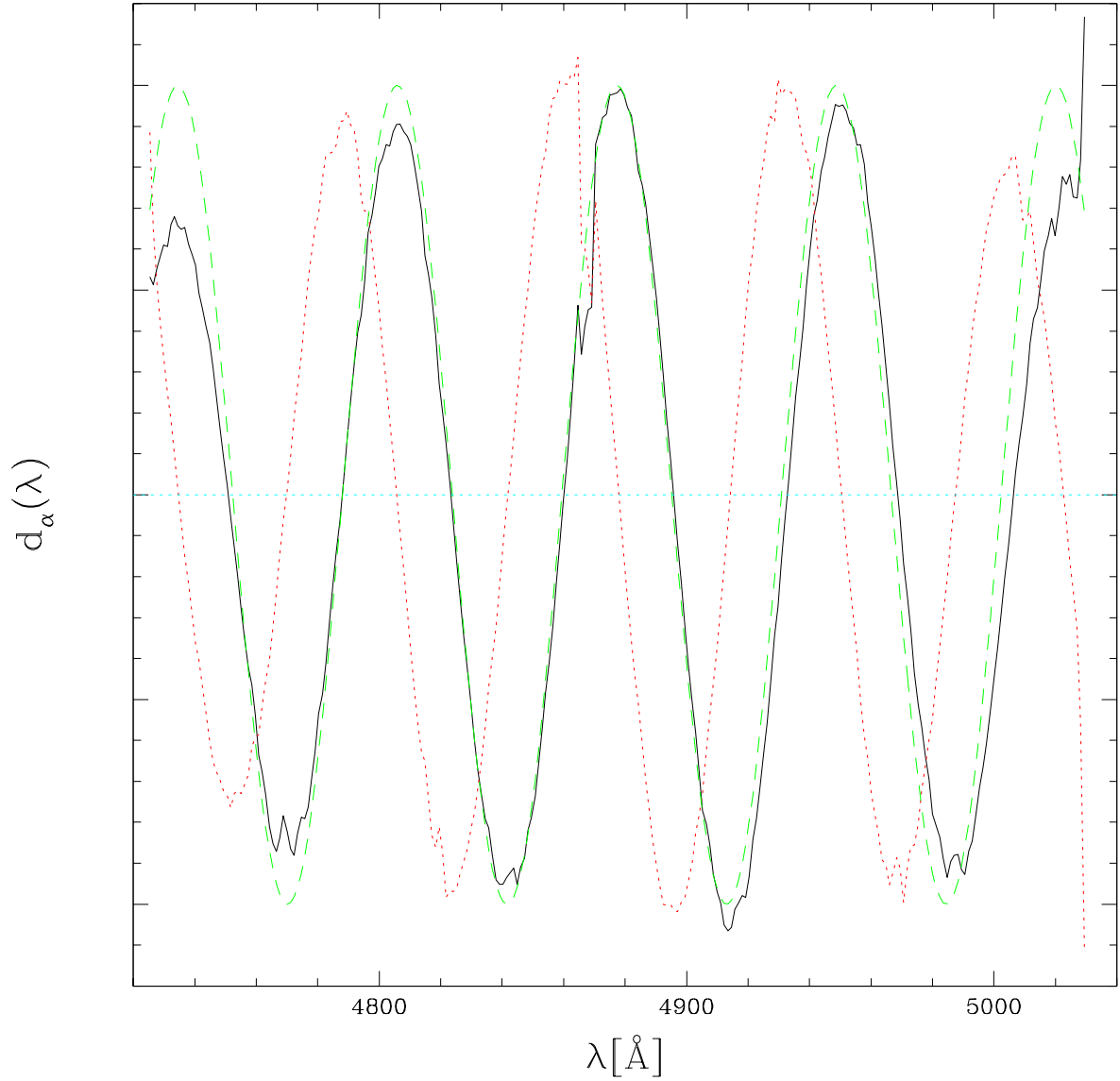}
\caption{Black solid and red dotted lines show the two primary modes onto
which the data is effectively projected when we estimate the power in our
bin centered on $k=0.00141~\ikms$, for the spectrum shown in 
Figure \ref{deltaFvector}a.  The 
horizontal axis scale in the figure is arbitrary.
For comparison, the dashed line shows a simple sine wave with 
$k=0.00141~\ikms$.
}
\label{demoestimator}
\end{figure}

How does our method compare to the straightforward Fourier transform (FT) 
method?
The basic FT method is to project the data vector, $\vdf$, onto a set of 
modes of the form $d_{\alpha \beta} = \exp(i~k_\alpha~\Delta v_\beta)$, 
and to simply compute the variance of the amplitudes of all the modes 
with $k$ in some bin, i.e.,
\begin{equation} 
\hat{P}_{\rm FT}\propto\sum_\alpha^{k_{\rm min}<\left|k_\alpha\right|<k_{\rm max}}
\left|\sum_\beta d_{\alpha \beta} \delta_{f,\beta}\right|^2~,
\end{equation}
where $k_{\rm min}$ and $k_{\rm max}$ define the bin, and the discrete 
spacing of $k_\alpha$ is somewhat arbitrary (the natural spacing is 
$\Delta k= 2\pi/L$, where $L$ is the length of the spectrum, but nothing
prevents one from choosing more finely spaced $k$s).
Our estimator, equation (\ref{Pestimator}), can be cast in a similar form, 
i.e., as a 
projection of the data vector onto a set of modes, and a sum of the
squares of the mode amplitudes.  We require that
the mode amplitudes are statistically independent, which makes their 
computation
equivalent to a computation of Karhunen-Lo\`eve eigenmodes (see, e.g.,
\cite{1997ApJ...480...22T}).
Figure \ref{demoestimator} shows the two most important modes for our bin with
$0.00126~\ikms < k < 0.00158~\ikms$, for the chunk of spectrum shown in
Figure \ref{deltaFvector}a.  
In this case two modes differing primarily by a phase shift, 
analogous to $\sin(k x)$ and $\cos(k x)$, contain most
of the information, because our bin width is approximately $2\pi/L$.
We see that the difference between our modes and a simple sine wave
is not dramatic -- there is a little bit of edge tapering (downweighting
the edges to make the effective window on the data more compact in 
Fourier space) and some straightforward downweighting of the most noisy
pixels.  Curiously, there seems to be an additional effect where pixels
adjacent to an edge are given extra weight, possibly as a way of 
compensating for missing data (this is seen more clearly in spectra
where a narrow gap is present in the middle of the data).
The picture is similar for bins with larger $k$, except of course that
there are increasingly many important modes as the width of our bins
increases (the bins have a fixed width $\Delta\log(k)$, but the relevant 
mode width is $\Delta k$).  For more discussion of the quadratic 
estimator see, e.g., \cite{1997PhRvD..55.5895T}.

The method we adopt is optimal for Gaussian fields and therefore 
guarantees that no other method can surpass it. 
An additional advantage is that within this formalism window 
and covariance matrices are automatically computed. 
For continuous spectra with few gaps and near uniform noise 
one does not necessarily expect an FT method to be significantly worse. 
In practice the noise level is slowly
varying across the spectrum, so averaging all the pixels
uniformly is not optimal and degrades performance.
Another advantage is that with our method each pixel pair 
has its own effective redshift and the correlations for a given pair are then 
interpolated to the redshift of interest using the appropriate 
evolution. In the FT method the whole spectrum is Fourier transformed first, 
so the redshift information is preserved only in an averaged sense, but 
a priori it is not clear how this average is defined. 

\subsubsection{Bootstrap Error Estimation \label{secbootstrap}}

While the Fisher matrix obtained during the estimation 
process would give the error matrix for \PX\ if the data were Gaussian, 
we cannot reliably assume this.  Our solution is to compute a bootstrap
error matrix by the standard procedure (Press et al. 1992).  From our 
data set of $N$ spectra, we form a bootstrap data set by selecting 
$N$ spectra at random, with replacement.  The covariance matrix of
\PX\ is taken to be 
$M^{i j}=\left< \Delta \tilde{P}^i_X  \Delta \tilde{P}^j_X\right>$,
where 
$\Delta\tilde{P}^i_X= \tilde{P}^i_X-\left<\tilde{P}^i_X\right>$,
$\tilde{P}^i_X$ is an estimate of the power in the $i$th bin from a
bootstrap data set, 
and $\left<\right>$ means average over bootstrap realizations.
We generally use 4000 realizations, after checking that this 
produces convergence in the result.  We assume that the error correlations
extend only one bin off diagonal in the $z$ direction, because the spectrum
of a single quasar practically never contributes to non-adjacent bins.  

We have no compelling reason to believe that this method of computing the
error bars will give rigorously correct results.  Considering
the large number of off-diagonal elements that must be estimated,
one worry is that a particular linear combination of the bins may
accidentally vary very little in our data set, so it will appear to have
an unrealistically small error.  Our tests on mock spectra
(\S \ref{sectestboot}) show no sign of this problem. 
Still, to be conservative
we apply one tweak to $M$ after it is computed, in an attempt to
inoculate it against the possibility.  We perform a singular-value 
decomposition on $M$, which produces a set of independent vectors and
their variances.  We then compute the variances of the same vectors
under the Gaussian approximation, using the Fisher matrix.  If the 
bootstrap variance is smaller than the Gaussian variance we replace
it with the Gaussian variance.  Finally, we transform back to $M$.
The tests on mock samples described below give us confidence that our
procedure is reliable. 

\subsection{Tests on Mock Data Sets  \label{sectestfake}}

We validate our procedure as implemented in code by applying it 
to mock data sets.  Many iterations of these tests were required
to produce results that show no serious problems in the error
estimation or the power spectrum estimation itself.  Testing 
the results on realistically created mock data is absolutely essential 
for measurements of such high precision.
In \S \ref{secmakefake} we describe our procedure for generating
the mock spectra.
We test our bootstrap error estimates in \S \ref{sectestboot}.
Finally, we test the power spectrum estimation procedure for 
systematic errors in \S \ref{sectestpow}.

\subsubsection{Generating Mock Spectra  \label{secmakefake}}

We generate mock spectra by combining the auxiliary information
we have for each observed spectrum (e.g., our continuum estimate, 
noise estimate, sky estimate, etc.) with a simplified version of 
the \cite{1992A&A...266....1B} model for the $\exp(-\tau)$ field, 
which results in realistic looking spectra.     

For each observed quasar we start with the term we divide
by before computing the power spectrum, 
$A_q~ \bC(\lr)~ (1+\delta_C)~ \bS(\lambda)$ 
(see equation \ref{datamodel}).  We multiply this by $\exp(-\tau)$ 
(generated as described below),
smooth the result using the resolution from the observed spectrum,
and sample the result onto the observed grid of pixels.  This produces
a noise free version of the flux we would observe coming from this
quasar.  We add flux from the sky as estimated for the observed 
spectrum, and transform the total flux to the number of photons
that would be expected in each pixel.  We generate a Poisson deviate
with this mean, add the appropriate Gaussian read-noise for each pixel,
transform back to the original flux units, and subtract the sky flux
estimate to obtain an observed (noisy) quasar spectrum.  The results 
of this procedure for each observed quasar are written into files in the same 
format as the observed spectra, so exactly the same code can be used 
to measure the power in the mock spectra.  

To generate the $\exp(-\tau)$ fields we use a simple model that
is arranged to give roughly the correct power spectrum as a function
of $k$ and $z$, and the correct mean absorption as a function of 
redshift.  For each observed spectrum, we start by generating a 
Gaussian random field, $\delta_{i,0}$,
on a very long, relatively finely spaced grid
(65536 cells with width $7 \kms$, to be precise), with power spectrum 
\begin{equation}
P_\delta(k) = \frac{1+[0.01~\ikms/k_0]^\nu}{1+(k/k_0)^\nu}
\exp[-(k R_\delta)^2]~,
\end{equation} 
where $k_0=0.001~\ikms$, $\nu=0.7$, and $R_\delta=5\kms$ [this $P_\delta$ was 
chosen after some experimentation because it produces a final flux power 
spectrum with approximately the same $k$ dependence as the observed \PF].  
An arbitrary cell in this grid is chosen to correspond to the redshift of
the quasar, and the evolution of the amplitude of the power spectrum with
redshift is imposed by the transformation $\delta_i=a(z_i)~\delta_{i,0}$
with $a^2(z_i)=58.6~[(1+z_i)/4]^{-2.82}$, where the form of $a(z)$ was 
chosen so that the final flux power spectrum would evolve like the observed
one.  Next we make the squared lognormal transformation 
$n_i=[\exp(\delta_i-\sigma_i^2/2)]^2$, where $\sigma_i^2$ is 
computed from the input power spectrum, including the amplitude factor
(the factor $\sigma_i^2/2$ in the exponential just fixes the mean of 
the lognormal field to 1).
We smooth the $n$ field with a Gaussian filter with rms width
$R_\tau=20 \kms$ and multiply it by a factor $0.374~[(1+z_i)/4]^{5.10}$ 
to produce a field $\tau$ (this redshift evolution factor produces roughly
the observed redshift evolution of $\bar{F}$).  
The mock transmitted flux in each grid cell 
is then $F_i=\exp(-\tau_i)$, which is sampled as described above.   

The procedure described above leads to realistic looking spectra
of the \lyaf. We have verified that it generates a bispectrum 
that is within a factor of 2 of the one measured in N-body simulations.
The main advantages of this procedure 
over the N-body simulation approach when generating the mock spectra 
are that it is faster, so one can make an arbitrary number of 
independent realizations, and that 
the simulated spectra can be of arbitrary length, important to eliminate 
any periodicity effects (this would be impossible with simulations, where
a typical box size is much shorter than the total length of a single 
spectrum).  Both of these advantages are critical for a high precision
test.
We determine the true $P_F(k,z)$ by a simple FFT of extremely long
$\exp(-\tau)$ fields (without redshift evolution).

\subsubsection{Tests of the Error Estimates  \label{sectestboot}}

Without accurate statistical errors it is difficult to identify systematic
problems, so we first test our bootstrap procedure for estimating
the errors.  
Note that there is no reason to expect bootstrap errors to be perfect 
(there is even some  
ambiguity in how exactly the bootstrapping should be done when the 
data do not consist of statistically identical objects).   
Regardless of systematic errors in the method, the only
difference between the power spectra measured from two mock data sets that
differ only in the random seed that was used to create them should be
the statistical errors that we estimate.  We test our error bars by
generating ten different sets of mock data and
computing $\chi^2$ for the differences between each of them and their
error weighted mean,
using the bootstrap error bars and the 
108 points in $0.0013<k<0.02~\ikms$, and $2.1<z<3.9$.
This is effectively a fit of 108 parameters to 1080 data points, with
972 degrees of freedom.  
The total $\chi^2$ is 939, perfectly consistent with a random fluctuation
around the mean, and strongly disfavoring an underestimation of the errors
by more than a couple percent.

\subsubsection{Tests of the Power Estimates  \label{sectestpow}}

\begin{figure}
\plotone{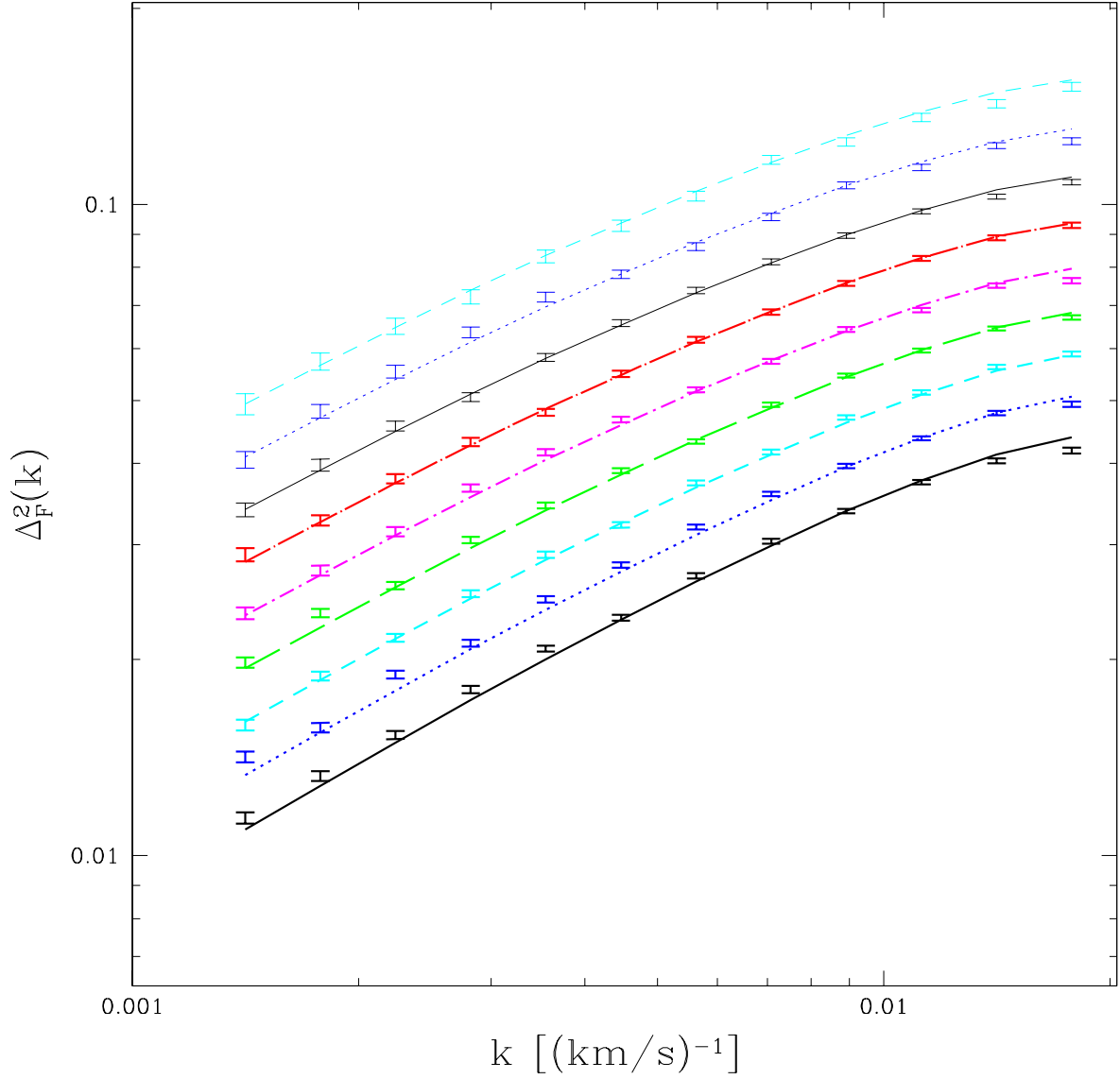}
\caption{
Error bands show the average power 
spectrum [$\Delta^2(k)\equiv\pi^{-1} k~P(k)$]
measured from ten sets of mock spectra.  Lines show the true power.  
Redshift bins, strictly from bottom to top, are:  
black, solid line, z=2.2, blue, dotted, 2.4, cyan, dashed,
2.6, green, long-dashed, 2.8, magenta, dot-dashed, 3.0, 
red, dot-long-dashed, 3.2, black, thin, solid line, 3.4, 
blue, dotted, thin, 3.6, cyan, dashed, thin, z=3.8.
}
\label{combinefake}
\end{figure}

We can now search for systematic
errors.  To enhance the statistical significance of any errors, we 
average our ten sets of mock spectra to form a single, more precise
measurement.  The result is shown in Figure \ref{combinefake}.
The results look reasonably good; however, we find an 
unacceptably bad $\chi^2=346$ for the comparison between 
our measured \PF\ and the true power spectrum (there are
108 degrees of freedom).  To quantify the 
systematic problem, we first assume the bias has the form 
$B(k)=B_0~[k/0.0067~\ikms]^\nu \equiv P_{measured}/P_{input}$ and
fit for the values of $B_0$ and $\nu$ that minimize $\chi^2$ in the 
comparison.  We find $B_0=1.0036\pm 0.0014$ and $\nu=-0.0173\pm 0.0013$ 
with $\chi^2=173$ for 106 degrees of freedom
[the pivot point $k_0=0.0067~\ikms$ was chosen to make the errors
independent; the amplitude coefficient would be larger if we were not 
already dividing by $1+3~\sigma_m^2$ as explained 
in \S\ref{secbandpowerest}].  
The combination of slope and amplitude errors corresponds to a 3.1\%
excess of power at our largest scale, $k=0.0014~\ikms$, and a 1.3\%
underestimate at $k=0.018~\ikms$.
We find some less significant dependencies by generalizing the fitting 
formula even more to
\begin{equation}
B(k,z)=B_0~a^\mu(z)~ 
\left(\frac{k}{k_0}\right)^{\nu+1/2~\eta\ln(k/k_0) + \zeta\ln[a(z)]}~,
\label{codebiaseq}
\end{equation}
where $a(z)=(1+z_0)/(1+z)$, with $z_0=2.85$.
The parameters are $B_0=1.0073\pm0.0016$, $\mu=0.049\pm0.012$, 
$\nu=-0.0195\pm0.0015$, $\eta=-0.0157\pm0.0038$, and
$\zeta=-0.026\pm0.012$, with $\chi^2=135$.
Where does this bias come from?
We expect some bias related to the division of each chunk of spectrum by 
its overall mean (not because of an integral constraint suppression of
large-scale
power -- our estimator should take care of that -- but because of 
statistical error in the estimate of the mean that we divide by).
When we measure the power without this division
by the mean, which we can only do using mock spectra, we find
significantly smaller 
corrections -- small enough to ignore when model fitting.

We expect that this bias should be present when we use real
observed spectra, so we will correct for it by dividing the 
measured power by $B(k,z)$. We describe its effect on the 
amplitude and slope of the power spectrum below (table \ref{modtab}). 

\subsection{Raw Power \label{secrawpower}}

\begin{figure}
\plotone{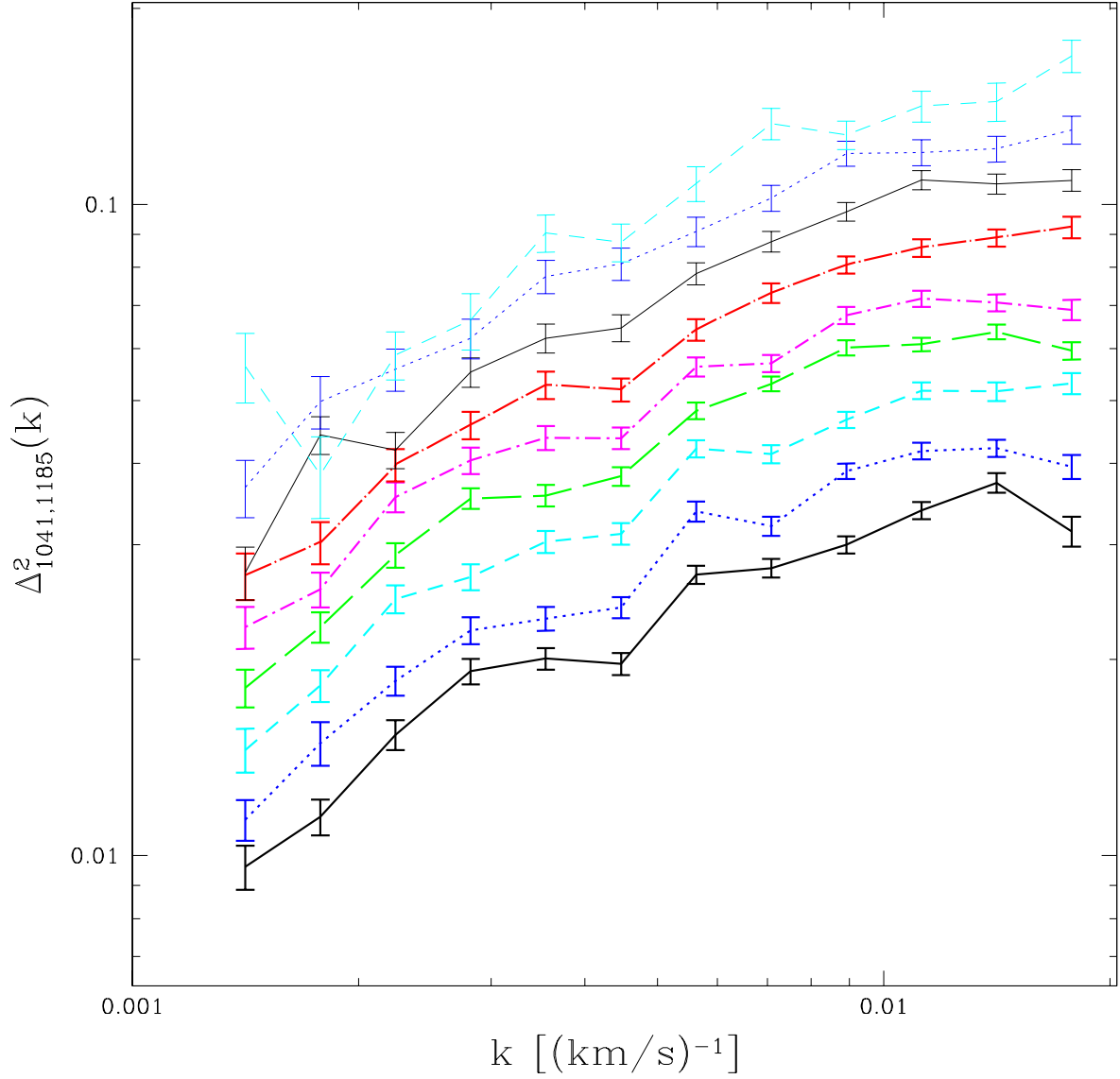}
\caption{
Error bars show the power spectrum measured from the observed
spectra in the wavelength range $1041<\lr<1185$.  The lines connect
the points to identify them and to guide the eye.
Redshift bins, from bottom to top (roughly) are:  
black, solid line, z=2.2, blue, dotted, 2.4, cyan, dashed,
2.6, green, long-dashed, 2.8, magenta, dot-dashed, 3.0,
red, dot-long-dashed, 3.2, black, thin, solid line, 3.4,
blue, dotted, thin, 3.6, cyan, dashed, thin, z=3.8.
}
\label{standardrawpower}
\end{figure}

Figure \ref{standardrawpower} shows the raw power measured in our
standard \lyaf\ rest wavelength range, $1041<\lr<1185$.  
All the figures
in this subsection show $P_{1041,1185}$, not the background subtracted
power $P_F$.
Our normalization convention is:
\begin{equation}
\left<\delta^2\right> = \int_{-\infty}^{\infty}\frac{dk}{2 \pi} P(k)~.
\end{equation}
We usually plot the dimensionless quantity $\Delta^2(k)\equiv \pi^{-1}k~P(k)$, 
the contribution to the variance per unit $\ln k$.

\begin{figure}
\plotone{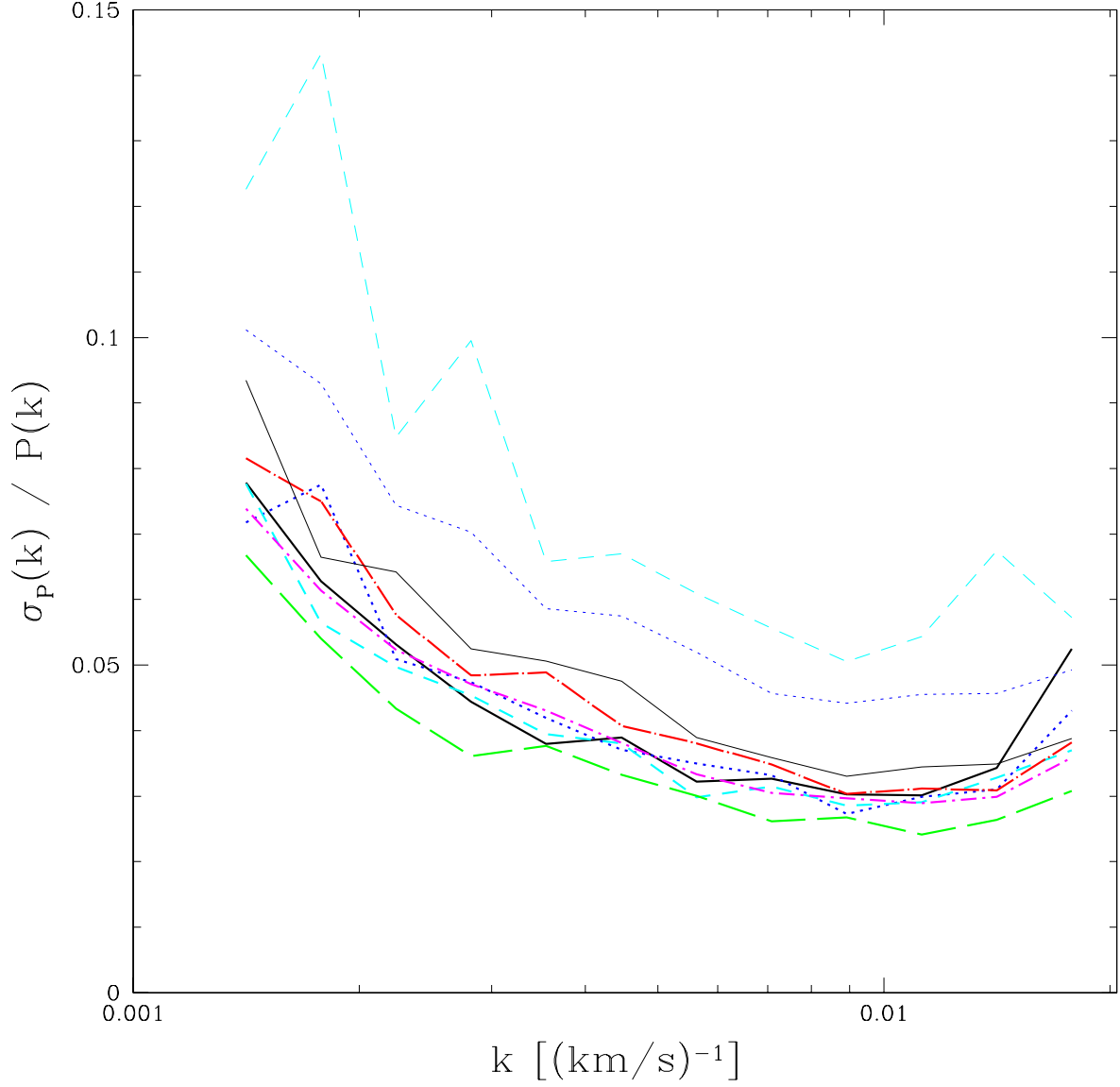}
\caption{
Lines connect the fractional errors on each measured $P_{1041,1185}(k,z)$
point, using the usual line-type and color and scheme 
(see Fig. \ref{standardrawpower} -- the highest two curves are the 
highest two redshifts, the lowest is $z=2.8$). }
\label{fractionalerrors}
\end{figure}

Figure \ref{fractionalerrors} shows the fractional errors on all of the
measured points.  
The errors are less than 5\% for most of
the points, and frequently as small as 3\%.  
If we were only estimating a single amplitude parameter
by combining all these points then its error would be 0.6\%.
An overall logarithmic slope would have an error $\pm 0.005$.
The errors on the largest
scales are increased somewhat by the diagonalization of the window
matrix.

\begin{figure}
\plotone{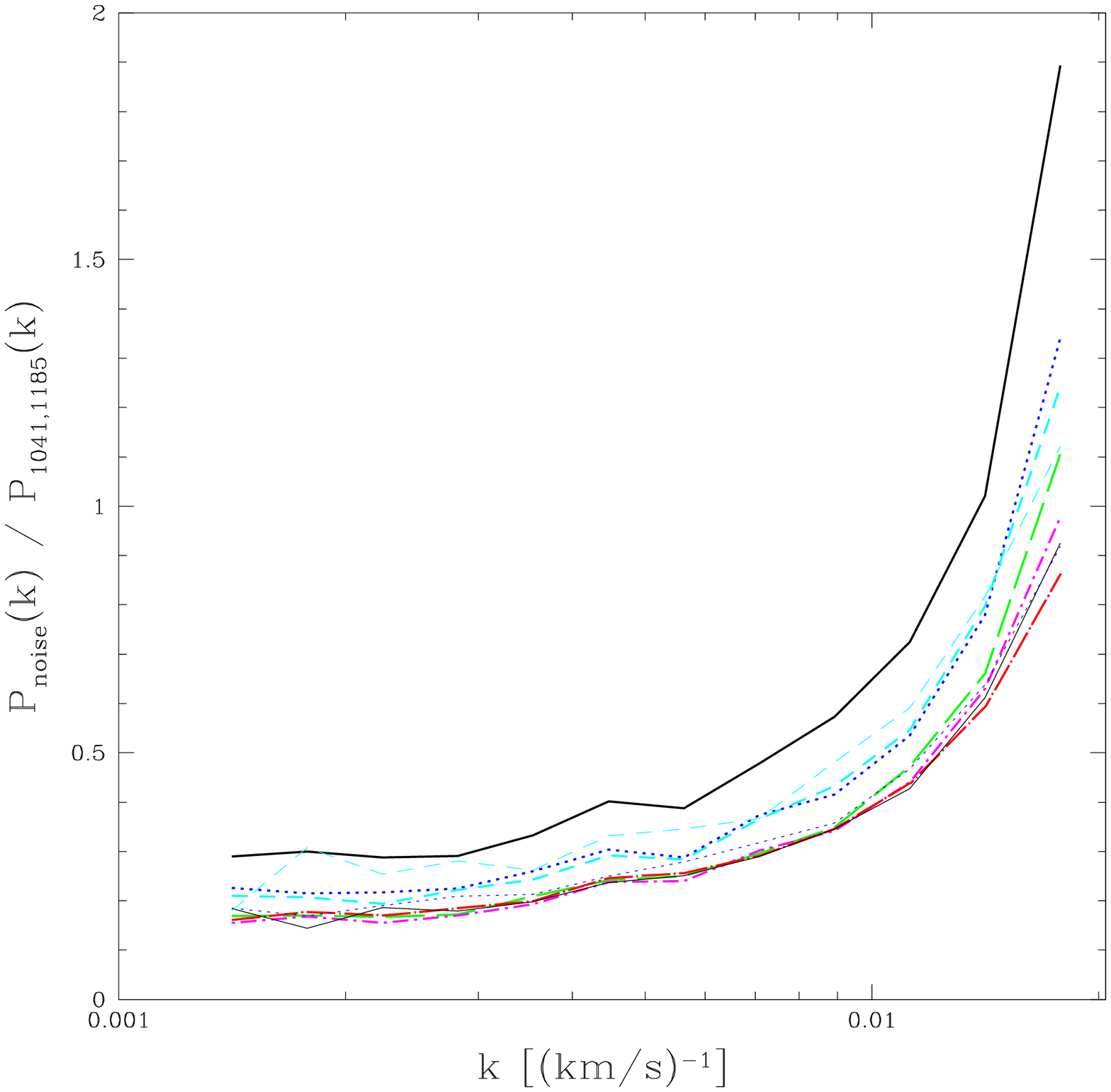}
\caption{
Lines, with types following the usual pattern (see 
Fig. \ref{standardrawpower}), connect the 
quantity $P_{\rm noise}/P_{1041,1185}$ for each measured 
point (the highest line is the lowest redshift).}
\label{noisefraction}
\end{figure}

Figure \ref{noisefraction} shows the ratio of subtracted noise power 
to measured signal power ($P_{1041,1185}$) for each point.
The noise power is significant (20-30\%) on all scales, but diverges 
at high $k$ where the resolution suppresses the absorption power.
The lowest redshift bin has the most noise, due to the 
lower \lyaf\ power combined with extra noise at the short wavelength
end of the spectra.

\begin{figure}
\plotone{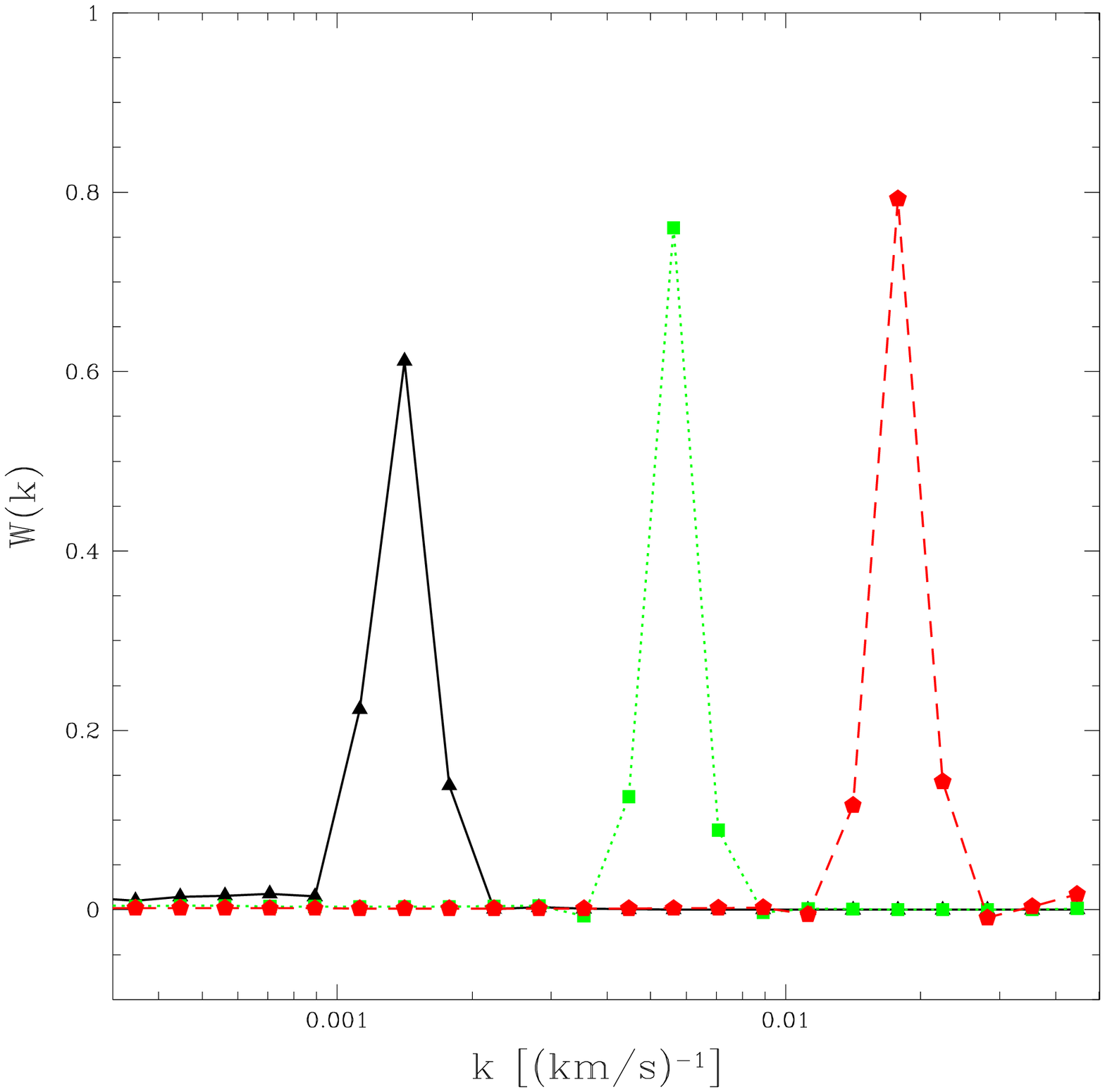}
\caption{
The window matrix for bands indicated by the maximum 
(before diagonalization).
}
\label{windowmatrix}
\end{figure}

Figure \ref{windowmatrix} shows our window matrix (at $z=2.6$), which 
we proceed to diagonalize.
The matrix is reasonably close to diagonal already, with large 
contributions only from adjacent bins.  It is useful to
diagonalize 
the matrix at this stage, rather than waiting until the
model-fitting stage, because this allows us to compute bootstrap 
errors directly for the final bins (the bootstrap error 
calculation and window matrix diagonalization do not perfectly
commute).

\begin{figure}
\plotone{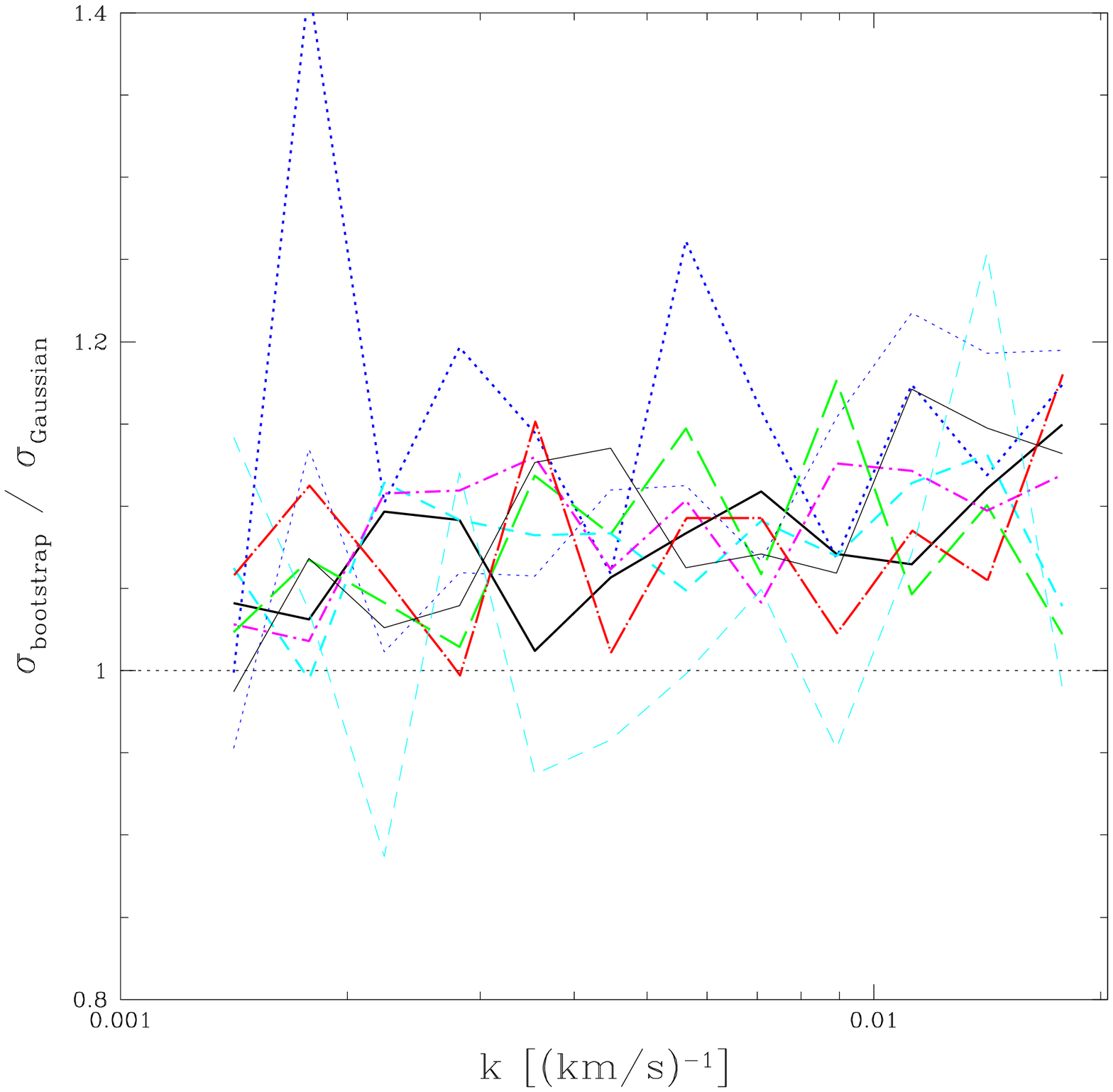}
\caption{
The ratio of bootstrap errors to the Gaussian estimate of the errors.
See Fig. \ref{standardrawpower} for the correspondence between lines
and redshift bins. 
}
\label{gausserrcomp}
\end{figure}

\begin{figure}
\plotone{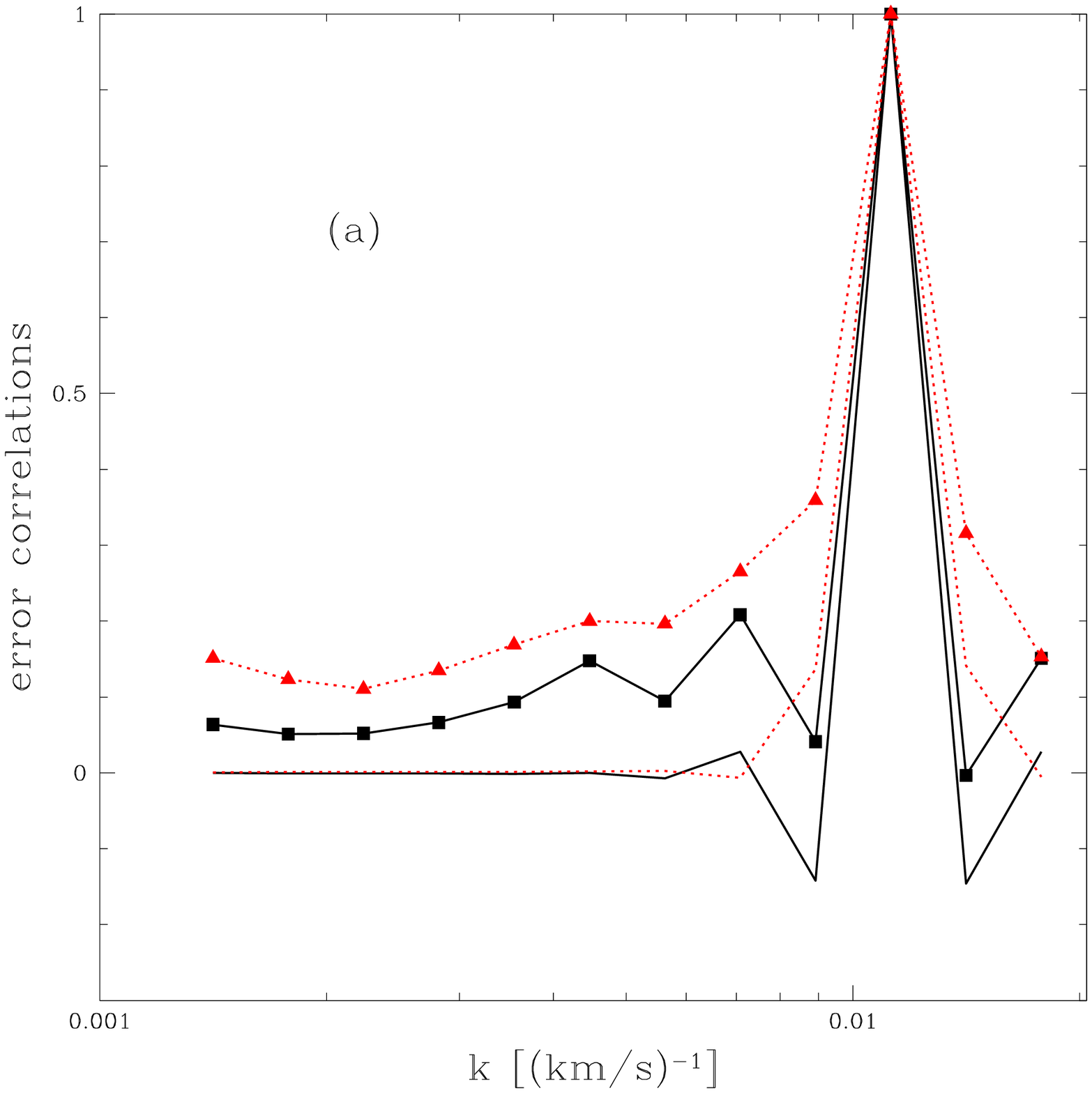}
\caption{
Examples of the correlations between the errors, 
$\left<\Delta P_i \Delta P_j\right>/\sigma_{P,i}~\sigma_{P,j}$.
The black solid lines and squares show the error correlation when
the window matrix is diagonalized.  The red dotted lines and 
triangles
show the correlations between points before diagonalization.  
The lines
marked by symbols are the bootstrap estimate, while the unmarked
lines are the Gaussian estimate.  
}
\label{errcorr}
\end{figure}

Figure \ref{gausserrcomp} shows the ratio of the bootstrap errors
to the errors estimated assuming the data are Gaussian.  
We did not apply the Gaussian floor to the 
bootstrap errors when making this figure.
Typically the bootstrap errors are 0-20\% larger than the Gaussian errors. 
Figure \ref{errcorr} shows examples of the estimated correlation 
between the errors, at $z=2.6$.
\begin{figure}
\plotone{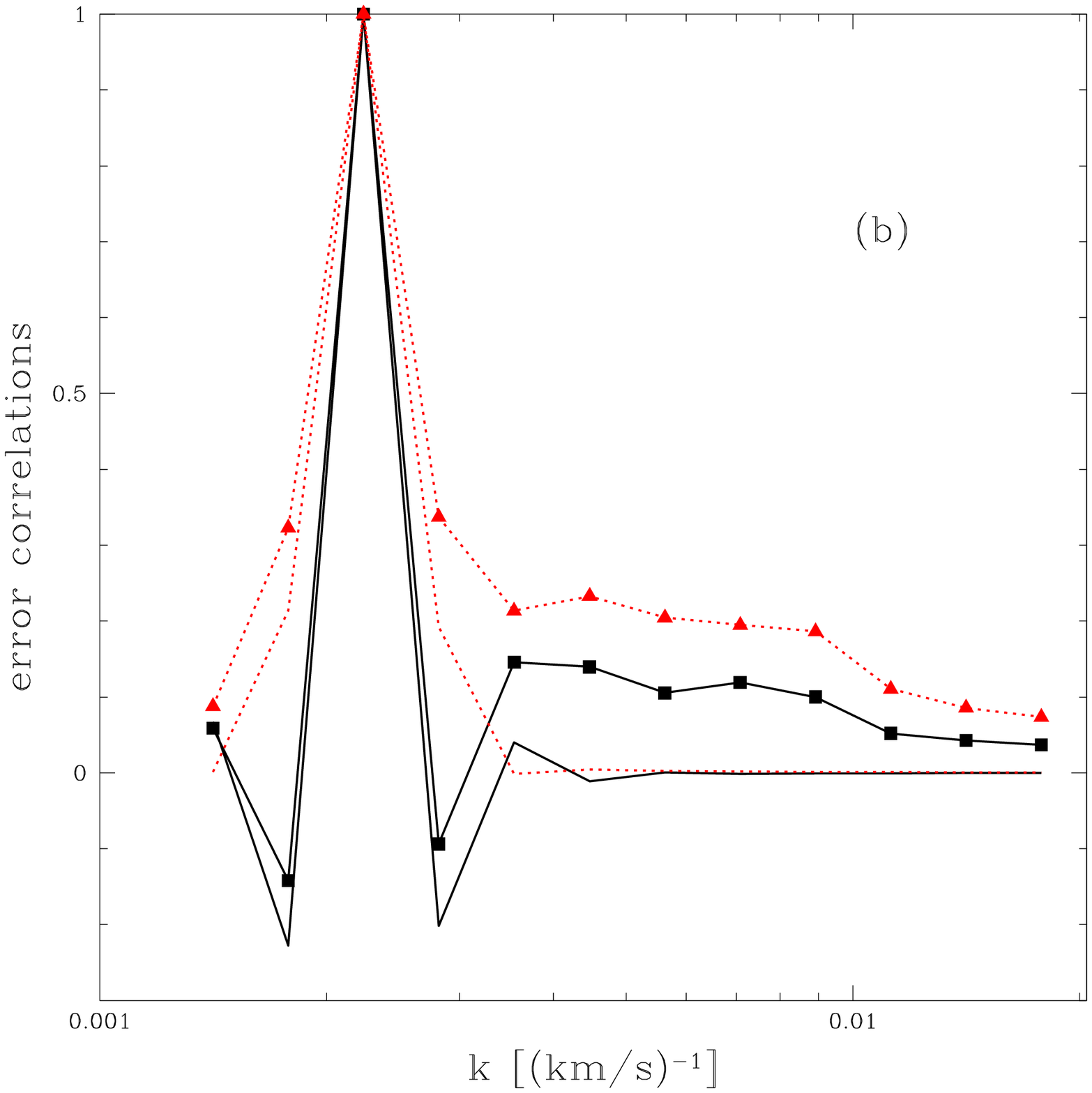}
\end{figure}
We note that diagonalizing the window matrix noticeably reduces the
error correlations.
The bootstrap errors are, in contrast to the Gaussian errors, 
noticeably 
correlated 
($\left<\Delta P_i \Delta P_j\right>/\sigma_{P,i}~\sigma_{P,j}\sim 0.0-0.2$
when $|i-j|>1$, where $i$ and $j$ label the bins) 
across the full $k$ range.  These differences between bootstrap and
Gaussian errors are not necessarily an indication of intrinsic 
non-Gaussianity in the absorption fluctuations.  
Possible alternative explanations for the differences include the 
uncertainty in the mean flux value that each chunk of spectrum is 
divided by and the uncertainty in the noise-subtraction term 
for each chunk, neither of which are included in the Gaussian 
estimate and both of which would increase the error in a way that
is correlated across $k$ bins.

\subsection{Background Subtraction \label{secbacksubtract}}

\begin{figure}
\plotone{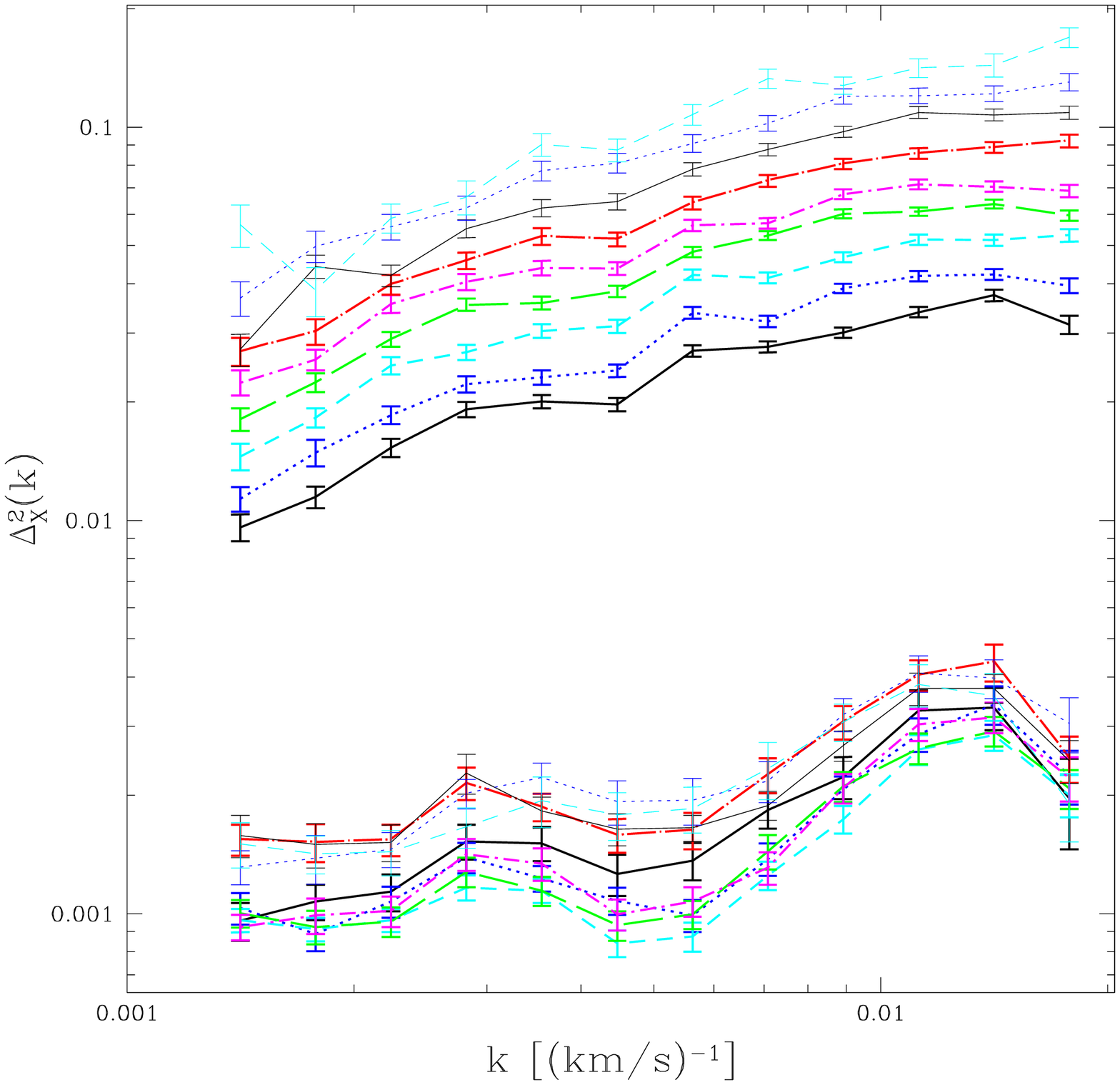}
\caption{The upper set of lines show $P_{1041,1185}$, the lower set of 
lines show $P_{1268,1380}$.  The colors and line types identify redshift
bins as defined in Fig. \ref{standardrawpower}.
}
\label{forestandredpower}
\end{figure}

\begin{figure}
\plotone{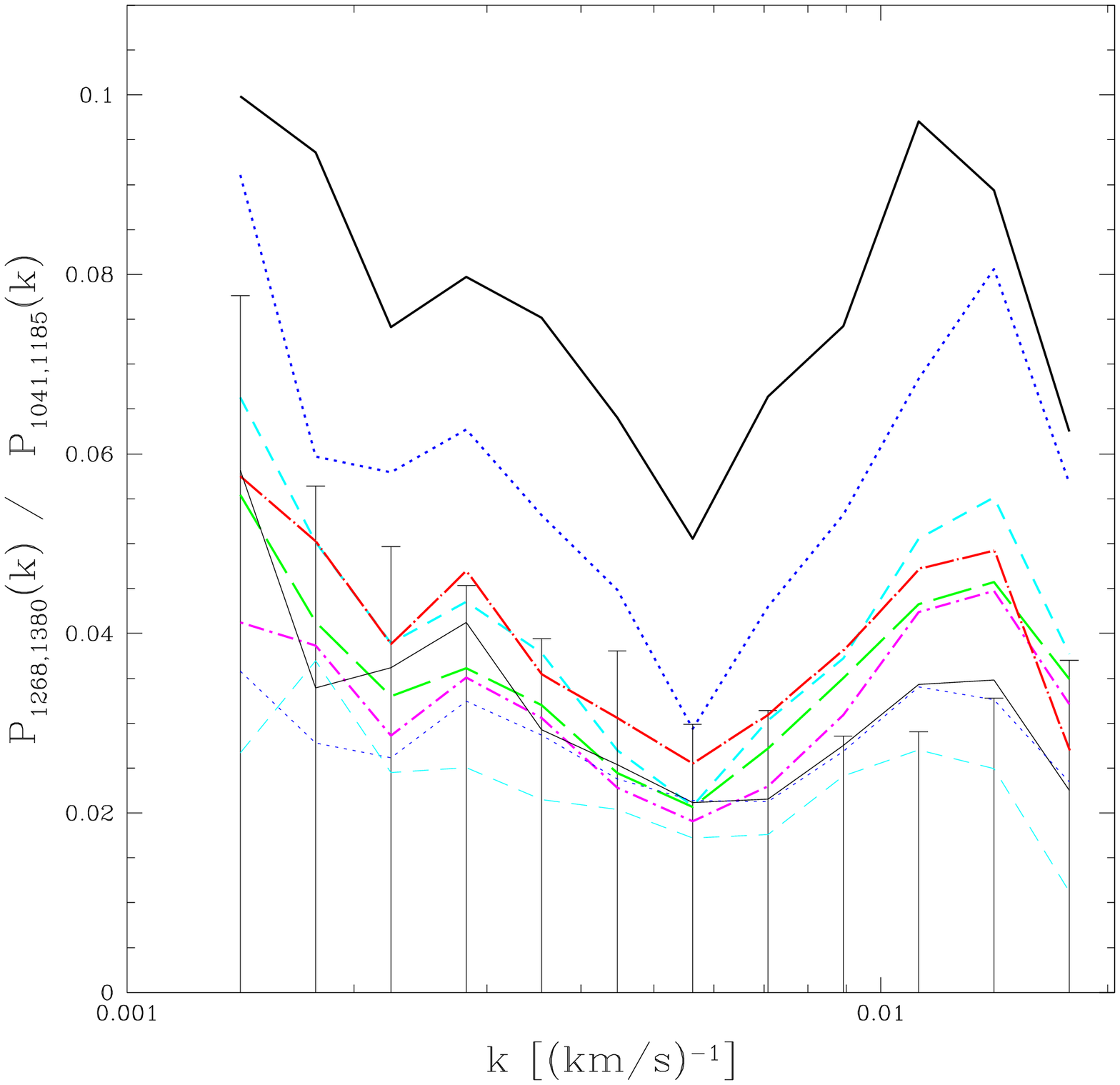}
\caption{
The lines show the ratio $P_{1268,1380}/P_{1041,1185}$.  
The uppermost (black solid) line is $z=2.2$, and
the next (blue dotted) is $z=2.4$ (see Fig. \ref{standardrawpower} 
for the rest of the line definitions).
For reference, the error
bars starting at zero show the fractional errors on 
$P_{1041,1185}(k,z=2.6)$, which are much larger than the errors on 
$P_{1268,1380}$ (we are simply plotting $\sigma_P(k)/P(k)$ as in 
Fig. \ref{fractionalerrors}, except that, for clarity, we show 
error bars starting at zero instead of a connected line).  
}
\label{ratio1268}
\end{figure}

Our background subtraction is the power in the wavelength
range $1268 < \lr < 1380$ \AA.  
Figure \ref{forestandredpower} shows $P_{1268,1380}$ 
and $P_{1041,1185}$ for comparison.
The bump at $k \sim 0.013~\ikms$ in $P_{1268,1380}$ 
is probably due to the CIV doublet at
separation $499~\kms$.  The bump at $k\sim 0.003~\ikms$ may be due to 
the SiIV doublet at separation $1933~\kms$.
Figure \ref{ratio1268} shows $P_{1268,1380}/P_{1041,1185}$.
We see that, even though the background power is a 
small fraction of the \lyaf\ power, it is quite significant when
compared to the small size of the errors on the \lyaf\ power.
It is important to remember that even a small overall
systematic error can be very significant if it covers many 
data points (e.g., a $1/2~\sigma$ error over 100 points shifts the
mean by $5~\sigma$).

\begin{figure}
\plotone{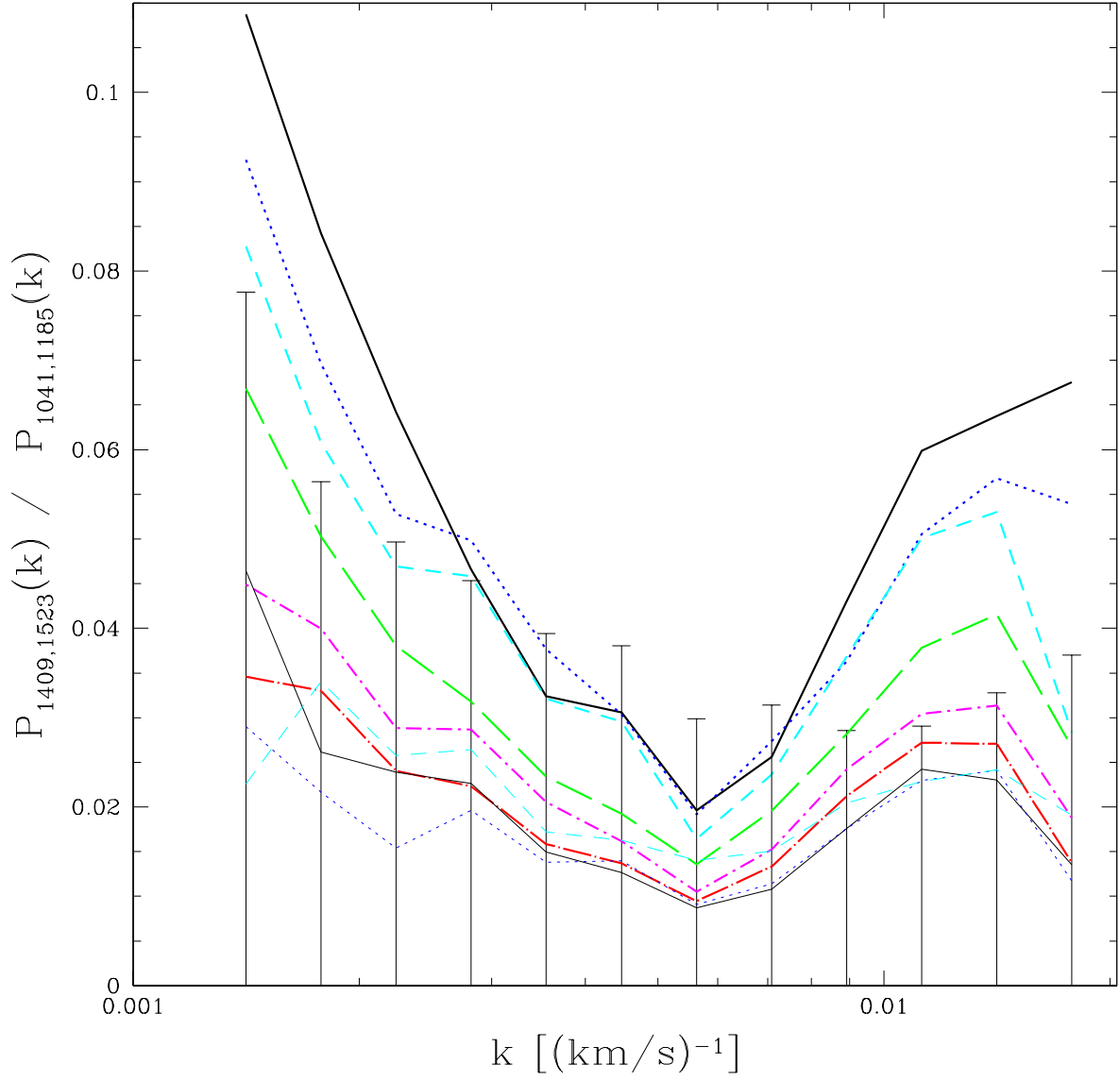}
\caption{The ratio of $P_{1409,1523}$ to $P_{1041,1185}$.
}
\label{ratio1409}
\end{figure}

\begin{figure}
\plotone{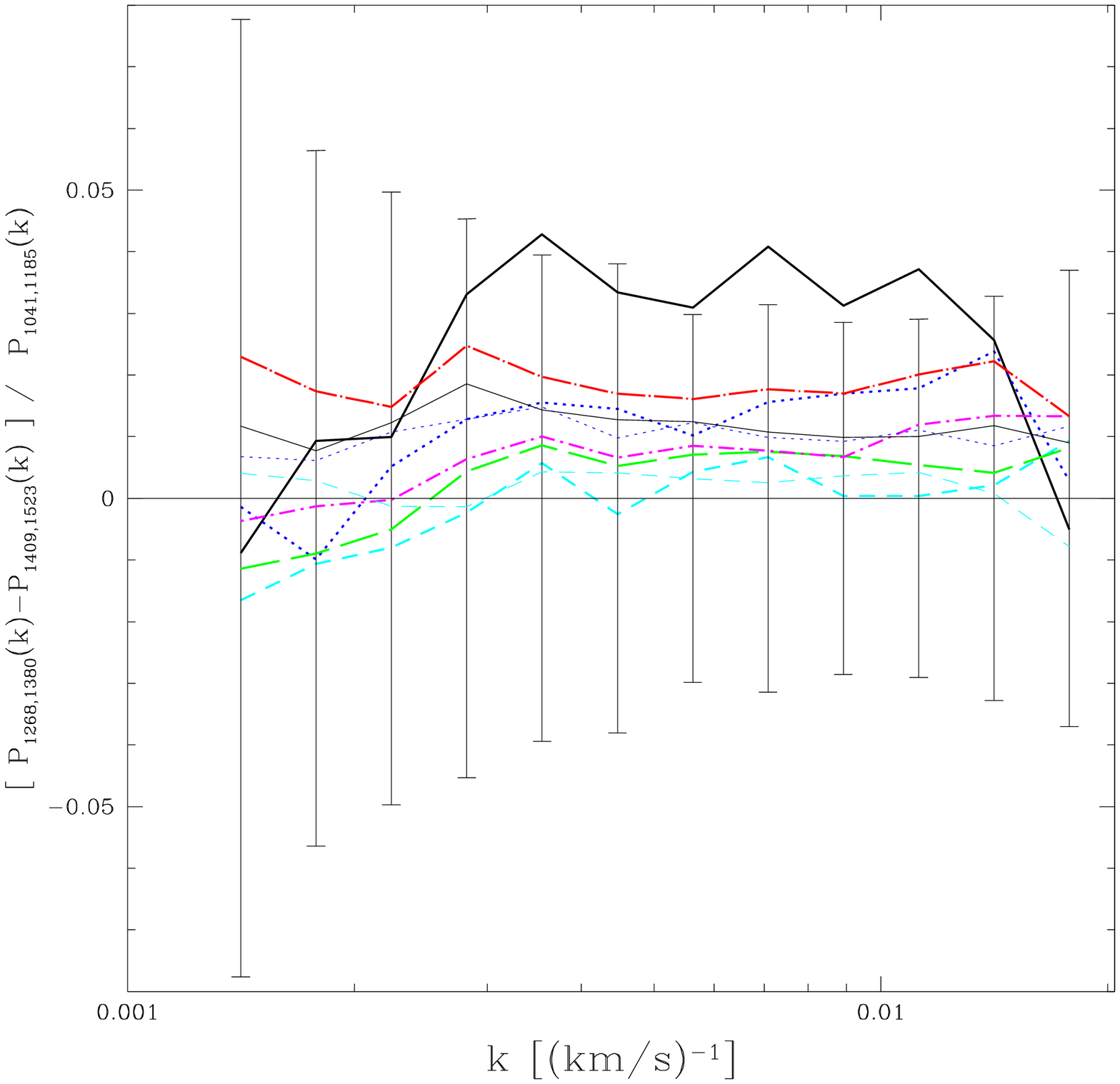}
\caption{
The difference between $P_{1268,1380}$ and $P_{1409,1523}$, divided
by $P_{1041,1185}$, with the fractional errors on $P_{1041,1185}(k,z=2.6)$
plotted as usual.
}
\label{diff12681409}
\end{figure}

\begin{figure}
\plotone{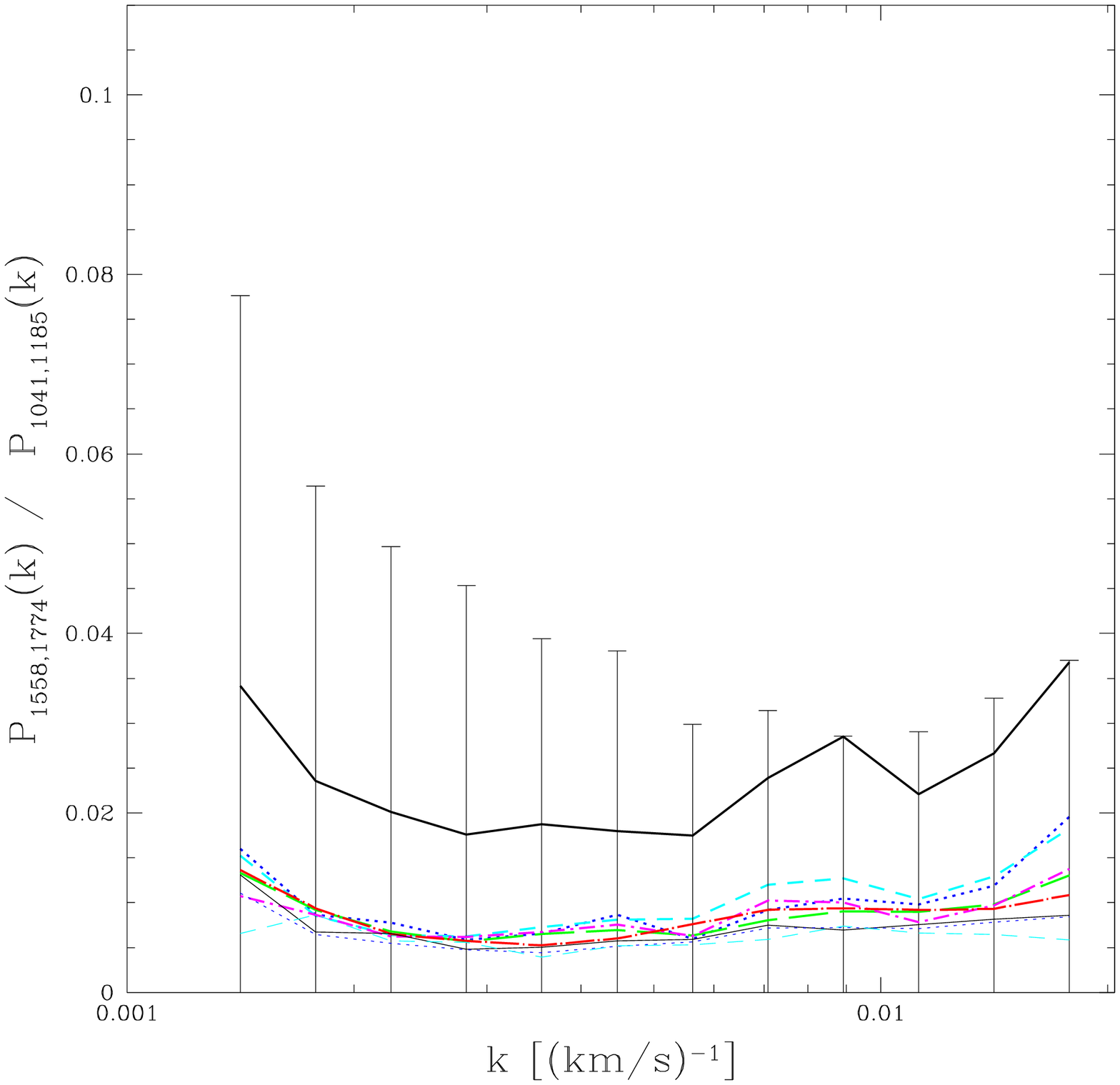}
\caption{The ratio of $P_{1558,1774}$ to $P_{1041,1185}$.
The uppermost (black solid) line is $z=2.2$.
}
\label{ratio1558}
\end{figure}

We are going to subtract the power in the range 1268-1380 \AA\
from the \lyaf\ power,
but it is informative to measure the power at other places in
the quasar rest frame for comparison.
The range 1409-1523 \AA\ includes CIV absorption (at 1548.2 and
1550.78 \AA) but excludes SiIV (at 1393.75 and 1402.77 \AA)
and shorter wavelength transitions.
Figure \ref{ratio1409} shows $P_{1409,1523}/P_{1041,1185}$.
If all of the power was coming from metal line absorption, the 
power in the range $1409<\lr<1523$ \AA\ should always be less than the 
power in the range $1268<\lr<1380$\AA.  As we see in 
Figure \ref{diff12681409}, which shows the difference in the background 
fractions, $(P_{1268,1380}-P_{1409,1523}) / P_{1041,1185}$, the 
power in  $P_{1268,1380}$ is greater than $P_{1409,1523}$ except on 
large scales.   
The difference on large scales suggests that there is tiny amount of 
power left in the quasar continua (in spite of our division by the mean
continuum), which is larger in the range 1409-1523 \AA\
than in the range 1268-1380 \AA.  
Finally, Figure \ref{ratio1558} shows $P_{1558,1774}/P_{1041,1185}$, past
the wavelength of CIV absorption.
The reduction of power relative to shorter wavelengths is dramatic,
but not surprising since CIV is the most common metal absorber.
It does suggest however that most of the power is due to metals and 
not continuum fluctuations, unless the continuum in the range 1558-1774 \AA\
has significantly less power relative to other intervals studied here
(which is admittedly not inconceivable).
It seems likely, although we are not certain, that the $z=2.2$ 
background power has a noticeable contribution from 
measurement-related problems, because the 
alternative is a very sudden increase in metal absorption power.

What is the upshot from these studies? 
The metal absorption appears to contribute a small but 
significant amount of power, which should also appear in the 
\lyaf\ region.  We subtract this power from the power measured
in the forest.
There is some indication of measurement-related problems
in our lowest redshift bin.      
The power contributed by deviations of the quasar continua from 
their mean appears to be small.

While the idea that $P_{1268,1380}$ contains almost
exclusively power due to simple
metal absorption seems plausible at this point, when we perform 
consistency checks in \S \ref{secsubsamples} we find evidence that this
is not the case.  Splitting the data set used to measure 
$P_{1268,1380}$ in half based on the noise level in each spectrum,
we find that the power in the halves is significantly different,
by as much as 50\% in some bins.  Splits based
on several other properties of the data (e.g., sky to quasar flux
ratio) also show significant differences, but we find that these
differences can all be accounted for by the difference in the basic
noise level in the subsamples.  Splits of
the \lyaf\ data set show similar trends in $P_{1041,1185}$ with the
splitting parameters, although the fractional differences are much
smaller.  While we don't know the source of this noise 
dependence, 
it is not hard to imagine relatively benign reasons for it.
For example, if sky subtraction is imperfect this
would add an increasing amount of power as the sky flux, and thus
noise level, increases relative to the quasar flux.  
The procedure we describe next would remove
this power. 

\begin{figure}
\plotone{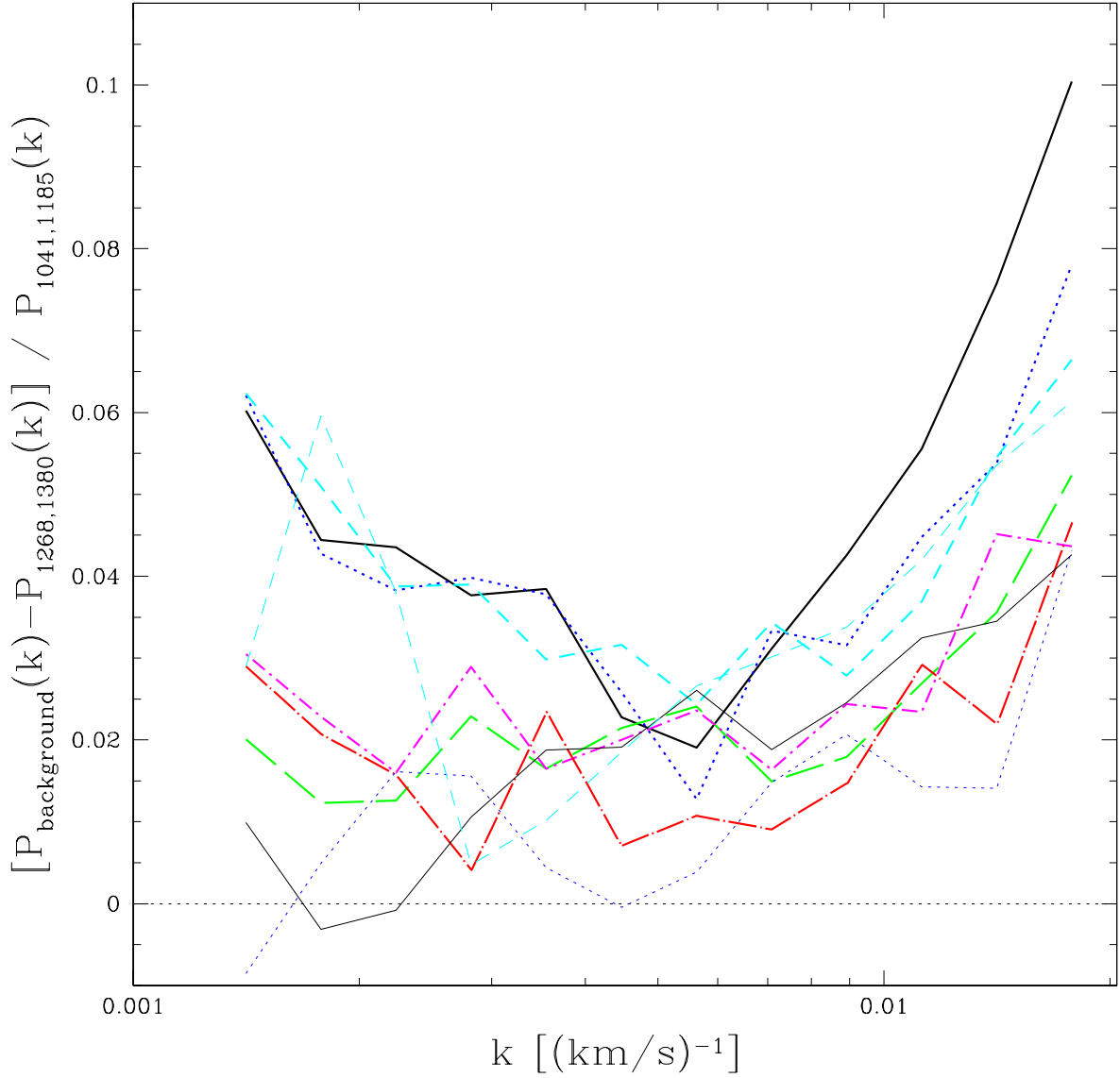}
\caption{The difference between the noise dependent background power that
we subtract through Equation (\ref{fitdepeq}) and $P_{1268,1380}$, 
relative to $P_{1041,1185}$, i.e., this is the extra fractional power
that is subtracted because we correct for the difference between the
typical noise level in the forest and in the range 1268-1380 \AA.
}
\label{noisedepbackground}
\end{figure}

Since we know that $P_{1268,1380}$ depends on noise it seems logical 
to subtract the value of $P_{1268,1380}$ corresponding to the level
of noise in the forest, rather than the best measured value of 
$P_{1268,1380}$, which is dominated in practice by data with less 
noise.
If we had simply misestimated the noise by an overall factor,
for example, the errors in $P_{1041,1185}$ and $P_{1268,1380}$
would cancel for this form of subtraction.
To implement this idea, we model the background subtraction term
as a linear function of the noise level,
\begin{equation}
P_{1268,1380}(k,z,\bar{\sigma}_w) = A(k,z) + B(k,z)~ \bar{\sigma}_w ~,
\label{fitdepeq}
\end{equation}
where $\bar{\sigma}_w$ is the mean noise level in the data computed in the 
same way as the mean flux level (this is the mean of the normalized
noise, i.e., after division by continuum and mean flux).
The choice of a linear relation is arbitrary but it does
the job (see \S \ref{secsubsamples}) better than the alternatives 
we tried.  
We compute $A(k,z)$ and $B(k,z)$ for each value of $k$ and $z$ 
using a linear fit to the full sample of spectra that 
probe $P_{1268,1380}(k,z)$, weighted by the Gaussian estimate of the 
error on each point for each spectrum.
When the time comes to subtract the background from $P_{1041,1185}$ to 
obtain $P_F$, we use $\bar{\sigma}_w$ computed in the $1041-1185$\AA\
wavelength range to compute the appropriate subtraction term. 
Figure \ref{noisedepbackground} shows the extra power subtracted through
Equation (\ref{fitdepeq}), beyond what we would subtract if we simply 
used $P_{1268,1380}(k,z)$ from Figure \ref{ratio1268}.  
It is typically
less than 4\% of the \lyaf\ power, but rises to 10\% at the 
highest $k$ for the lowest redshift.  

The reader who is paying attention may complain that we have no 
compelling reason
to believe that this source of noise-dependent power that we do not 
understand depends on noise in the same way inside and outside the 
\lyaf\ region.  This would be true, except that when we follow this
prescription for background subtraction the differences in 
$P_{1041,1185}$ between subsamples are removed (see \S \ref{secsubsamples}).  
This would be a 
remarkable coincidence if our model for the subtraction was not 
substantially correct.  

Note that the background power has much smaller (absolute) statistical 
errors 
than $P_{1041,1185}$, mostly because there is simply less power, but
also because there are more quasars probing a given redshift interval.

\section{Consistency Checks  \label{secconchecks}}

In \S \ref{secfitting} we describe how we use fits of theoretical models 
to the \PF\ results to help understand the importance of any systematic 
errors.
We plot the correlation function of the \lyaf\ in \S \ref{secxisi3} and
use it to identify a significant contribution to \PF\ from SiIII 
absorption.  
In \S \ref{secmodifications} we examine the effects of modifications of
our procedure.
In \S \ref{secsubsamples} we break the data set up in many
different ways to look for dependencies of \PF\ that should not exist.
In \S \ref{seccontinuum} we discuss the possibility that continuum
fluctuations contribute significant power.
Finally, we compare our results to past measurements in \S \ref{comppastsec}.

\subsection{Rudimentary Fitting of Theoretical Models \label{secfitting}}

The ultimate purpose of measuring the \lyaf\ power spectrum is to determine
cosmological parameters by comparing the observed \PF\ to the predictions
for different cosmological models.  For the $\Lambda$CDM models supported
by present observations, the universe is nearly Einstein-de Sitter at 
$z>2$, so cosmology influences the \lyaf\ almost entirely through 
the linear
theory power spectrum of the mass fluctuations, $P_L(k,z)$, at $z\sim 3$ and 
$k\sim 0.01~\ikms$ (roughly 1 comoving h/Mpc, depending somewhat on 
the model). We usually 
parameterize $P_L(k,z)$ by its amplitude, 
$\Delta^2(k_p,z_p)\equiv k_p^3 P_L(k_p,z_p)/2 \pi^2$, slope
$n_{\rm eff}(k_p,z_p)\equiv \left. d\ln P_L/d\ln k \right|_{k_p,z_p}$,
and curvature 
$\alpha_{\rm eff}(k_p,z_p)\equiv \left. d n_{\rm eff}/d\ln k \right|_{k_p,z_p}$,
where we use $k_p=0.009~\ikms$ and $z_p=2.6$ as the pivot points.

A full discussion of the details of theoretical modeling of \PF\ using
numerical simulations is beyond the scope of this paper.  Furthermore, the
theory of the \lyaf\ is perhaps 
less certain than the observations, so we want to
present the observational results un-tarnished by theoretical 
interpretation.  However, it is very useful to interpret the possible 
systematic errors in the appropriate context of cosmological model fitting.
In other words:  without model fitting, it is difficult to know how
important a given change in \PF\ is.

In this paper we take a cautious approach to the 
theoretical model 
fitting -- we perform fits to different estimates of \PF\ computed 
using modifications
of the extraction procedure or different subsamples of the data, however,
we do not give the central result, only the deviations in the 
results from the value obtained from our preferred \PF.  
These deviations in fitting results should give the reader a useful 
indication of the importance of systematic effects in the data,  
regardless of the reader's opinion of the theory.

The simulations and fitting procedure that we use are described in 
McDonald et al. (2004), where we present the final result.
We use a $\Lambda$CDM transfer function, and perform
the fit with $\Delta^2$ and $n_{\rm eff}$ as free parameters (because 
$\alpha_{\rm eff}=dn_{\rm eff}/d\ln k$ is not tightly constrained by the present \lyaf\ data
alone, we fix the primordial running $\alpha=dn/d\ln k$, not to be confused 
with $\alpha_{\rm eff}\simeq -0.2$, to zero).
Unless otherwise specified, we perform the fits using the 108 \PF\ 
points in the ranges $0.0013 < k < 0.02~\ikms$ and $2.1<z<3.9$.
We allow for some error in our noise estimate by permitting the 
noise subtraction terms to vary independently in each redshift bin
by 5\% (9 extra free parameters to fit for, constrained by Gaussian 
likelihood function with this rms width).  
We also allow a single overall parameter describing the squared
resolution error to vary with rms constraint $(7 \kms)^2$. 

The \lyaf\ model in the simulations is controlled by the externally 
constrained functions 
$\bF(z)$, the mean absorption, $T_{1.4}(z)$, the temperature at overdensity
1.4, $(\gamma-1)(z)$, the logarithmic slope of the temperature-density 
relation, and a reionization parameter that
we will call $\xrei$.  $\bF(z)$ is described in our fits by the 10
parameter formula $\bF_i = {\mathcal F} {\tilde F}_i$, 
where $i$ labels our 9 redshift bins, ${\tilde F}_i$ gives the 
arbitrarily normalized 
$z$ dependence and ${\mathcal F}$ is an overall normalization.
We have performed a preliminary analysis using the formalism 
in \S\ref{secmeandiv} to measure $\bF(z)$ from SDSS data and we use
this to constrain the parameters ${\tilde F}_i$ (the error on each 
redshift bin is $\sim 0.005$).  Because the SDSS analysis does not 
constrain the overall normalization, we leave ${\mathcal F}$
free except for the additional constraint that we require $\bF_i$ interpolated
to $z=(3.9,3.0,2.4)$ to match the HIRES constraints $\bF=(0.458\pm0.034,
0.676\pm0.032,0.816\pm0.023)$ 
(see \cite{2000ApJ...543....1M} -- we have modified the numbers slightly and 
increased the errors
to allow for systematic uncertainties, as discussed in \cite{2003MNRAS.342L..79S}).
We parameterize $T_{1.4}(z)$ and $(\gamma-1)(z)$ by quadratic functions of $z$ 
(3 parameters each)
with the external constraints 
$T_{1.4}=(20100\pm3400,20300\pm2400,20700\pm 2800)$K
and $\gamma-1=(0.43\pm0.45,0.29\pm0.3,0.52\pm 0.14)$ at $z=(3.9,3.0,2.4)$ 
(see \cite{2001ApJ...562...52M} -- we added 2000 K in quadrature to the 
temperature errors
to allow for systematic errors).  
Note that there are other, sometimes more precise, measurements 
of $\bF$ \citep{2003ApJ...596..768S,2003AJ....125...32B} 
and the temperature-density relation 
\citep{2000MNRAS.318..817S,2000ApJ...534...41R} in
the literature -- our
choice of \cite{2000ApJ...543....1M} and \cite{2001ApJ...562...52M} for
this example is simply for convenience.
The redshift of reionization and post-ionization temperature of the gas 
influence \lyaf\ predictions because the smoothing of the gas on small 
scales depends on its pressure history.  At the level of precision we
care about, this dependence can be captured by a single parameter.
In our modeling, we use $\xrei$ to interpolate between two reasonable
boundaries, reionization heating of the gas to 25,000 K at $z=7$ or to
50,000 K at $z=17$, both of which are consistent with our temperature
constraints $T_{1.4}(z)$.  However, 
in this paper we fix $\xrei$, because
it is weakly constrained by the data and the hard lower limit we have to
impose on the redshift of reionization
leads to non-Gaussian errors on the power spectrum parameters we
are interested in (this is a problem of presentation, not of principle).

\begin{figure}
\plotone{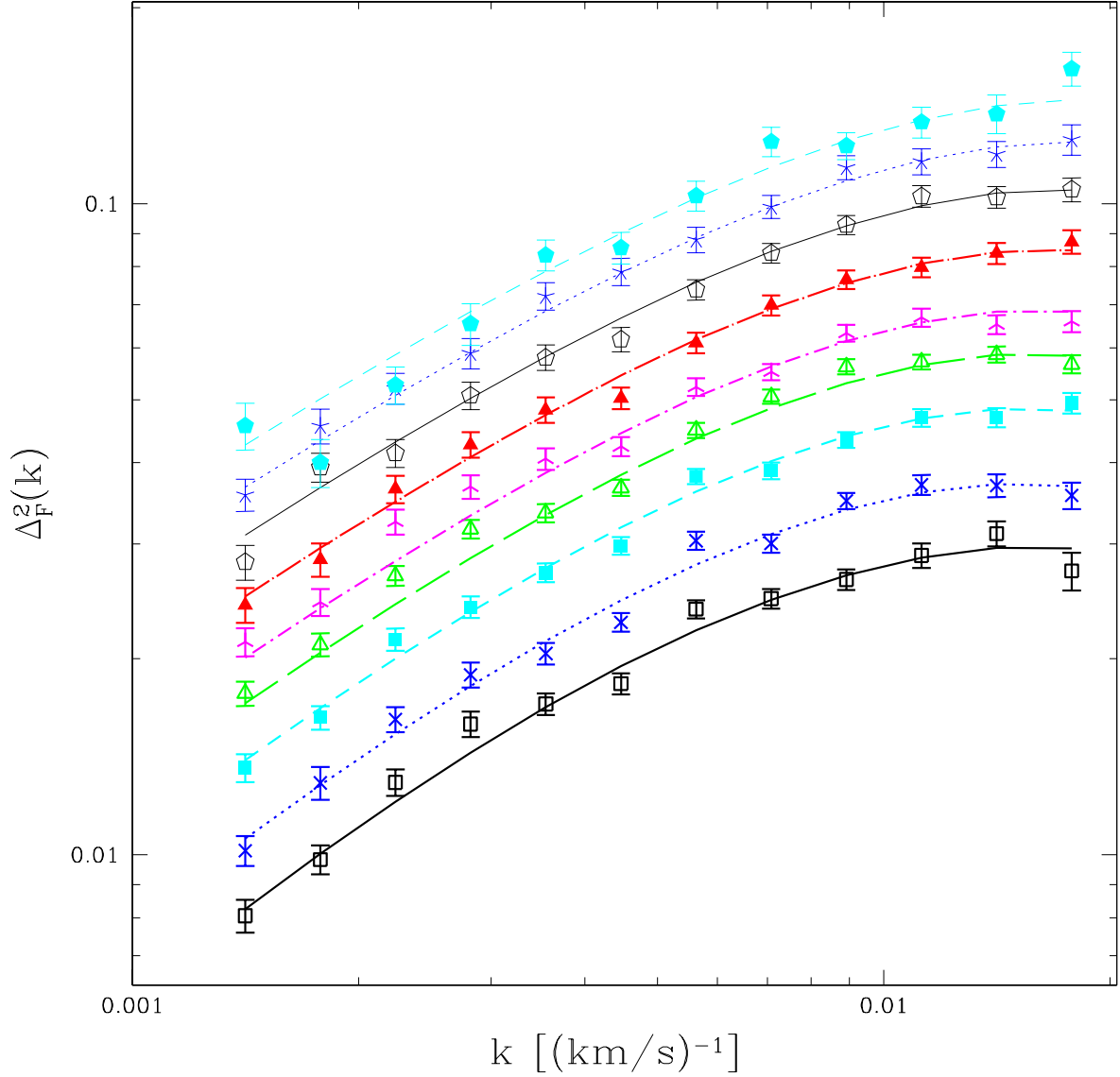}
\caption{
Points with error bars show the observed \PF.  Lines show our first attempt to 
fit a theoretical model, which is not a good fit to the data.  From bottom to top ---
z=2.2:  black, solid line, open square;
z=2.4:  blue, dotted line, 4-point star (cross);
z=2.6:  cyan, dashed line, filled square;
z=2.8:  green, long-dashed line, open triangle;
z=3.0:  magenta, dot-dashed line, 3-point star;
z=3.2:  red, dot-long-dashed line, filled triangle;
z=3.4:  black, thin solid line, open pentagon;
z=3.6:  blue, thin dotted line, 5-point star;
z=3.8:  cyan, thin dashed line, filled pentagon.
}
\label{naivefit}
\end{figure}

Figure \ref{naivefit} shows our first fit to the standard \PF\ results.
The value of $\chi^2=193.7$ is much too high for approximately 106 degrees
of freedom (we are marginalizing over a large number of nuisance parameters, 
but these generally are externally constrained so they do not reduce the
number of degrees of freedom).  Including $\alpha_{\rm eff}$ as a free
parameter does not improve the fit significantly.
It appears that much of the disagreement
comes from bumps in the power spectrum, e.g., at $k\sim 0.003~\ikms$.
This motivates us to look at the correlation function of the flux. 

\subsection{The Correlation Function and the SiIII Cross-Correlation 
\label{secxisi3}}

\begin{figure}
\plotone{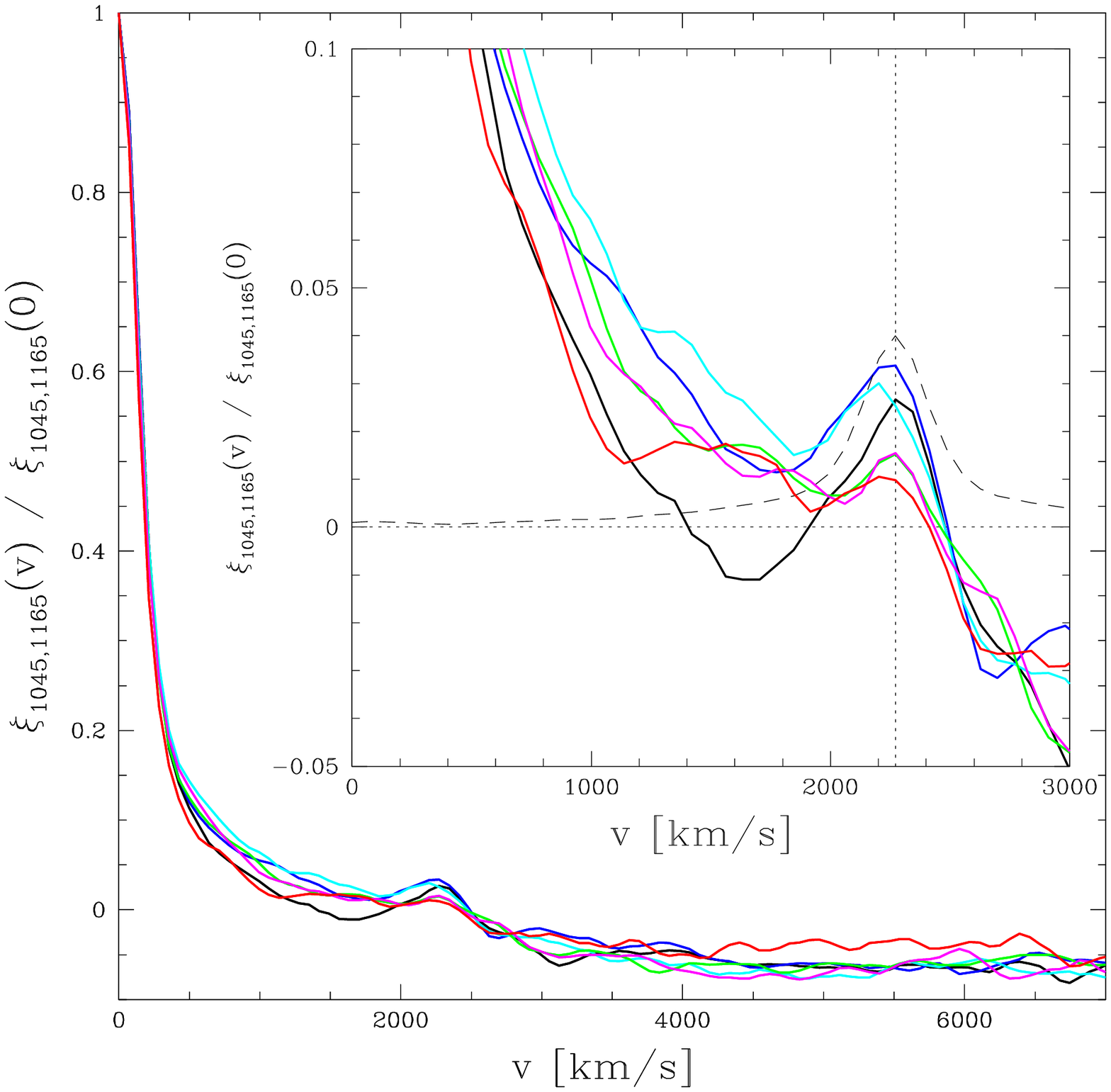}
\caption{The normalized correlation function, $\xi(v)/\xi(0)$, in 
the \lyaf\ region, 
uncorrected for resolution.  
In the inset panel, the solid lines show an expanded view 
of the SiIII-\lya\ cross-correlation bump, the dashed line shows 
$0.04~\xi(v-2271)/\xi(0)$
for comparison, and the vertical dotted line marks $v=2271\kms$.
Note that there is no evidence for any other metal with wavelength close
to \lya\ transition being important. In particular, we see no 
bump at $\sim5600\kms$ or $\sim 6700\kms$, corresponding to 
NV-\lya\ velocity differences.
This correlation function 
should not be used for any quantitative science, as
we have not corrected for resolution effects,
have not checked carefully for systematic errors, and have not
given statistical errors.  
}
\label{xiwinset}
\end{figure}

Sometimes features in the power spectrum are easier to understand
by looking at the correlation function, 
$\xi(v) = \left<\delta(x)\delta(x+v)\right>$ ($v$ is as usual a 
stand-in for wavelength differences, as is $x$ in this case).
We show the normalized correlation function, $\xi(v)/\xi(0)$ for the first 
six redshift bins in Figure \ref{xiwinset}.
The correlation function shows the expected behavior -- positive for
small $v$, negative for large $v$ -- except for an obvious bump
at $v \sim 2200\kms$.  We focus on this bump in the inset panel
of Figure \ref{xiwinset}.
The most likely explanation seems to be cross-correlation between 
\lya\ and SiIII absorption.  SiIII absorbs at $\lambda=1206.50$\AA, so
the SiIII absorption by gas at some point along the line of sight
will appear in the spectrum separated by 2271 km/s from the \lya\ absorption
by the same gas.  We see that the bump in $\xi(v)$
appears at this separation, and note that the features that 
ruined our power spectrum fit appear at the expected
multiples of $k=2\pi/2271=0.0028~\ikms$. 
Note that this is the only correlation seen; another metal 
correlation one might expect to see is 
NV ($\lambda=1238.8, 1242.8$\AA), but there is no apparent
feature at the corresponding velocity differences ($\Delta v\sim5600,6700$km/s), 
as seen in figure \ref{xiwinset}. 

What should we do about this SiIII-\lya\ cross-correlation,
since the poor $\chi^2$ suggests that it is too significant to ignore?
Our first guess might be that the SiIII-\lya\ correlation is 
a simple offset version of the \lya-\lya\ correlation, i.e., 
something like $\xi_{{\rm SiIII-Ly}\alpha}(v)\propto
\xi_{{\rm Ly}\alpha-{\rm Ly}\alpha}(|v|-2271\kms)$.
The simplest way to model this is to assume that the SiIII structure 
is equal to that of the \lyaf\ up to an overall normalization, 
$\delta_F=\delta(v)+a~\delta(v+v_3)$, 
where $\delta(v)$ is for \lya\ only
and $v_3=2271\kms$. The corresponding correlation function is
\begin{equation}
\xi_F(v)=(1+a^2)~\xi(v)+
a~ \xi(v+v_3)+
a~\xi(v-v_3),
\end{equation}
with corresponding power spectrum
\begin{equation}
P_F(k)=(1+a^2)~P(k)+2~ a~ \cos(v_3 k)~ P\left(k\right),
\label{Pwsi3}
\end{equation}
where unsubscripted $\xi$ and $P$ are understood to mean \lya-\lya.
For our first fit to \PF\ accounting for SiIII using equation (\ref{Pwsi3}),
we assume $a=f/[1-\bar{F}(z)]$, with $f$
as a single extra free parameter of the fit.  We find a remarkable improvement
in $\chi^2$, from 193.7 to 130.9.  
We find $f\sim 0.011$ ($a \sim 0.04$, 
depending on the redshift). The small value suggests that the relative 
contribution of SiIII to the autocorrelation is $a^2<0.004$, which 
will not affect our fit results significantly 
(see \S \ref{secmodifications}). 
We thus only need to estimate the cross-correlation term.
We also tried allowing a power law $1+z$ dependence for $f$, 
but the improvement in fit, $\Delta \chi^2=1.1$, was not
significant.

\begin{figure}
\plotone{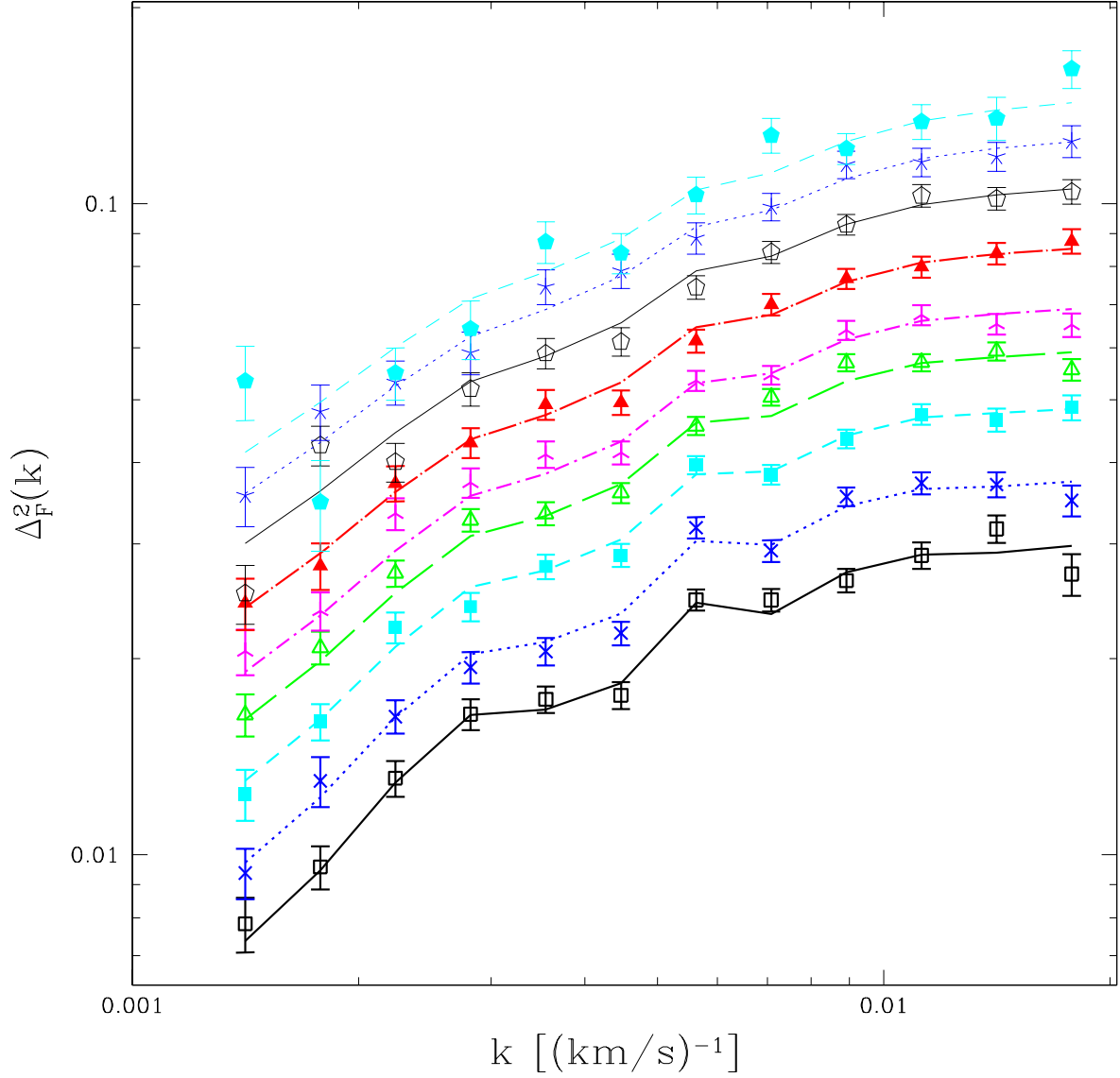}
\caption{
Points with error bars show the observed \PF.  Lines show our 
best fit after including SiIII absorption approximately in the 
theory.  See Fig. \ref{naivefit} for line and point definitions.
}
\label{bestfit}
\end{figure}

In the inset panel of Figure \ref{xiwinset} we plot scaled 
$\xi(v-2271)$, to show how the
shape of the bump compares to the zero-lag correlation.  
It is difficult to
compare the shapes by eye, because of the slope of the underlying  
correlation, but it appears that this model explains 
the cross-correlation reasonably well. 
We can allow for a change in scale using the slightly more general form
\begin{equation}
\xi_F(v)=\xi(v)+
a~ \xi\left[s \left(v+v_3\right)\right]+
a~\xi\left[s \left(v-v_3\right)\right]~.
\label{si3scalechange}
\end{equation}
Allowing $s$ to vary freely only improves $\chi^2$ 
by 0.7 (note that the logarithmic $k$-binning that we use may 
reduce our ability to constrain these parameters).
The error bars on other parameters may increase when we include
$z$ dependence of $f$ and allow $s$ to be free, 
so to be conservative
one probably wants to leave them free even though they are not 
needed.  In our standard fit in this paper, we allow $f$ to 
have $z$ dependence but fix $s=1$.
We show the improved fit to \PF\ in Figure \ref{bestfit}.

\subsection{Modifications of the Procedure \label{secmodifications}}

\begin{figure}
\plotone{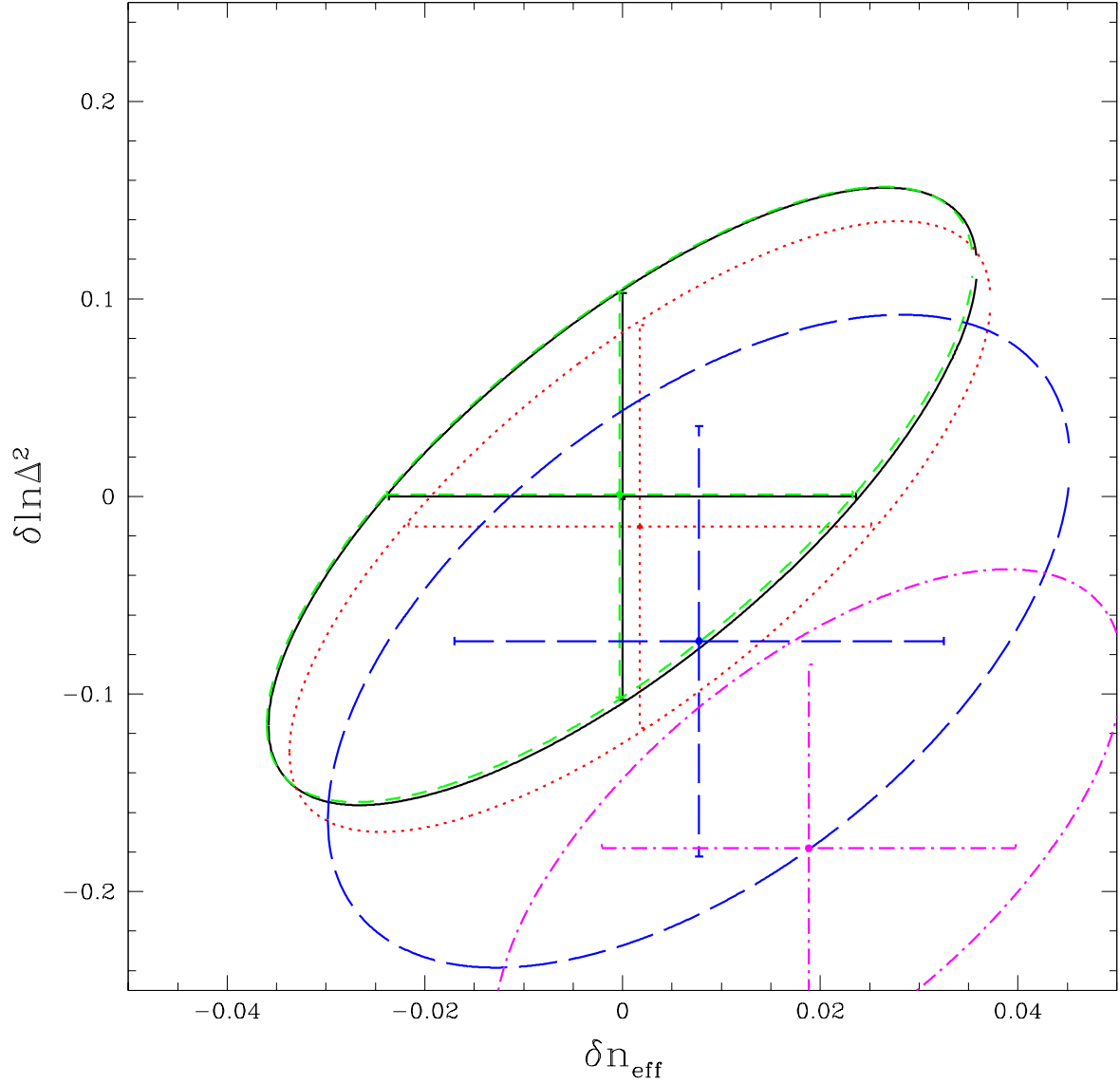}
\caption{
Fit results for variant noise treatments.
The error bars show the $1~\sigma$ error on each parameter.  
The ovals show $\Delta \chi^2=2.3$.  Standard case (with 5\% noise amplitude
freedom in each redshift bin):  black, solid lines.  
No individual noise estimate for each quasar:  red, dotted lines.
Noise amplitude freedom 0.5\% (50\%):  green, short-dashed lines 
(blue, long-dashed lines).  The magenta, dot-dashed line shows the 
result using the pipeline noise estimates.
}
\label{fitcontnoisevar}
\end{figure}

The pipeline developed for this analysis 
includes many improvements and corrections
that were added throughout the development. 
It is worth taking a step back to ask how important the various 
corrections are for the final result. 
Table \ref{modtab} lists various modifications of our procedure (described
individually below), and quantifies their effects on the fit results.  
In each case we show the 
change in the best fit $\Delta^2$ and $n_{\rm eff}$ relative to our standard 
fit, and their error bars for comparison to the standard fit.  We give
$\chi^2$ to indicate the goodness of fit of the theory to the modified
measured \PF.  We reiterate that we are not asserting the correctness
of the theory that we use in the fitting -- we give these $\chi^2$ values
and other fitting results only to show trends.
We list $\Delta \chi^2$ between the standard procedure
best fit and the variant best fit, to give an indication of how significant
the deviation is (this is necessary because the errors are correlated so
simply knowing $\Delta^2$ and $n_{\rm eff}$ and their errors does not give the
full picture -- see Figure \ref{fitcontnoisevar} for an example of the full 
error contours).  
Because the statistical fluctuations between these 
different fits should be small, a $1~\sigma$ difference (or even less) 
should be interpreted as ``significant'', however, since we believe that our
standard fit is better or more conservative than all of the variants,
our systematic error should generally be smaller than the deviation shown.
Note that where applicable the changes in procedure are only applied to
the $P_{1041,1185}$ calculation, not the $P_{1268,1380}$ result that
is used in the background subtraction (small changes in $P_{1268,1380}$ 
have no effect on the final results).

\begin{figure}
\plotone{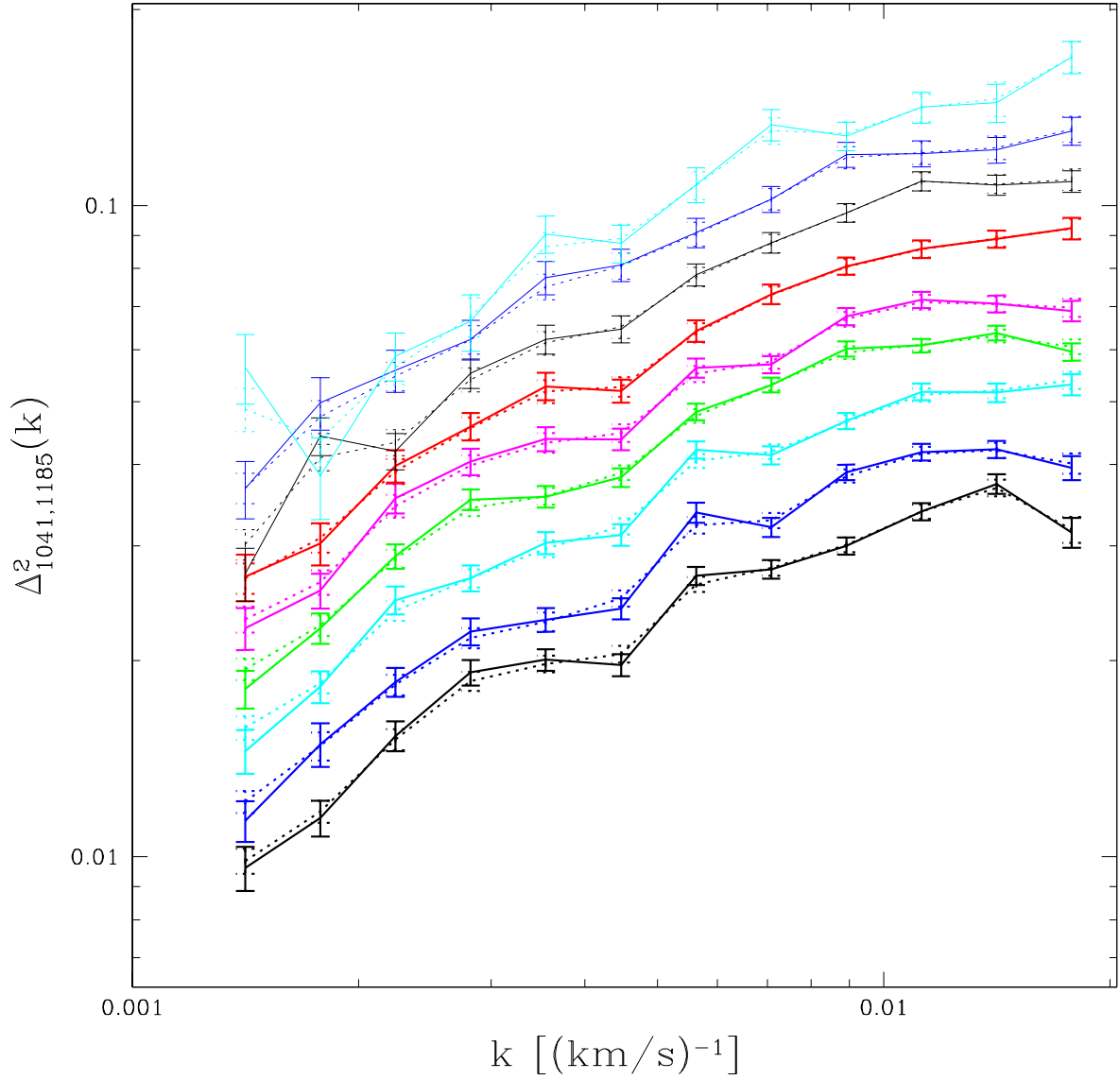}
\caption{
Dotted lines connect the power spectrum points before diagonalization 
of the window matrix.  Solid lines show the points after 
diagonalization. }
\label{compinvertwin}
\end{figure}

\begin{figure}
\plotone{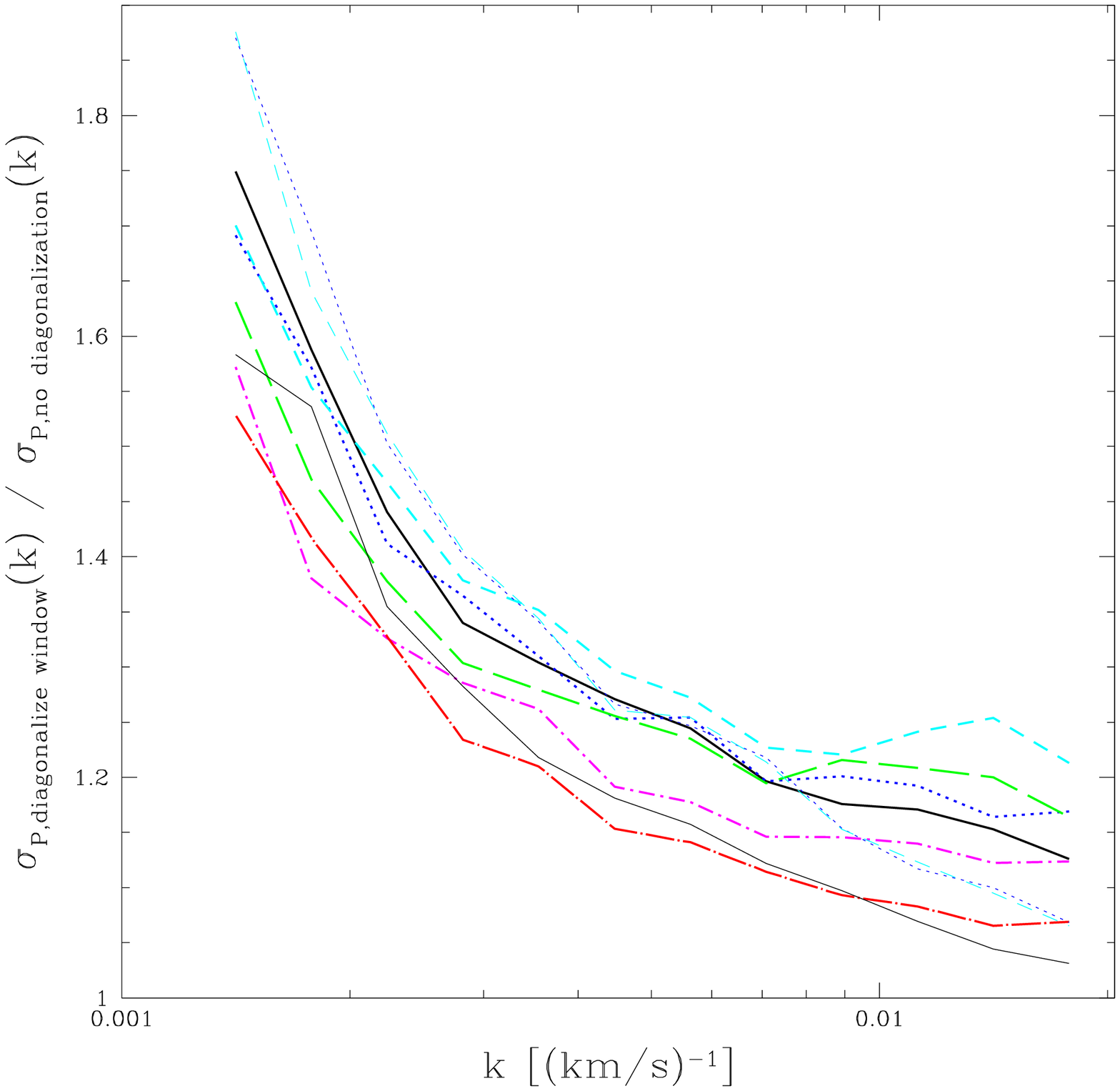}
\caption{
Ratio of the diagonal errors on our $P_{1041,1185}(k,z)$ points (connected) 
after the window matrix diagonalization to before the window matrix 
diagonalization, with the usual line definitions.}
\label{comperrinvertwin}
\end{figure}

Our first variant is to not diagonalize the window matrix.
Figure \ref{compinvertwin} shows the measured power spectrum before
and after diagonalization.
Figure \ref{comperrinvertwin} shows the ratio of the diagonal errors 
after diagonalization of the window matrix to before 
diagonalization.
Not diagonalizing the window matrix does lead to a significant change in 
the fitted parameter values, and the error on $n_{\rm eff}$ 
decreases by about 12\%.  We are, in effect, using information from a wider
range of scales, but this forces us to use theory results outside their 
range of validity (e.g., at low $k$ we need to extrapolate beyond the
size of our simulation boxes).  
Note that the change in error on $n_{\rm eff}$, from
0.024 to 0.021 implies that we should expect a random difference between
the two results with typical size $(0.024^2-0.021^2)^{1/2}=0.012$, i.e.,
what might seem like a surprisingly large part of the difference between 
the results could be random. 

As discussed above, the correction for SiIII-\lya\ correlation is very 
important to the goodness of our fit. It is less
important for the best fit values, changing them only  
by $0.8~\sigma$ for $n_{\rm eff}$ and $0.4~\sigma$ 
for $\Delta^2$.  
Normally we allow a power law dependence on redshift for the amplitude 
of the SiIII-\lya\ correlation, but removing this freedom makes almost no 
difference.  Allowing the correlation scale 
for the SiIII-\lya\ correlation to be different than for \lya-\lya\ 
(freeing $s$ in eqn. \ref{si3scalechange} -- we usually fix this in
this paper for technical reasons) has only a very small 
effect.  Including the SiIII-SiIII autocorrelation term (the $a^2$ 
part of eqn. \ref{Pwsi3}) in the 
fit has essentially no effect.

\begin{figure}
\plotone{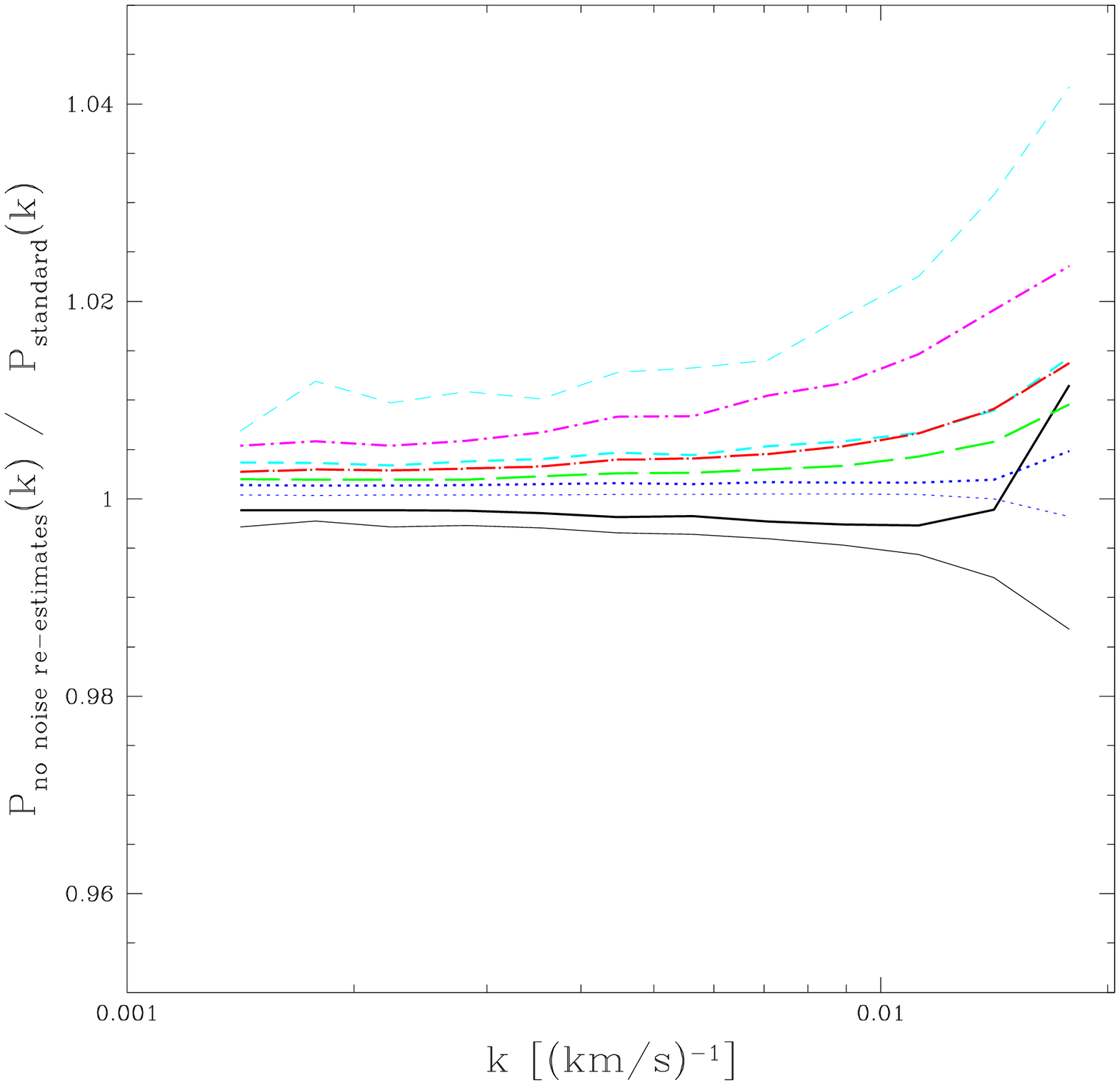}
\caption{
Ratio of $P_{1041,1185}$ computed without quasar-by-quasar 
noise re-estimates (a constant 16\% extra noise power was assumed
instead) to the standard case.
}
\label{nofitnoiserat}
\end{figure}

For our standard fit, we allow variation in the noise amplitude at
each redshift, represented by a multiplicative parameter subject 
to a 5\% rms Gaussian constraint. 
Our fitting procedure then marginalizes over this component.  
Reducing this constraint to 0.5\% (i.e., fixing
the noise) produces no change in our fit result, and does 
not even reduce the error bars noticeably.  Leaving the noise essentially
free makes a noticeable difference in
the fit results, decreasing the amplitude by about $2/3~\sigma$, 
increasing its error by 20\%,
and decreasing $\chi^2$ to 123.8. 
Changes at this level are expected when we remove the constraints on 
some parameters, and do not imply that the constraints were too small
(i.e., we are effectively removing 9 data points from the fit so we
generally expect a decrease in $\chi^2$, increase in the error bars, 
and some corresponding drift in the parameter values). 
Removing our spectrum-by-spectrum noise estimation makes
very little difference.  
Finally, we note that if we did not correct the noise as discussed in
\S \ref{secexpcomb}, the results would change significantly.
As an example, we show the fit  
results in the $\delta \ln \Delta^2-\delta n_{\rm eff}$ plane in Figure 
\ref{fitcontnoisevar}, for our standard case and these noise-related 
variants.
We show the ratio of the power without individual noise estimates
for each quasar to our standard case in Figure \ref{nofitnoiserat}.

Our requirement that the principal
components of the error matrix have at least the Gaussian level of variance
makes no difference to the fit results, although it improves $\chi^2$ for 
the fit a little bit.  Simply using the Gaussian error matrix instead
of bootstrap errors makes no difference to the fit results but   
increases $\chi^2$ significantly. 
Ignoring the bootstrap 
error correlations increases the error on $n_{\rm eff}$ by about 12\%, and 
significantly reduces $\chi^2$. 

We normally use the range $1268<\lr<1380$\AA\
for our background subtraction (i.e., subtract $P_{1268,1380}$).
Removing the background subtraction entirely reduces the inferred amplitude
by $2\sigma$, and the slope by $1\sigma$, and results in a very large $\chi^2$
(the error on $n_{\rm eff}$ also decreases significantly, but this is mostly
because of the change in the best fit values, not because of uncertainty 
in the background
subtraction).  
(Note that removing the background subtraction, which increases \PF, 
decreases the inferred amplitude because the fitted $\bF$ decreases more 
than enough to offset the increase in \PF.)
Clearly the background cannot be ignored.
Using $P_{1409,1523}$ instead produces a somewhat disturbingly large
0.028 ($1.1~\sigma$) increase in 
$n_{\rm eff}$.  We expect the longer wavelength range to give 
a less accurate estimate of the background power, because some metals are
missing, but further investigation shows that most of this difference is  
probably caused by CIV BALs adding power to the $1409<\lr<1523$\AA\ region. 
As we see in Table \ref{modtab}, removing the adjustment for noise dependence
of the background (see Equation \ref{fitdepeq})  brings the two background 
regions closer together (this 
is reflected in Figure \ref{diff12681409}).  Adding the 147 BAL quasars 
identified by our automated algorithm leads to a huge discrepancy (0.094 in 
$n_{\rm eff}$) when
we use the $P_{1409,1523}$ background, but only when we adjust for
noise level (without this the discrepancy for $n_{\rm eff}$,
not shown in the table, is only 0.029).  
Note that the BAL cut makes essentially no difference to our standard fit
using the $P_{1268,1380}$ background.
All of these differences are easy to understand:  First, BALs are known 
to be strongest in CIV absorption (\cite{2002ApJS..141..267H}), so it 
is not surprising that we see the effects of BALs primarily in this
wavelength region.  Second, both our original by-eye 
and subsequent automated removal of BALs inevitably identify the features
more easily in less noisy data, so the power from BALs naturally shows
up when we intentionally use the noisier spectra for the background 
power.  The fact that removing the 147 most obvious BALs has essentially
no effect on our basic result gives us confidence that any remaining 
BAL features in the \lyaf\ and $1268<\lr<1380$\AA\ regions are not 
significant.
  
To investigate the effect of a systematic uncertainty in the spectral
resolution, we include in our fits an overall factor of the 
form $\exp(\alpha k^2)$ multiplying the power spectrum, 
where $\alpha$ is a free parameter.  In our standard fit we impose an
external constraint on $\alpha$, $\pm(7\kms)^2$ rms.  
Essentially removing this freedom has no effect on the fit, while 
leaving $\alpha$ essentially free increases the error on the amplitude
by 40\%, and increases $n_{\rm eff}$ by $2/3\sigma$ (the change in 
fitted amplitude is certainly consistent with drift from the 
increased error).
As we show in Figure \ref{skypower5579}, our standard fit should be
conservative.

\begin{figure}
\plotone{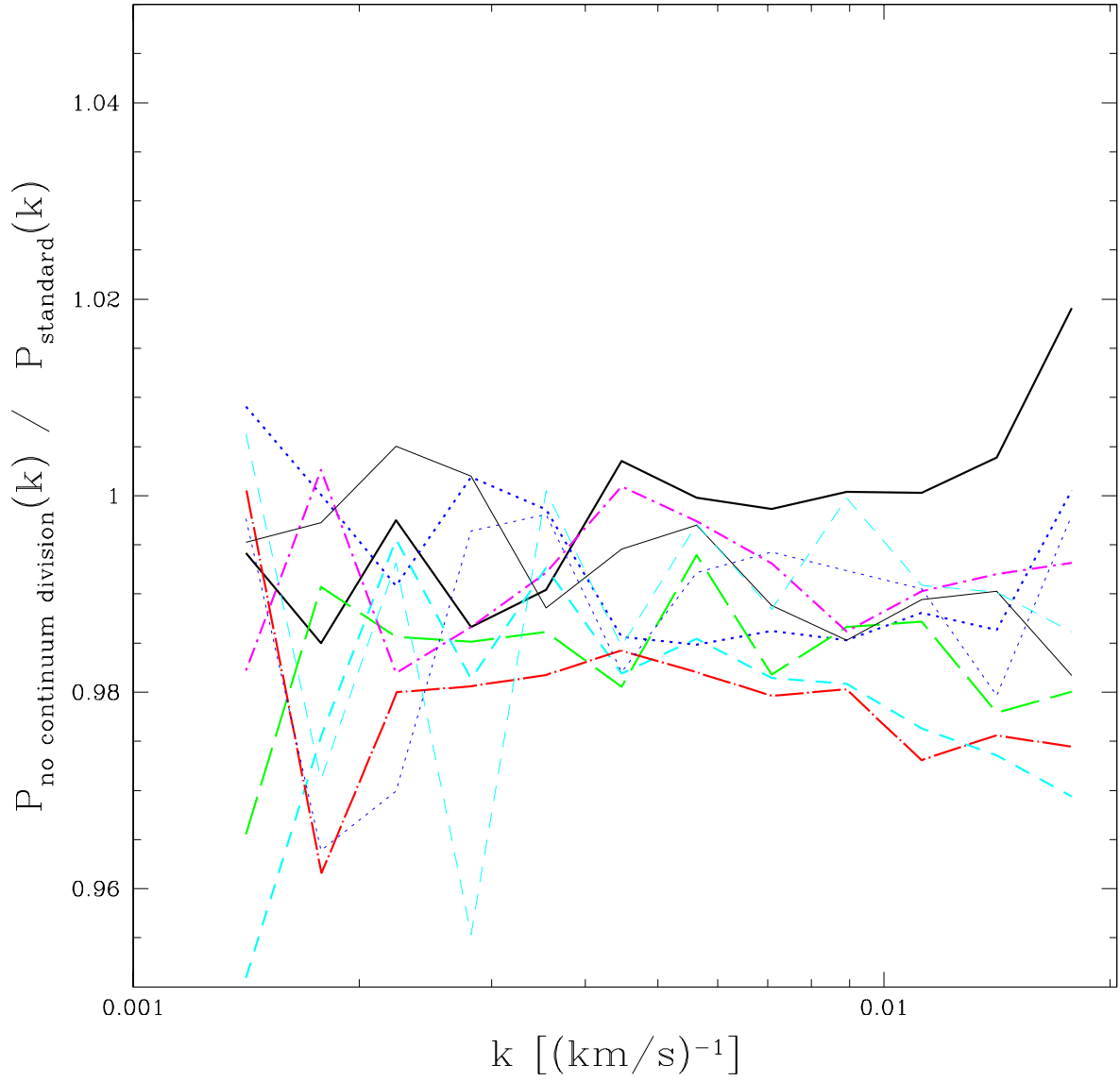}
\caption{
Ratio of $P_{1041,1185}$ computed without dividing by the continuum 
estimate to the standard case.
}
\label{nocontdivrat}
\end{figure}

\begin{figure}
\plotone{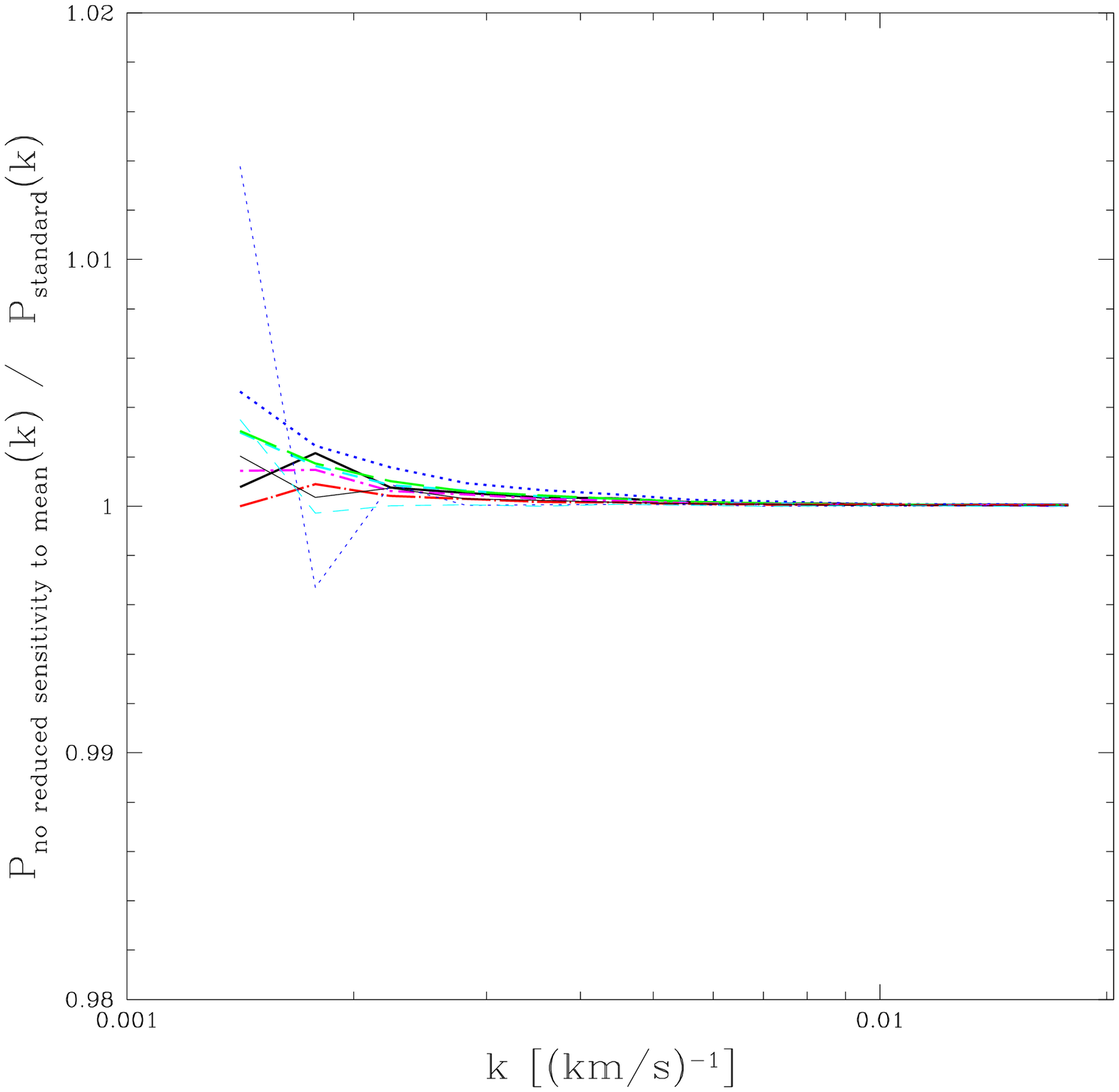}
\caption{
Ratio of $P_{1041,1185}$ computed without the large constant added
to the weight matrices to make the results less sensitive to the
mean of the chunks, relative to the standard case.  
}
\label{meanerrcovaddrat}
\end{figure}

Simply dividing each chunk of spectrum by its mean instead of also
dividing by
the mean continuum before estimating the power from the chunk 
makes little difference to the fit results.
Division by the mean continuum actually increases 
the measured 
flux power  by $\sim 0-2$\%, as we show in Fig. \ref{nocontdivrat}.  
We have performed a preliminary PCA analysis to try to model
fluctuations around the mean continuum.  When we use continua
for each quasar composed of 13 PCA eigenvectors, our results 
change only a little ($n_{\rm eff}$ is reduced by $0.4 \sigma$),
and $\chi^2$ increases, probably an indication of the unsatisfactory
level of noise that we know remains in our estimates.
As we see in Figure \ref{meanerrcovaddrat},
the modification of adding a large constant to the weight matrix to
make our measurement less sensitive to the mean of each chunk has 
little effect (the effect is larger on larger scales).

\begin{figure}
\plotone{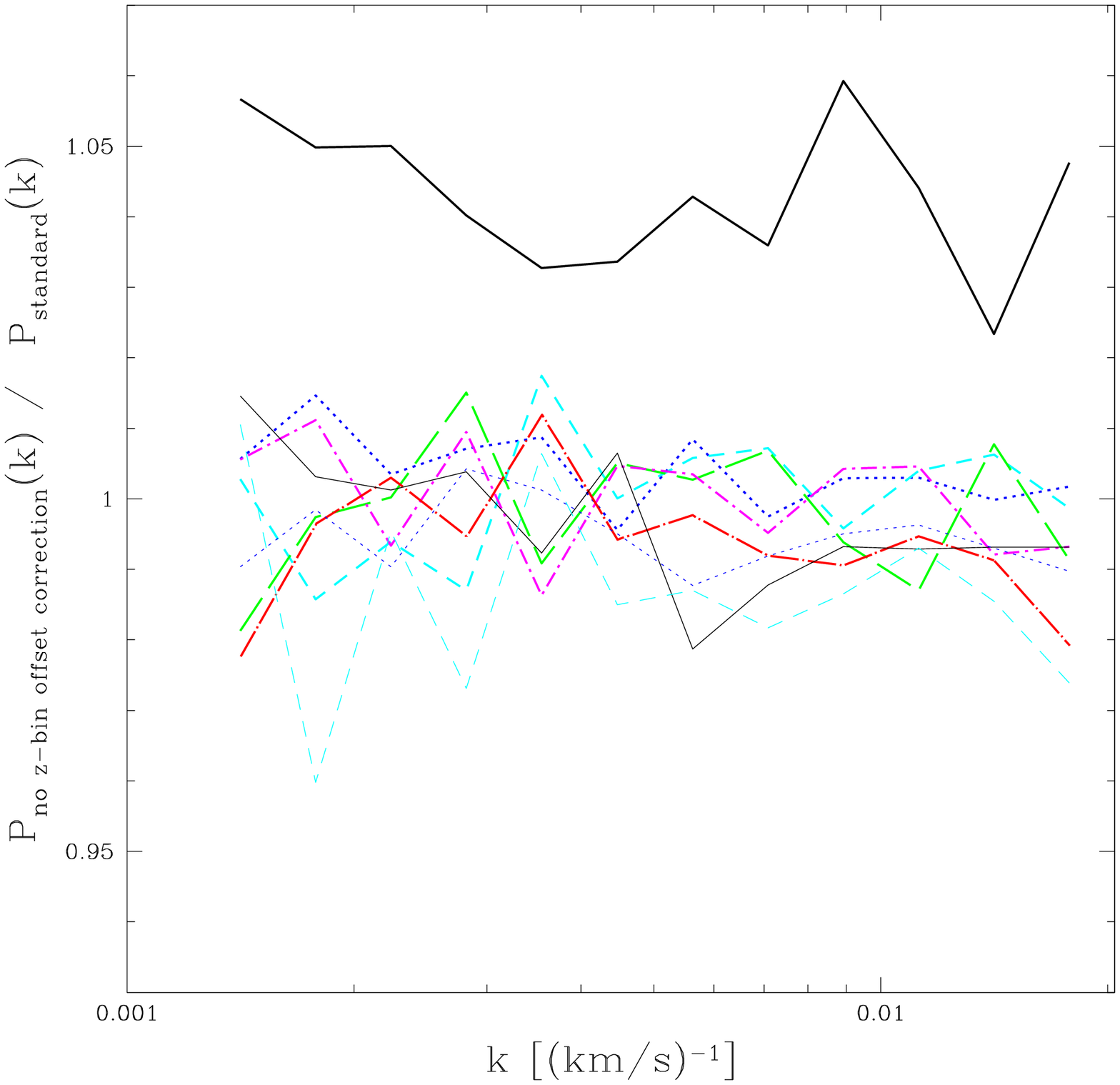}
\caption{
Ratio of $P_{1041,1185}$ computed with no correction for
the offset between the defined center of each redshift bin and
the center of weight of the data to the standard case.
}
\label{nobinzcorrat}
\end{figure}

The line ``no bin-redshift correction'' in Table \ref{modtab} refers to
removing the correction for evolution in the power across the width of the 
redshift bins (see eqn. \ref{DFalpha}).  We see (Fig. \ref{nobinzcorrat}) 
that this correction 
mostly affects the lowest redshift bin (where the low-$z$ edge
of the bin is empty of data) and has little effect on the fit (not
surprisingly, leaving out this correction increases $\chi^2$).

\begin{figure}
\plotone{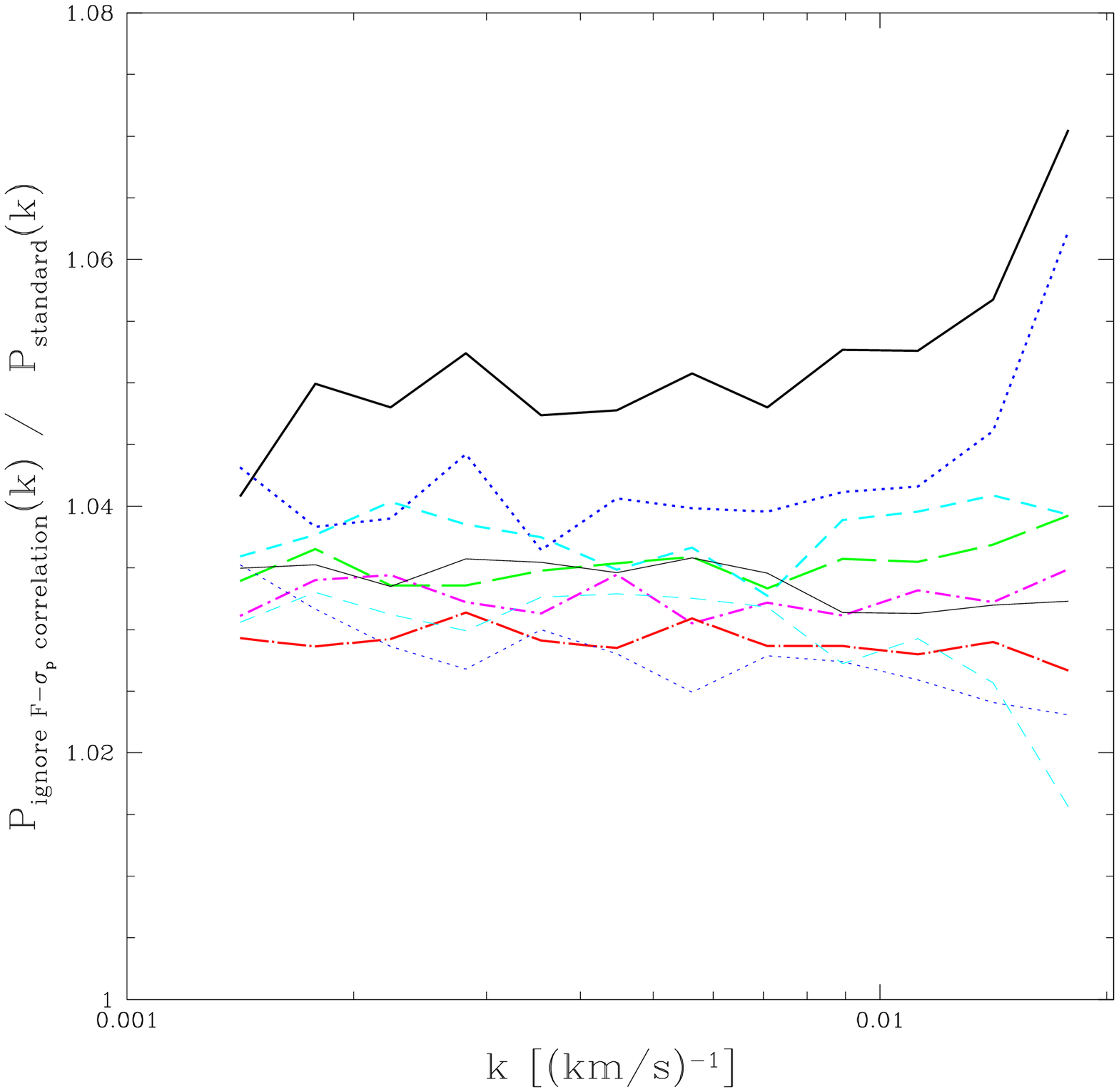}
\caption{
Ratio of $P_{1041,1185}$ computed without accounting for the
correlation between the noise amplitude and the observed flux 
to the standard case.
}
\label{nosopherrrat}
\end{figure}

\begin{figure}
\plotone{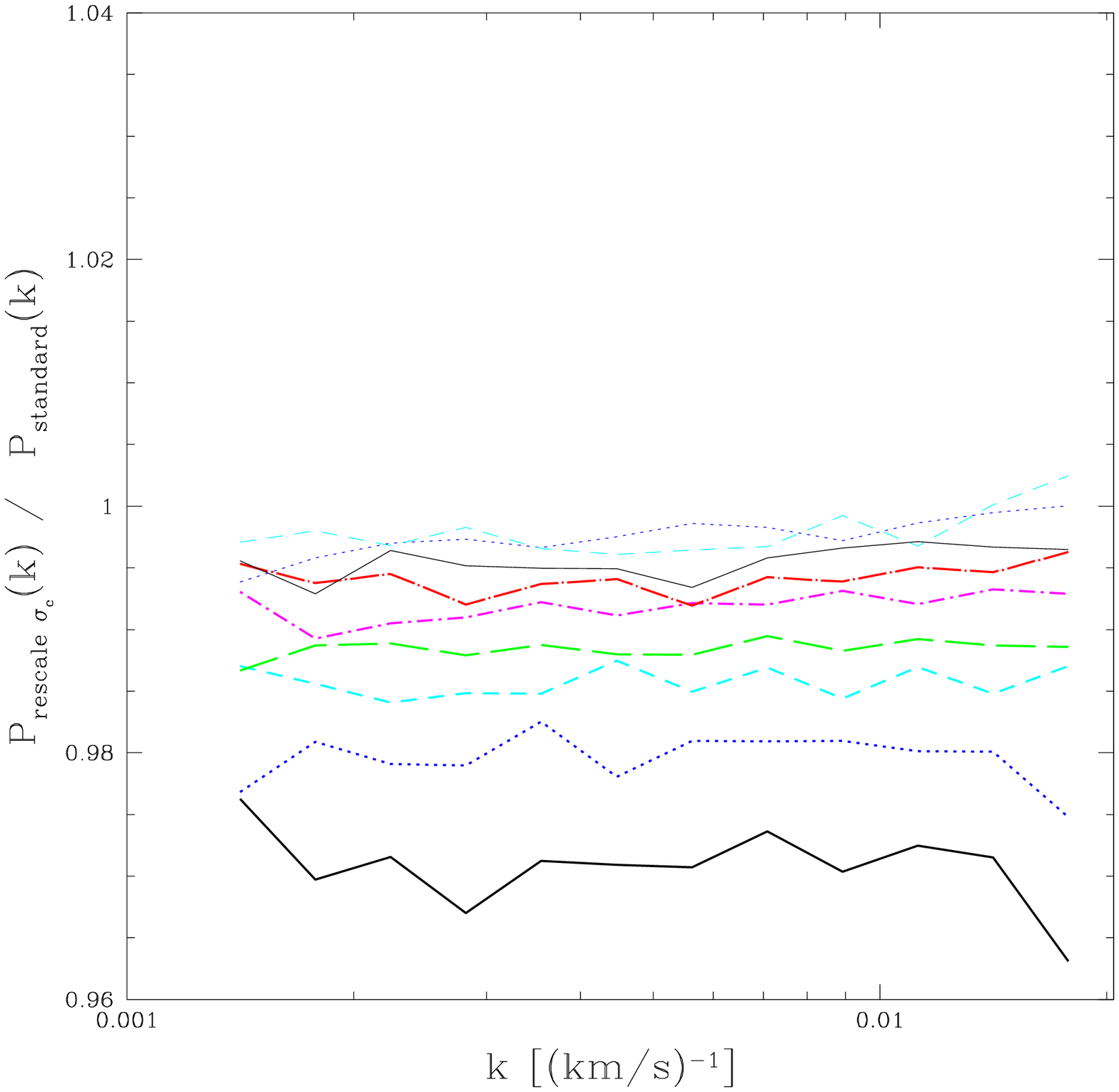}
\caption{
Ratio of $P_{1041,1185}$ computed using an alternative estimate of
the fraction of the noise that is due to photon counting noise  
associated with flux from the quasar (see text)
to the standard case.
}
\label{nobelievephotrat}
\end{figure}

The line ``ignore $F-\sigma_p$ correlation'' in Table \ref{modtab}
shows the change in the fitted parameters if we naively use the given 
noise estimates for 
weighting without accounting for the fact that there is a correlation 
between the flux estimate and the noise amplitude estimate for each pixel.
Figure \ref{nosopherrrat} shows that the bias is a fairly constant 3-5\% 
increase in the flux power.
The difference is not actually caused by the change in weighting used in the 
power spectrum estimation -- instead, the power is biased high because the 
estimation of the mean that each chunk of spectrum is divided by is biased 
low because low-flux pixels have smaller noise estimates.  
Ignoring this effect does not change our fit results.  
Normally we base our estimation of the amount 
of the noise that is due to quasar flux on the separate estimates we have
from the spectral reduction pipeline for the flux, sky, and read-noise 
contributions, however these estimates do not add up to the total noise
reported by the pipeline.  If we rescale the individual numbers to make
them consistent with the total (not necessarily the correct thing to do)
we see that the fit results are not changed significantly 
(the line ``rescale $\sigma_c$'' -- we use $\sigma_c$ to refer to noise
computed using the 
separate flux, sky, and read-noise estimates), although the power 
does change by as much as 3\% (Fig. \ref{nobelievephotrat} -- 
this difference would be a bit larger if we did not directly measure and 
correct for the cross-correlation between the noise amplitude and flux).

Finally, our power spectrum extraction code has a bias 
(partially related
to division by the mean of each chunk of spectrum),
that we correct for by dividing the result by 
Equation \ref{codebiaseq}. 
Removing
this correction decreases the estimated $n_{\rm eff}$ by about
$1/2~\sigma$ and increases the amplitude by a similar 
amount.  The combined change is actually quite significant
because it is transverse to the degeneracy direction for 
these parameters.

To summarize, most of the effects described above 
are small relative to the 
statistical errors on the final estimated parameters. 
We understand the cases where the difference is significant,
and expect that our standard method will be much more accurate
than the difference between it and the variants (we show these
variants to help the reader better understand our measurement).
These tests give us confidence
that the 
final results are very robust to small changes in the analysis 
pipeline. 

How sensitive are these conclusions to our assumptions about 
the nuisance parameters, $\bF$, $T_{1.4}$, and $\gamma$, 
i.e., if these constraints improve in the future, will we need 
to worry more about systematic errors in \PF?
We investigate this by first fixing all the parameters in the 
fit (including removing the noise amplitude uncertainty, 
resolution uncertainty, and freedom in the SiIII correction), 
so the only uncertainty is on \PF. 
Table \ref{modtab} (the ``fixed nuisance parameters'' line) shows 
that the error on the
amplitude improves dramatically, by a factor of 5.  The
error on $n_{\rm eff}$ improves by a factor of 2.  So in 
principle the amplitude error can be improved a lot relative to
potential systematic errors, and $n_{\rm eff}$ improved as
well \citep[see also ][]{2003MNRAS.344..776M}.  
The next
line (``optimistic $\bF$''), where we assume the HIRES 
constraint on $\bF$ 
is improved by a factor of 5, and the SDSS constraint by a 
factor of 2,
shows that $\Delta^2$ is substantially degenerate with $\bF$ (as
expected), but $n_{\rm eff}$ is less degenerate.  
Improving the
constraints on $T_{1.4}$ and $\gamma$ by factors of 5, in addition
to the improved constraints on $\bF$, leads to
little further improvement.  
Finally, for comparison,
we tried simply reducing the errors on \PF\ by a factor of $\sqrt{3}$, 
and found that the error on $n_{\rm eff}$ decreases by almost the 
same factor (1.6), but the error on $\Delta^2$ decreases less 
(a factor of 1.2).  SDSS will collect a factor of $\sim 3$ 
more data than we have in the present sample.
We conclude that the error on $n_{\rm eff}$ 
can easily be reduced by simply gathering more data, while 
improvements on $\Delta^2$ can be made by improving the errors on
$\bF$.  

\begin{deluxetable}{lccccccc}
\tablecolumns{8}
\tablecaption{Effect of uncertainties or modifications of
the \PF\ measurement procedure on the inferred linear 
power spectrum  \label{modtab}}
\tablehead{ \colhead{Variant\tablenotemark{a}} & 
\colhead{$\delta \ln \Delta^2$} 
& \colhead{$\sigma_{\ln \Delta^2}$}
& \colhead{$\delta n_{\rm eff}$} 
& \colhead{$\sigma_{n_{\rm eff}}$} 
& \colhead{$\Delta \chi^2$ \tablenotemark{b}}
& \colhead{$\chi^2$} \tablenotemark{c}}
\startdata
Standard fit & 0 & 0.10 & 0 & 0.024 & 0 & 129.7 \\
No window diagonalization & -0.06 & 0.10 & -0.024 & 0.021 & 1.4 & 129.5  \\
No SiIII correction & 0.04 & 0.10 & 0.017 & 0.021 & 0.7 & 193.7 \\
$z$-independent SiIII & 0.00 & 0.10 & 0.0 & 0.024 & 0.0 & 130.9 \\
variable width SiIII & -0.02 & 0.11 & -0.006 & 0.025 & 0.1 & 129.0  \\
Include SiIII-SiIII term & -0.02 & 0.10 & -0.003 & 0.023 & 0.0 & 132.1 \\
$\sigma_{\rm noise power}=0.5$\% & 0.00 & 0.10 & -0.000 & 0.024 & 0.0 & 130.3 \\
$\sigma_{\rm noise power}=50$\% & -0.08 & 0.12 & 0.008 & 0.025 & 1.1 & 123.8 \\
No individual noise re-estimation & -0.02 & 0.10 & 0.002 & 0.023 & 0.1 & 128.1 \\
Believe pipeline noise & -0.20 & 0.11 & 0.019 & 0.021 & 9.6 & 129.9 \\
No Gaussian floor on errors & 0.01 & 0.10 & 0.002 & 0.024 & 0.0 & 133.1  \\
Gaussian errors & 0.02 & 0.10 & 0.001 & 0.023 & 0.1 & 151.7 \\
Ignore error correlations & 0.04 & 0.11 & 0.002 & 0.027 & 0.2 & 117.2 \\
No background subtraction & -0.20 & 0.10 & -0.022 & 0.019 & 3.6 & 169.6 \\
Background 1409-1523\AA\ & 0.05 & 0.10 & 0.028 & 0.025 & 1.5 & 133.2  \\
No background noise matching & -0.07 & 0.10 & -0.004 & 0.021 & 0.7 & 142.3 \\
Previous, but use 1409-1523\AA\ & -0.06 & 0.10 & 0.008 & 0.022 & 1.7 & 143.1 \\
No automated BAL cut & 0.01 & 0.10 & 0.003 & 0.023 & 0.0 & 127.0 \\
Previous, but use 1409-1523\AA\ & 0.14 & 0.10 & 0.094 & 0.025 & 16.0 & 156.6 \\
$(70\kms)^2$ resolution error & -0.11 & 0.14 & 0.015 & 0.024 & 1.4 & 126.4 \\
$(0.7\kms)^2$ resolution error & 0.02 & 0.10 & -0.003 & 0.024 & 0.3 & 130.4 \\
No continuum division & 0.02 & 0.10 & 0.002 & 0.024 & 0.1 & 132.2 \\
PCA continuum division & 0.02 & 0.10 & -0.010 & 0.024 & 0.8 & 139.1 \\
No reduced sensitivity to mean & -0.00 & 0.10 & -0.002 & 0.024 & 0.0 & 130.5 \\
No bin-redshift correction & -0.02 & 0.10 & -0.006 & 0.023 & 0.1 & 137.2 \\
Ignore $F-\sigma_p$ correlation & 0.00 & 0.10 & 0.002 & 0.024 & 0.0 & 128.7 \\
rescale $\sigma_c$ & 0.00 & 0.10 & 0.002 & 0.024 & 0.0 & 129.0 \\
No code bias correction & 0.06 & 0.10 & -0.013 & 0.024 & 3.0 & 132.3 \\
8000 bootstrap sets & 0.00 & 0.10 & 0.000 & 0.024 & 0.0 & 128.5 \\
fixed nuisance parameters & 0.083 & 0.021 & -0.025 & 0.012 & --- & --- \\ 
optimistic $\bF$ & 0.068 & 0.062 & 0.009 & 0.019 & --- & --- \\
optimistic $T_{1.4}$, $\gamma$ & 0.002 & 0.082 & -0.014 & 0.021 & --- & --- \\
optimistic $\bF$, $T_{1.4}$, $\gamma$ & 0.002 & 0.051 & -0.016 & 0.018 & --- & --- \\
\PF\ errors divided by $\sqrt{3}$ & -0.051 & 0.081 & -0.000 & 0.015 & --- & --- \\
\enddata
\tablecomments{$z_p=2.6$, $k_p=0.009~\ikms$.}
\tablenotetext{a}{The meaning of each variant is 
explained in \S\ref{secmodifications}.}
\tablenotetext{b}{$\Delta \chi^2$ of the fitted parameters relative to 
the standard parameters, using the errors from the variant fit.}
\tablenotetext{c}{$\chi^2$ for the fit 
(essentially unrelated to $\Delta \chi^2$). }
\end{deluxetable}
\clearpage

\subsection{Subsamples of the Data \label{secsubsamples}}
Another way to test for systematic errors is to search for 
internal discrepancies between the different subsamples of 
the same data. Of course, there are only a finite 
number of possible 
subsamples we can try, so this test cannot be fully exhaustive.
In addition, with many such tests performed one must worry that some 
will give an apparently statistically significant deviation 
just by random chance. 
Table \ref{subtab} shows results of splitting the data 
into roughly equal weight subsamples, defined by various properties 
of the spectra that, at least at first glance, should not be 
correlated with the measured power.  In practice, we rank
the spectra by the property of interest and split the sample into halves
by requiring that the Gaussian errors on the $k=0.007~\ikms$ point are 
equal for the two halves (the bootstrap errors will not be precisely equal).   
We list the probability
of obtaining $\chi^2$ greater than the value computed by differencing the 
power spectra (these differences include the different background subtraction
computed using eqn. \ref{fitdepeq} for different noise levels).  
We also list the fitting parameter results for each 
subsample, and give the probability for obtaining the observed level of
difference between the fits.  Because these subsamples are basically 
independent, deviations within the error bars are expected and are not an 
indication of systematic errors.  
We describe these subsample splits below.

\begin{deluxetable}{lcccccc}
\tablecolumns{7}
\tablecaption{Comparison between subsamples of
the data \label{subtab}}
\tablehead{ 
\colhead{Split} 
& \colhead{$P_{>\chi^2}$\tablenotemark{a}}
& \colhead{$\delta \ln \Delta^2_<$\tablenotemark{b}} 
& \colhead{$\delta n_{\rm eff,<}$} 
& \colhead{$\delta \ln \Delta^2_>$} 
& \colhead{$\delta n_{\rm eff,>}$} 
& \colhead{$P_{>\chi^2}$} \\
& (points) &&&&& (fit)
}
\startdata
$\lr$ & 40\% & $-0.03\pm0.12$ & $-0.027\pm0.031$ & $0.04\pm0.11$ & $0.018\pm0.029$ & 51\% \\
noise & 10\% & $-0.01\pm0.11$ & $0.020\pm0.028$ & $-0.01\pm0.12$ & $-0.001\pm0.030$ & 76\% \\
noise (raw)\tablenotemark{c} & 0.0006\% & $-0.05\pm0.11$ & $0.027\pm0.029$ & $-0.19\pm0.13$ & $-0.010\pm0.026$ & 61\% \\
sky & 5.9\% & $-0.02\pm0.11$ & $-0.001\pm0.028$ & $0.02\pm0.12$ & $0.014\pm0.030$ & 93\% \\
$\sigma_w-\sigma_c$ & 34\% & $0.02\pm0.11$ & $0.029\pm0.030$ & $-0.01\pm0.12$ & $-0.011\pm0.030$ & 49\% \\
read noise & 94\% & $0.08\pm0.11$ & $0.020\pm0.030$ & $-0.05\pm0.12$ & $-0.015\pm0.028$ & 68\% \\
cont. $\chi^2/\nu$ & 33\% & $-0.04\pm0.11$ & $0.017\pm0.028$ & $0.06\pm0.12$ & $0.001\pm0.030$ & 40\% \\
resolution & 73\% & $0.08\pm0.11$ & $0.036\pm0.031$ & $-0.08\pm0.12$ & $-0.025\pm0.028$ & 32\% \\
flexure & 14\% & $0.10\pm0.11$ & $0.040\pm0.030$ & $-0.11\pm0.11$ & $-0.033\pm0.026$ & 19\% \\
alignment & 29\% & $0.09\pm0.11$ & $0.031\pm0.030$ & $-0.09\pm0.11$ & $-0.031\pm0.027$ & 29\% \\
exp. $\chi^2/\nu$ & 65\% & $0.01\pm0.11$ & $-0.015\pm0.029$ & $-0.00\pm0.12$ & $0.010\pm0.030$ & 63\% \\
error on mean  & 56\% & $-0.07\pm0.10$ & $-0.014\pm0.030$ & $0.07\pm0.11$ & $0.018\pm0.031$ & 67\% \\
error on $A_q$ & 40\% & $0.06\pm0.09$ & $0.020\pm0.027$ & $-0.01\pm0.12$ & $-0.014\pm0.030$ & 68\% \\
\enddata
\tablenotetext{a}{Probabilities may not be fully reliable 
because we have not demonstrated that $\chi^2$ is properly distributed.  }
\tablenotetext{b}{The subsample fit results cannot be combined to produce the 
result of the fit to the full data set because the underlying nuisance
parameters were not required to be the same.}
\tablenotetext{c}{The ``noise (raw)'' line shows the comparison without 
accounting for the noise dependence of the background.} 
\end{deluxetable}

\begin{figure}
\plotone{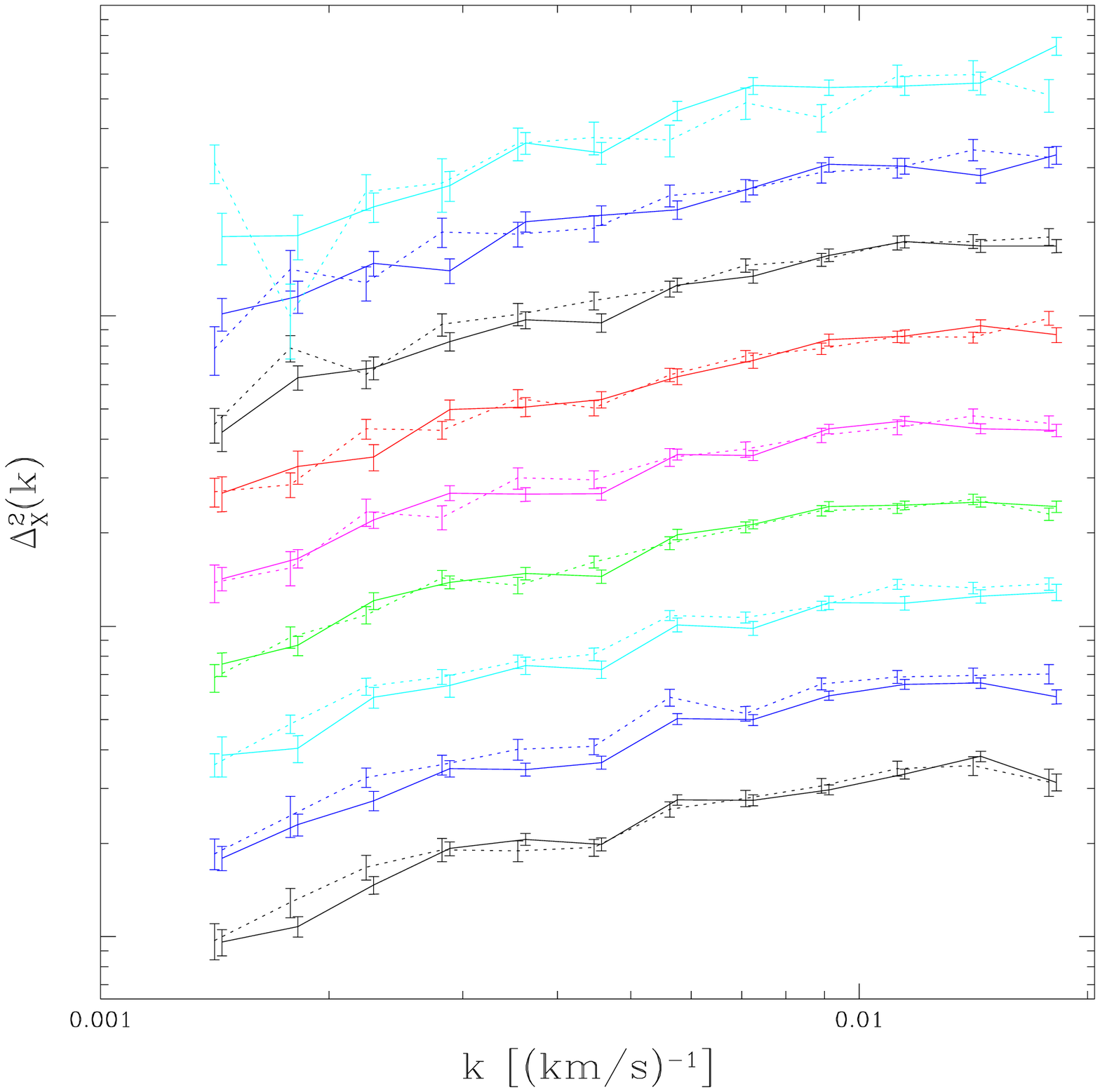}
\caption{Comparison of $P_{1041,1113}(k,z)$ (connected by
dotted lines) to
$P_{1113,1185}(k,z)$ (connected by solid lines, and shifted
slightly to the right).  
The different redshifts have
been shifted vertically by arbitrary amounts on this 
logarithmic plot ($z$ increases from bottom to top). 
}
\label{compspechalves}
\end{figure}

The power we measure should be independent of the rest frame region
of the quasar continuum in which it is measured.  
Figure \ref{compspechalves} shows $P_{1041,1113}(k,z)$ 
and $P_{1113,1185}(k,z)$ to test this expectation.
The results look pretty similar, but to compare them quantitatively,
we compute $\chi^2=(\mathbf{P}_<-\mathbf{P}_>)^T 
(\mathbf{C}_<+\mathbf{C}_>)^{-1} (\mathbf{P}_<-\mathbf{P}_>)$,
finding $\chi^2=111.0$  for 108 points.  The agreement appears perfect.
To compare the two in a different way, we perform separate fits of the 
linear mass power spectrum parameters
$\Delta^2$ and $n_{\rm eff}$ to \PF\ computed from $P_{1041,1113}(k,z)$
and $P_{1113,1185}(k,z)$.  The results,  
given in the first line of
Table \ref{subtab}, are consistent within the expected errors.
This test provides some evidence that power from continuum fluctuations
is not an important contribution to the total, beyond what we would 
expect from looking at the red side of the \lya\ emission line.
It is possible that the two halves of the forest could have 
significant extra continuum power, but if they do it has to be the
same in each half.

\begin{figure}
\plotone{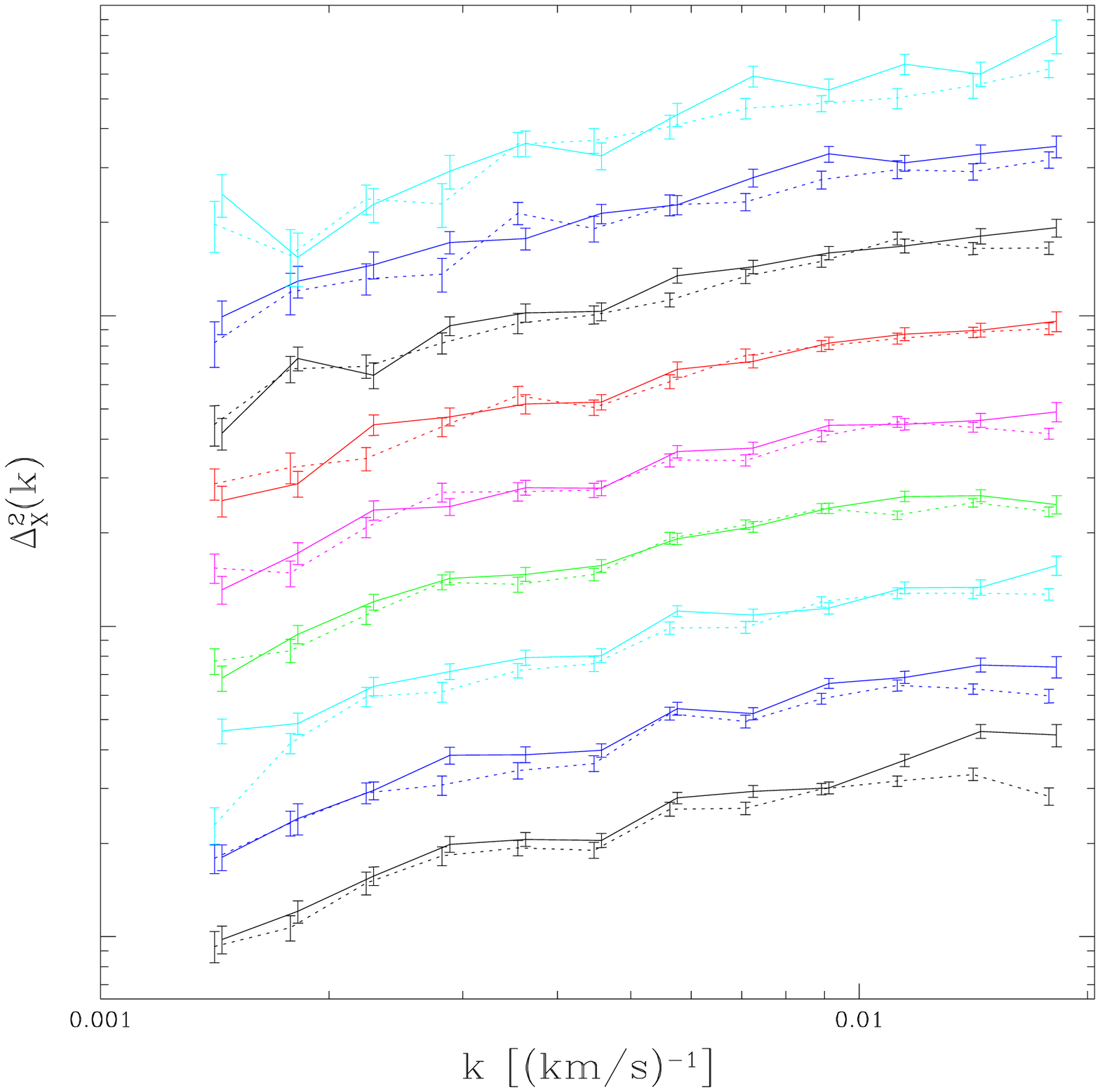}
\caption{
$P_{1041,1185}$ split by noise in the \lyaf\ region.  The dotted line connects
the low noise results, while the high noise results are
offset slightly to the right.  
The results at different redshifts have
been shifted vertically by arbitrary amounts ($z$ increases from bottom to top). 
}
\label{splitsigma}
\end{figure}

\begin{figure}
\plotone{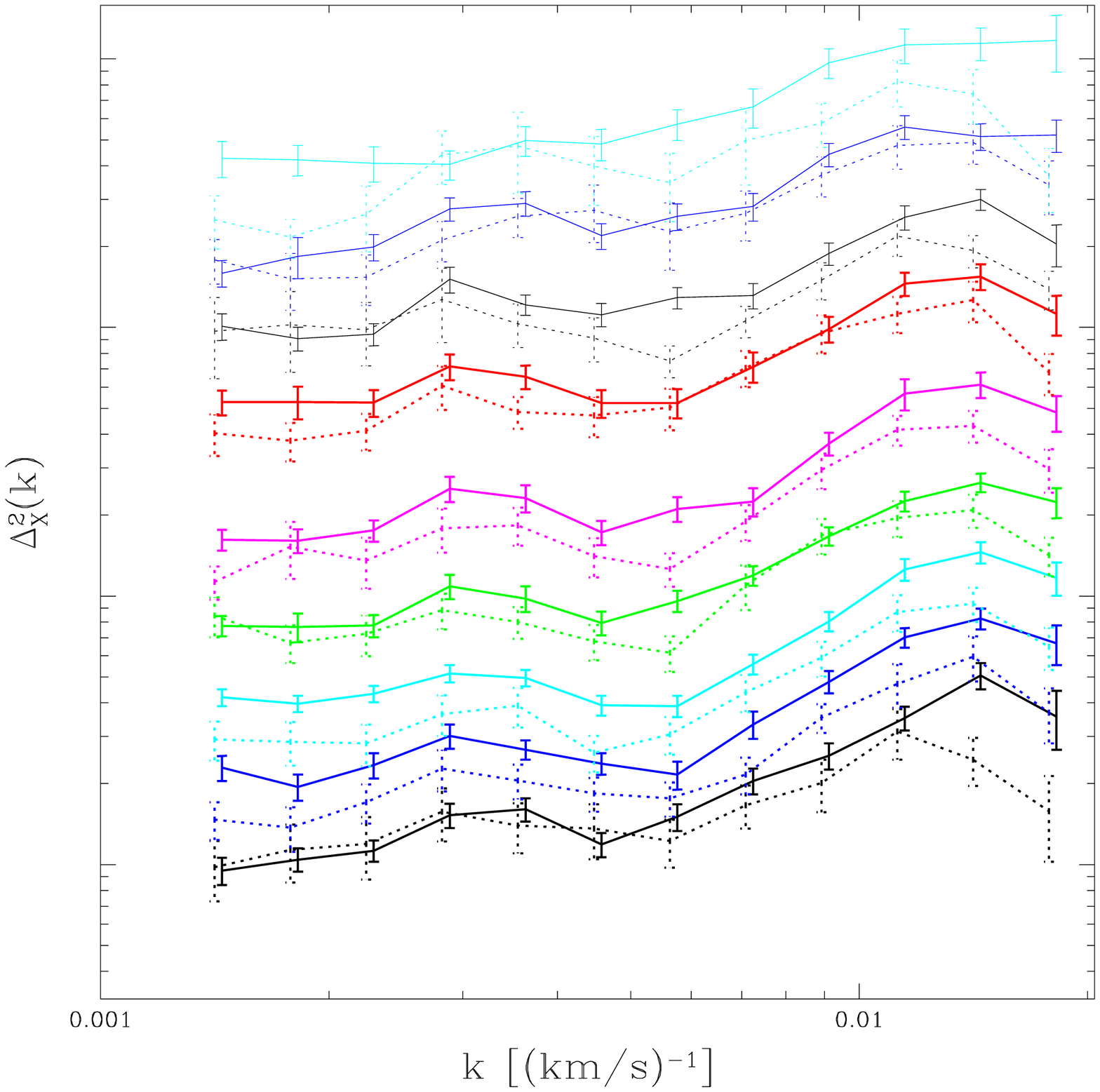}
\caption{
$P_{1268,1380}$ split by noise in the same region.  The dotted line connects
the low noise results, while the high noise results are
offset slightly to the right.  
The results at different redshifts have
been shifted vertically by arbitrary amounts.
}
\label{splitsigma1268}
\end{figure}

We compute the weighted mean of the rms noise for each chunk of spectrum
as we use it to estimate the power spectrum.  A split based simply on 
this noise level, illustrated in Figure~\ref{splitsigma}, produces a small but
unambiguously significant discrepancy in 
the raw measurement of $P_{1041,1185}$, $\chi^2=185$,
though the fit parameters agree within their errors (Table~\ref{subtab}, 
line 3).  
This discrepancy in power is the motivation for, and is largely removed by, 
our noise-dependent 
background subtraction procedure defined by equation~(\ref{fitdepeq}).
Figure~\ref{splitsigma1268} shows the power $P_{1268,1380}$ that is
used for background subtraction, 
again subsampled by noise level.
There is a clear difference in power, which is not isolated to a few
wavenumbers or redshifts.  Once the $P_{1268,1380}$ power is subtracted
according to equation~(\ref{fitdepeq}), we obtain $P_F$ estimates and
corresponding fit parameters from the high and low noise subsamples
that agree within the errors (Table~\ref{subtab}, line 2).
Since the fit parameters agree even without noise-dependent background
subtraction, it appears that the discrepancy in raw
$P_{1041,1185}$ power does not mimic a change in cosmological parameters,
and our ultimate conclusions would therefore not change even if we
did not implement it.  Nonetheless, the origin of the difference remains
somewhat mysterious, since we went to great effort to estimate noise
correctly.

Two more splits that yield discrepant $P_{1041,1185}$ 
but show no sign of trouble after the 
noise-dependent background subtraction
are based on the ratio of the mean 
sky flux to the mean quasar flux
and on $\sigma_w-\sigma_c$, the difference between the pipeline estimate of 
the noise and the sum of our estimates of the quasar flux, sky flux,
and read-noise components of the noise.  
We are not sure what $\sigma_w-\sigma_c$ means, since we do not 
understand the source of noise misestimation in the standard pipeline.
Even without noise dependent background subtraction,
the fit results did not differ significantly in either of these cases.
They are almost surely symptoms of the same noise-related problem 
discussed above.

The split based on read-noise in the spectra shows good agreement
between the \PF\ measurements, even without noise dependent background
subtraction as does a split based on 
how well the mean continuum matches the quasar spectrum outside
the \lyaf, quantified by computing $\chi^2/\nu$ for the difference
between the continuum and spectrum 
(``cont. $\chi^2/\nu$'' in Table \ref{subtab}).  
Several other splits that show little or no sign of trouble are based on:
the mean value of $\chi^2/\nu$ computed for each pixel when combining 
exposures (this was the comparison that motivated
our spectrum-by-spectrum noise re-estimation), the mean resolution, 
the movement of the spectrum relative to the detector
pixel grid during the observation (``flexure''), the alignment of
the pixels in the different exposures for the same spectrum (closely
related to flexure), the error on the overall normalization of 
the spectrum, $A_q$ (see eqn. \ref{datamodel}, this error is set by
a combination of the noise level outside the 
forest and the length of spectrum observed outside the forest), 
and the error on the means computed for the forest chunks (differences
at fixed $z$ are related
to the length of the chunk and the noise in the forest).

Overall, the agreement between our subsamples is excellent, both
for the \PF\ results and the fit results.  In some cases this 
agreement relies on the noise-dependent background subtraction,
which we would like to understand better (in no case does the
fit agreement rely on this).  

\subsection{Continuum Power \label{seccontinuum}}

\begin{figure}
\plotone{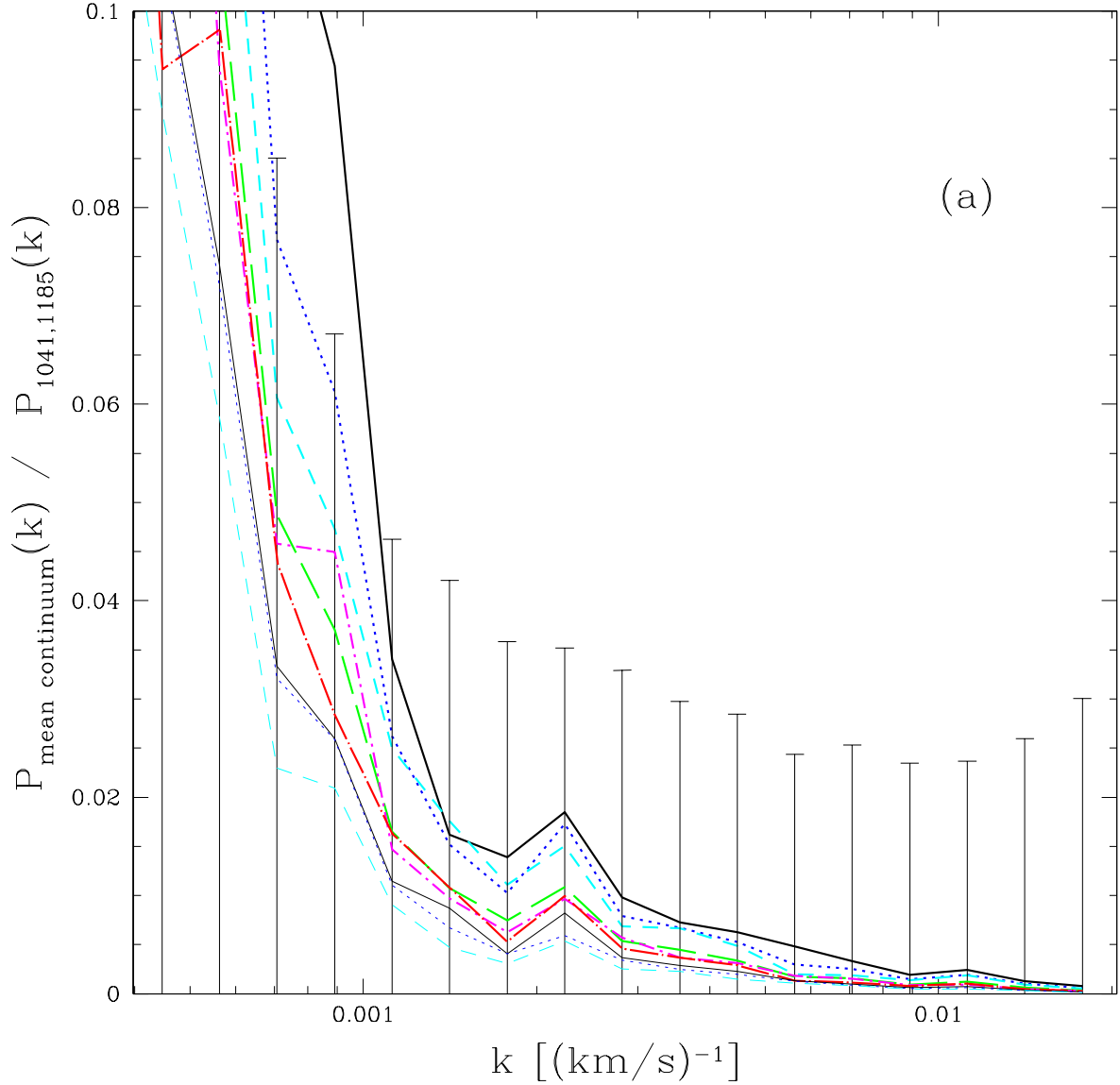}
\caption{Power in the mean continuum relative to the \lyaf\ power,
for various rest wavelength intervals:  (a) $1041<\lr<1185$\AA, 
(b) $1268<\lr<1380$\AA, (c) $1409<\lr<1523$\AA, (d) 
$1558<\lr<1774$\AA.
The error bars show the fractional error on $P_{1041,1185}$ 
(without diagonalizing the window matrix because 
the diagonalization works poorly at the lowest $k$s that we show).  
}
\label{meancontinuumpower}
\end{figure}

The power in the mean continuum, for the 4 different rest frame 
regions identified in Figure \ref{fullspecexamp},
is shown in Figures \ref{meancontinuumpower}a-d,
relative to the \lyaf\ power (the mean continuum power was measured by 
replacing the quasar flux in each pixel by the mean continuum level
at that pixel). 
\begin{figure}
\plotone{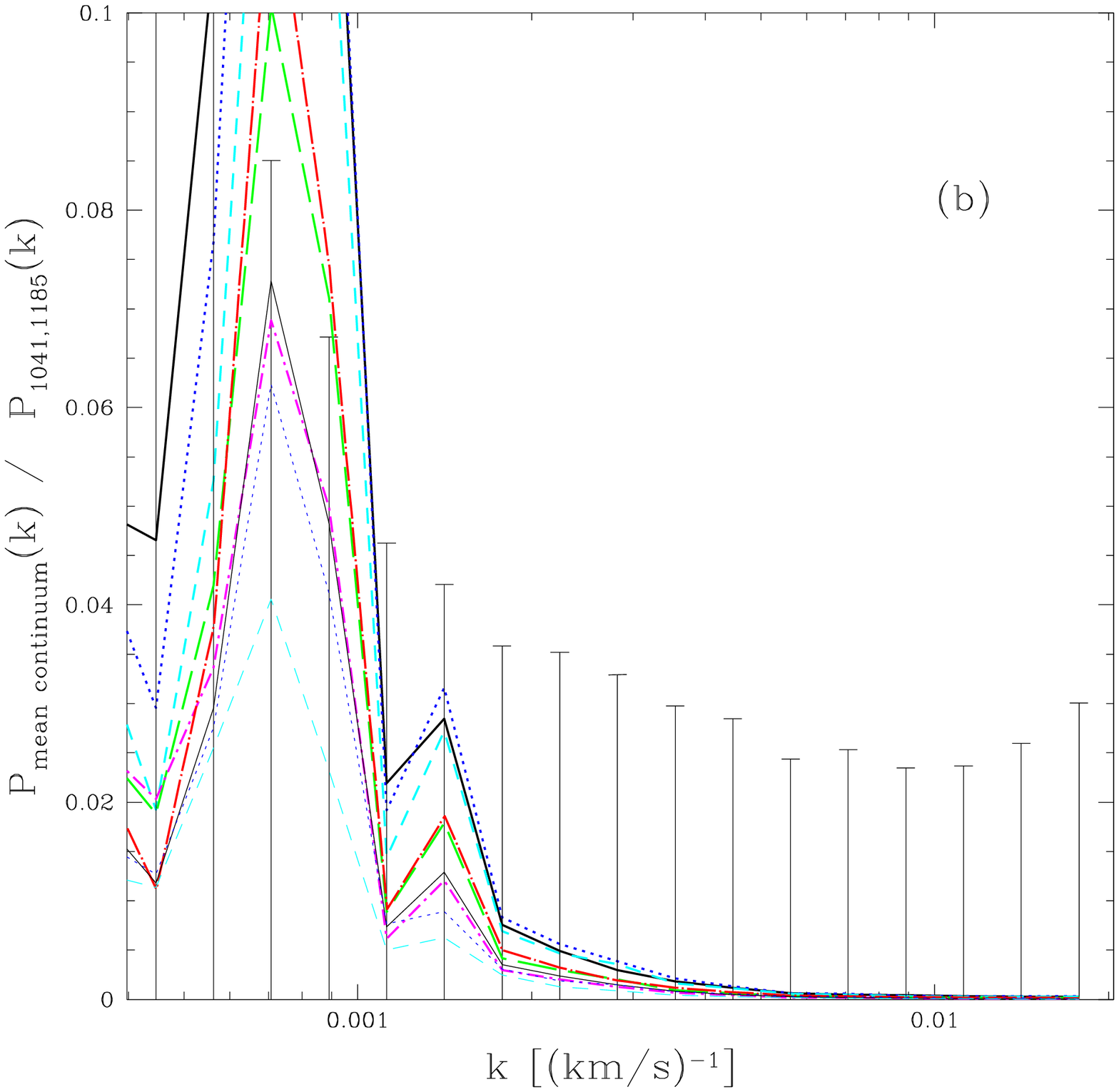}
\end{figure}
\begin{figure}
\plotone{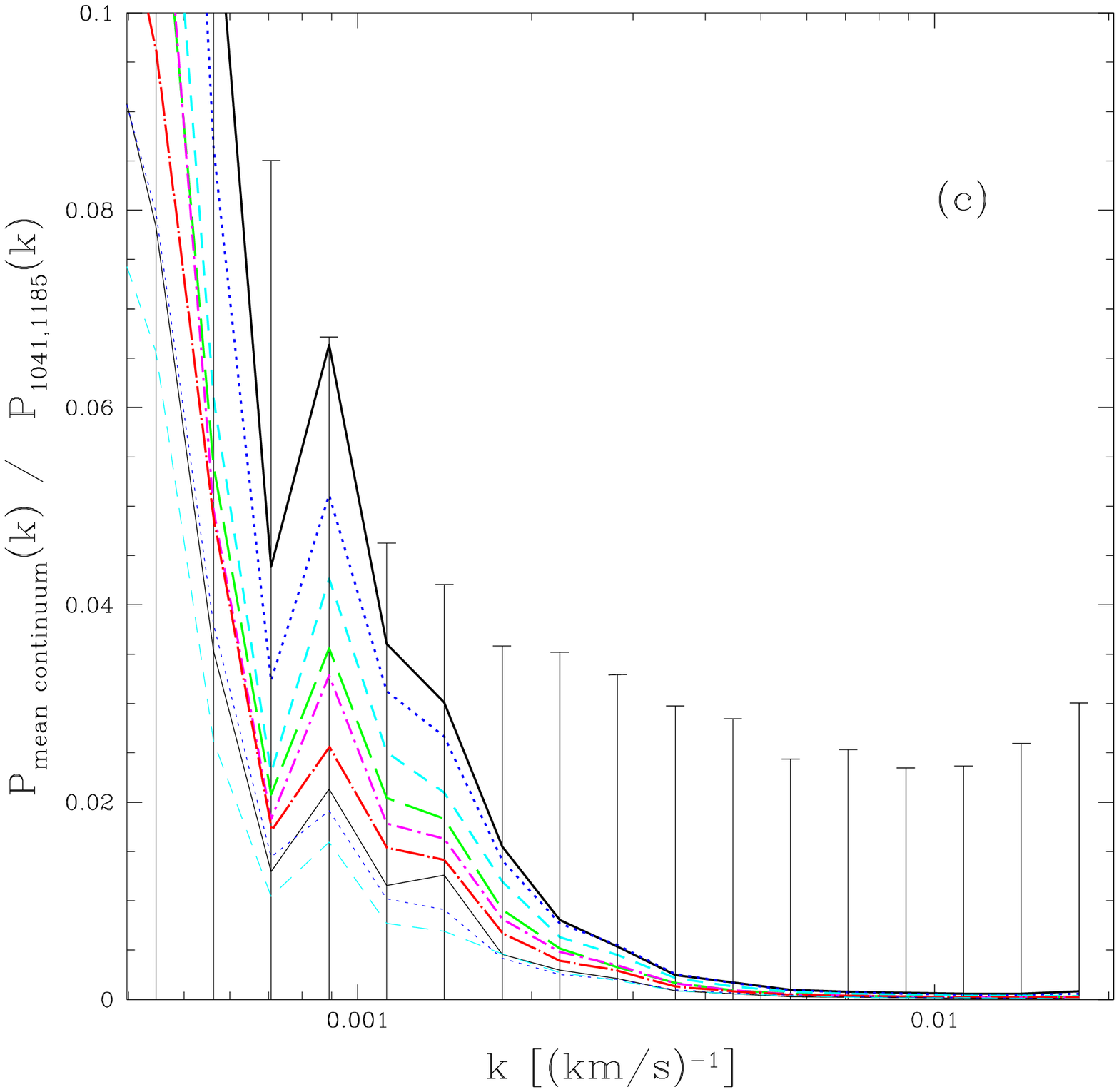}
\end{figure}
\begin{figure}
\plotone{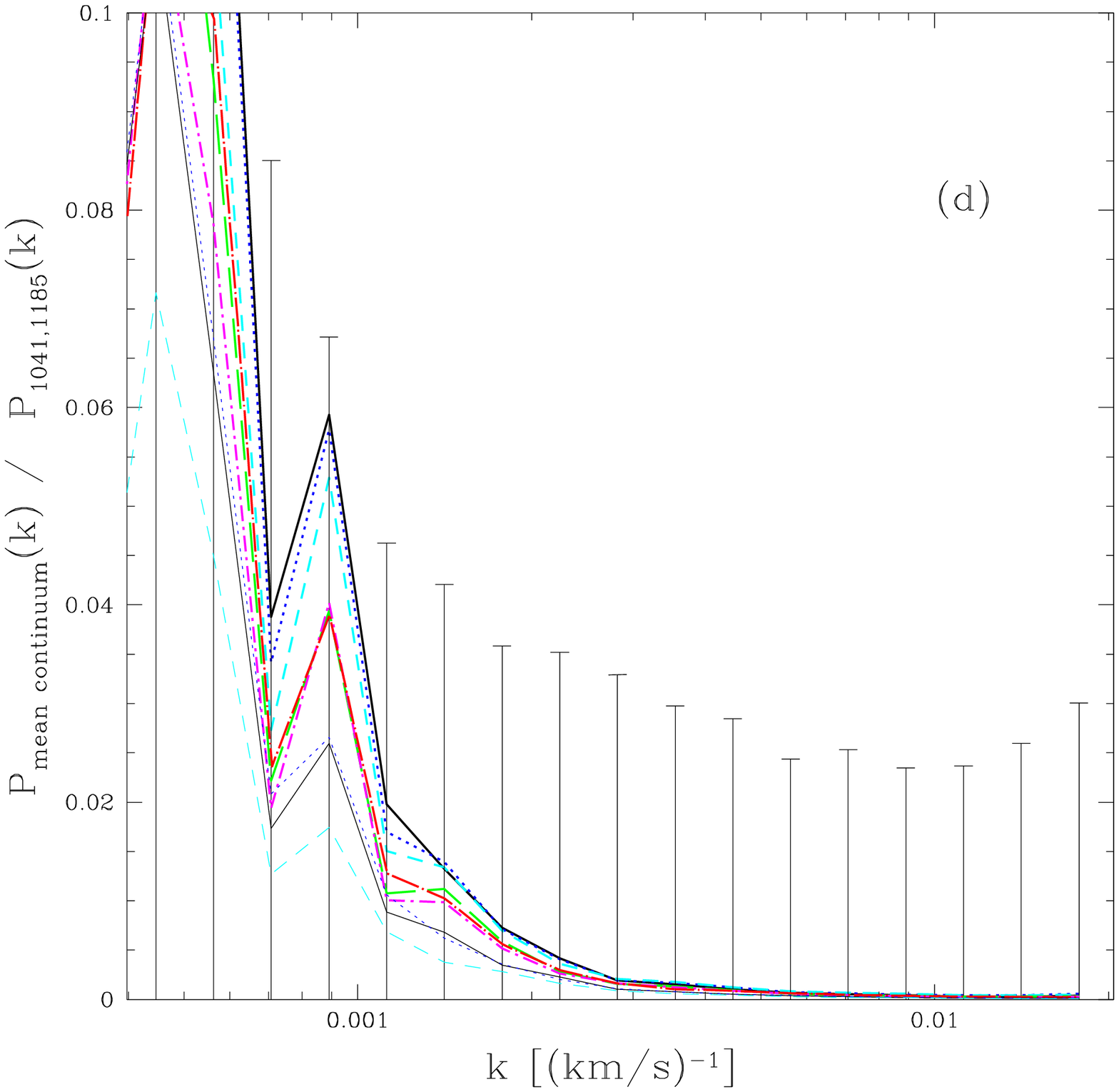}
\end{figure}
The mean continuum is very well behaved over the $k$ range that we use 
($0.0013-0.02~\ikms$),
but its fluctuations quickly become significant at $k\lesssim 0.001~\ikms$.  
What little power the mean continuum shows in our chosen $k$ range should 
be removed when we divide the spectra by the continuum; it is only 
fluctuations around the mean that matter.

We summarize our strong, but maybe not airtight, argument for believing
that continuum fluctuations are not corrupting our measurement as follows:
\begin{itemize}
\item The power in the mean continuum is small.
\item The results for the \PF\ measurement in two halves of the forest 
      region, $P_{1041,1113}$ and $P_{1113,1185}$, agree.
\item The power in the background regions, $P_{1268,1380}$ and 
      $P_{1409,1523}$, agree at the level we care about in the
      low-noise data, as long as BALs (which mostly affect the latter 
      region) are removed.
\item Division by preliminary 13 eigenvector PCA estimates of the 
      continua (i.e., including fluctuations around the mean) does
      not change the results. 
\end{itemize}
To be quantitatively important despite these arguments,
the power in quasar-to-quasar continuum fluctuations 
in the forest must be substantially larger than the power 
in the mean continuum itself, the continuum fluctuations in the forest
must be substantially 
different from those in the background regions (despite 
those regions being similar to each other and the two halves of
the forest being similar to each other), and our PCA analysis must
be substantially flawed.
Further study is warranted, but a big effect seems unlikely.

\subsection{Comparison with Past Measurements \label{comppastsec}}

There are three \PF\ measurements already in the literature,
\cite{2000ApJ...543....1M}, \cite{2002ApJ...581...20C}, 
and \cite{2004MNRAS.347..355K}, all using at least some high 
resolution data.  
Each uses its own set of redshift
bins, so to compare we need a way to interpolate our 
results to these redshifts.  
We do this by performing our standard cosmological fit to
all of the data (at first -- later we will remove some of 
the past results).  This gives us a set of best fit model 
parameters that can be used to compute the power at 
any $k$ and $z$. 
Within the range of $k$ where we have SDSS measurements, the
fit is always dominated by the SDSS points.  The fitted
curves always match the SDSS results to much better 
than the size of the errors on the past results, meaning that,
for the the purpose of comparison to the past results, the 
curves are simply a faithful interpolation between the SDSS
points.  At $k>0.02\ikms$, the fit is effectively a weighted
average of the past results, although the constraint that it
must match SDSS at lower $k$ has some influence (our simulation
predictions do not allow for sharp features in \PF.)
 
\begin{figure}
\plotone{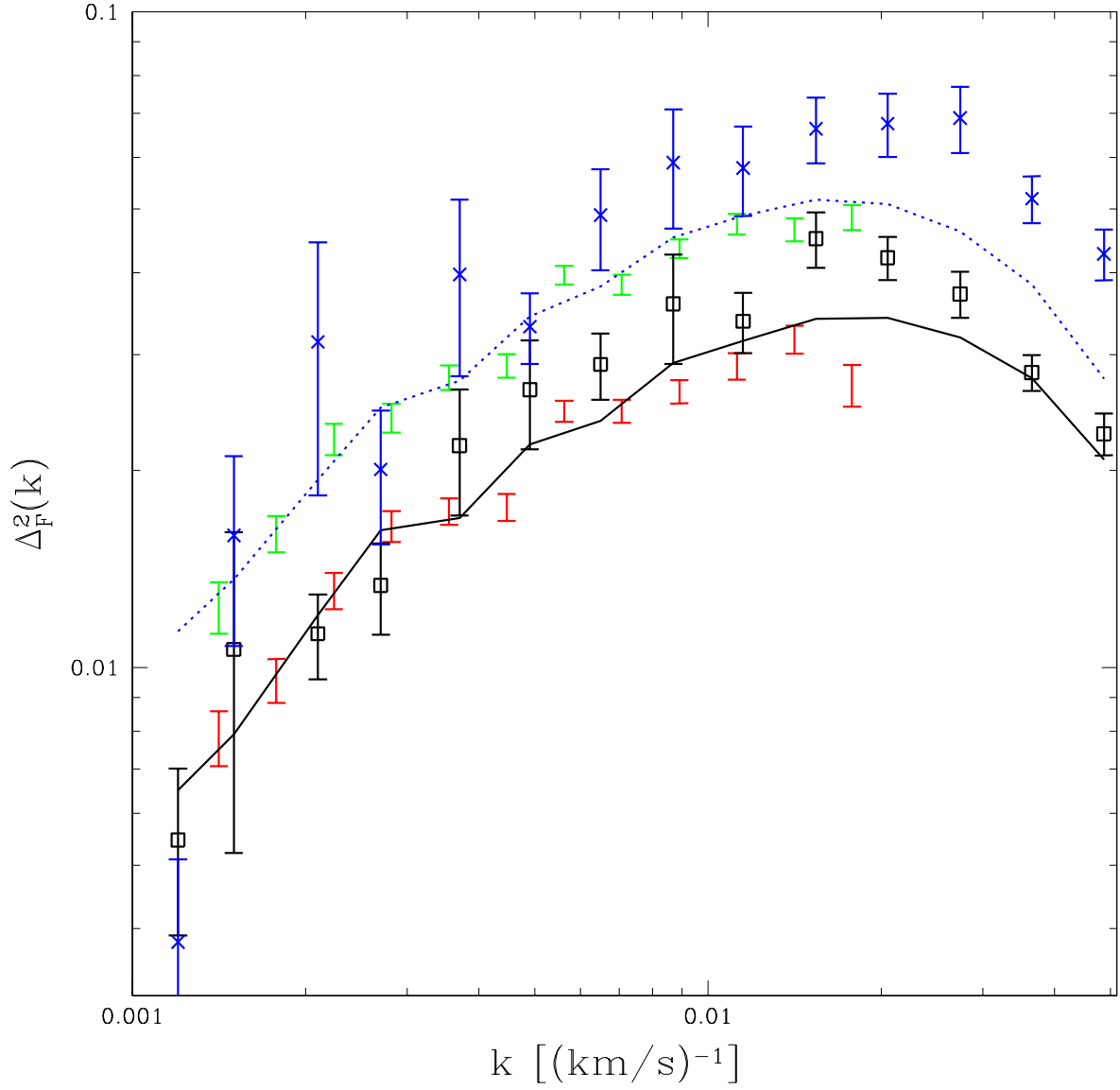}
\caption{
The black, solid line and open squares 
(blue, dotted line and crosses)
show the fit prediction and measured points from 
\cite{2004MNRAS.347..355K} at $z=2.18$ (2.58).
Red (green) error bars show our SDSS measurement 
at $z=2.2$ (2.6).
}
\label{kimcompwsdss}
\end{figure}

We first
perform a fit to all the data with $k<0.05\ikms$ and $z>2.1$, 
finding an atrociously bad $\chi^2=392$ for $\sim 238$ dof.
Removing \cite{2000ApJ...543....1M}, reduces $\chi^2$ by 
53.4 (for 39 data points), removing \cite{2002ApJ...581...20C} 
reduces $\chi^2$ by 85.2 (for 65 points), and 
removing \cite{2004MNRAS.347..355K} reduces $\chi^2$ by 123.3 
(28 points).  Clearly there is gross disagreement between 
\cite{2004MNRAS.347..355K} and the other results.
Figure \ref{kimcompwsdss} shows the 
\cite{2004MNRAS.347..355K} points at $z=2.18$ and $z=2.58$
(from their Table 5) along with the fit prediction for them.
Note that we include SiIII contamination in the model as 
described in \S\ref{secxisi3}, so the model curves are not
perfectly smooth.
We see large discrepancies, as we expect from the bad $\chi^2$.
The point at $k=0.0012\ikms$, $z=2.58$ is $5.9\sigma$ below
the prediction (as well as any reasonable extrapolation of the
other \cite{2004MNRAS.347..355K} points), and the points at 
increasingly high $k$ are 
generally too high (at the highest $k$, this is a reflection
of disagreement with the other high resolution data, but actually
the agreement is not much better if we only include SDSS in the fit,
because no model can fit the highest $k$ SDSS points and then 
climb to match the higher $k$ \cite{2004MNRAS.347..355K} points).
To reassure the reader that we are not playing games with the fitted
curves, we also plot the SDSS points at $z=2.2$ and 2.6.

\begin{figure}
\plotone{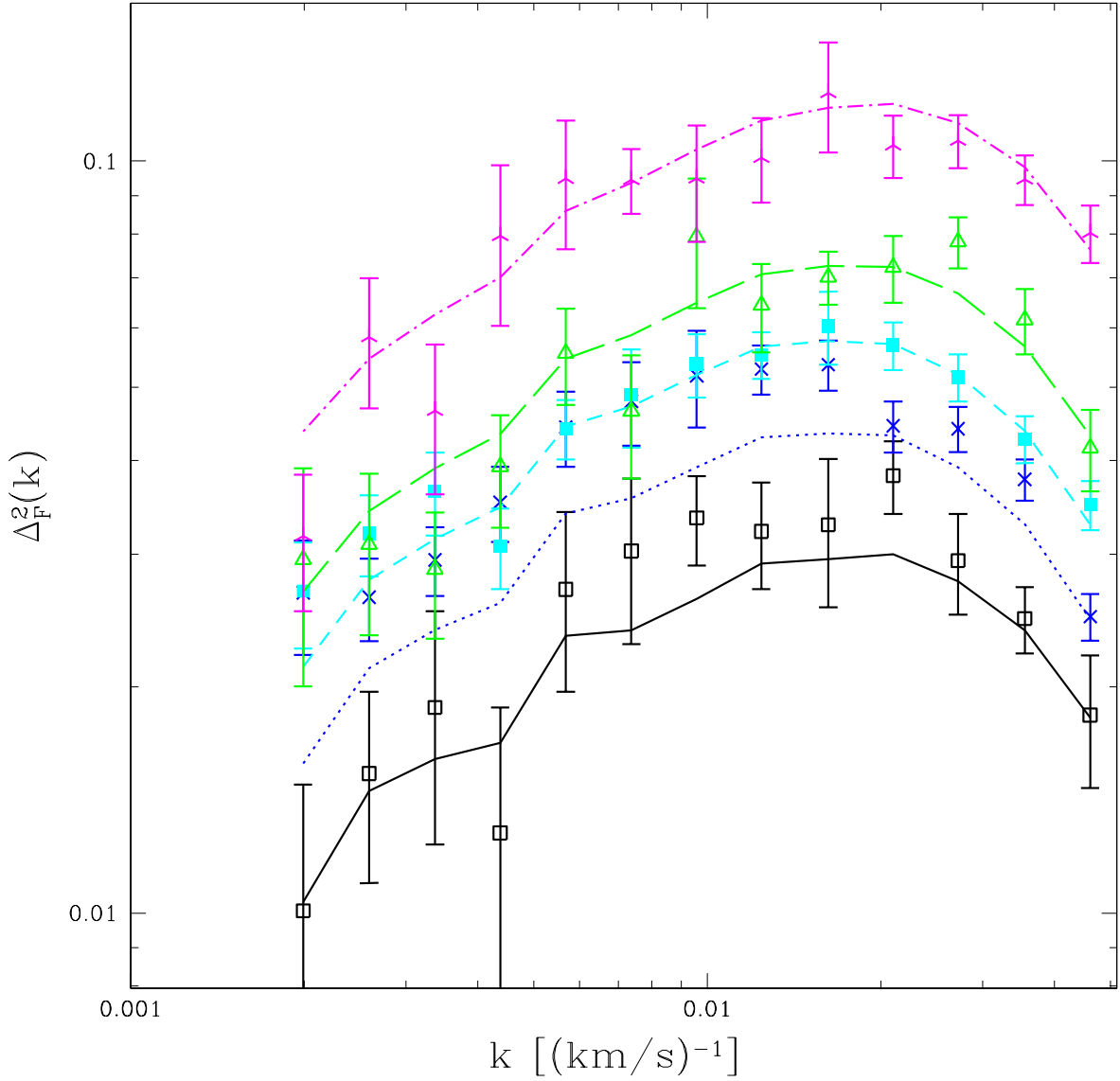}
\caption{
Measured points and fit prediction for the
\cite{2002ApJ...581...20C} results.
{}From bottom to top (roughly) ---
z=2.13:  black, solid line, open square; 
z=2.47:  blue, dotted line, 4-point star (cross); 
z=2.74:  cyan, dashed line, filled square; 
z=3.03:  green, long-dashed line, open triangle; 
z=3.51:  magenta, dot-dashed line, 3-point star.
}
\label{croftcomp}
\end{figure}

\begin{figure}
\plotone{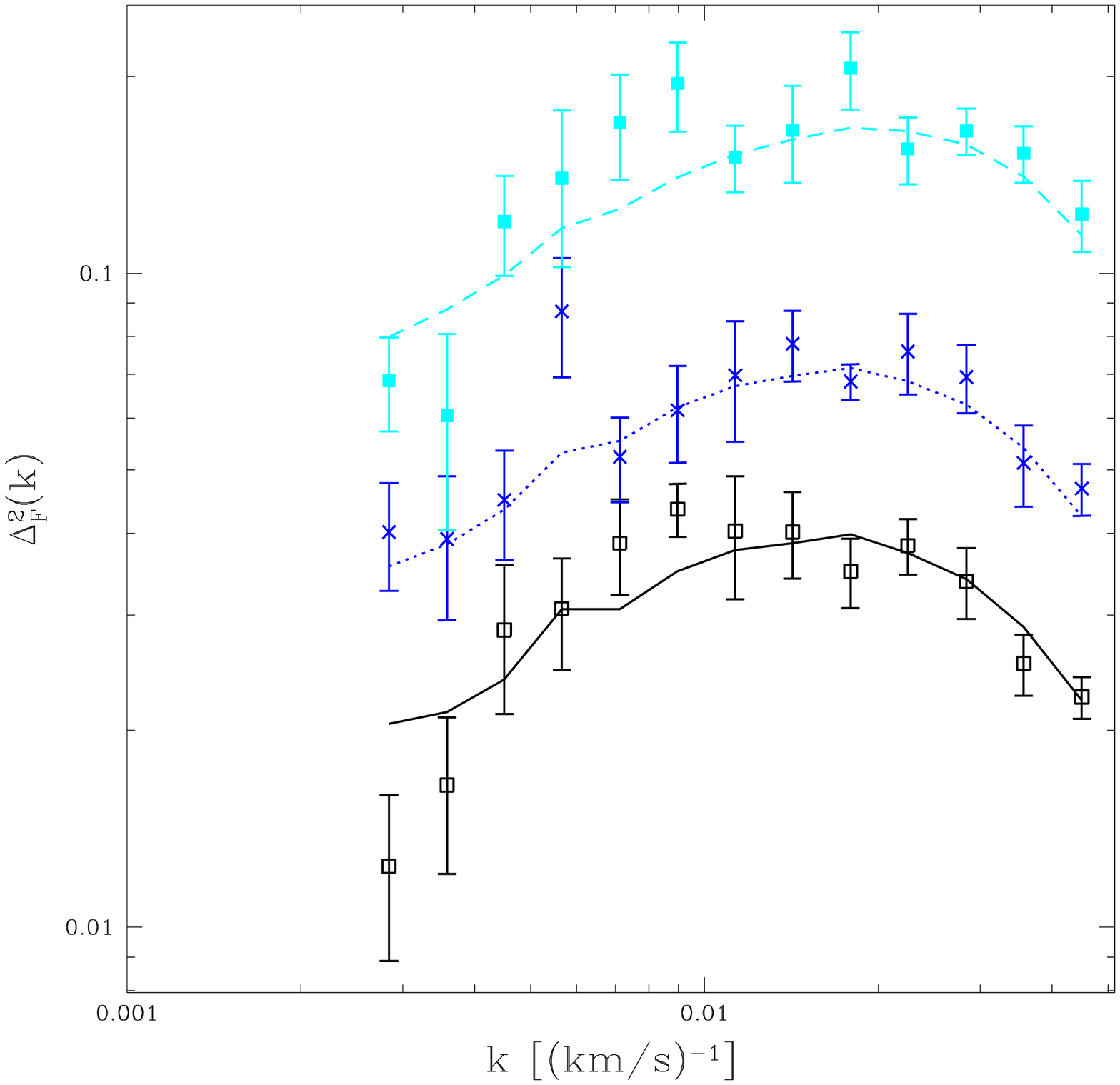}
\caption{
Measured points and fit prediction for the
\cite{2000ApJ...543....1M} results.
>From bottom to top (roughly) ---
z=2.41:  black, solid line, open square; 
z=3.00:  blue, dotted line, 4-point star (cross); 
z=3.89:  cyan, dashed line, filled square; 
}
\label{mecomp}
\end{figure}

Since the \cite{2004MNRAS.347..355K} results clearly have some
problem, unless the other three measurements are all wrong  
(we will see that, with one exception, the other three agree with 
each other), we eliminate them from the rest of the comparison.
A fit to SDSS, \cite{2000ApJ...543....1M}, and 
\cite{2002ApJ...581...20C} gives $\chi^2=269$ for $\sim 210$ dof
(still a bad fit).  
Removing \cite{2000ApJ...543....1M}, 
reduces $\chi^2$ by
44.3 (39 points), while removing \cite{2002ApJ...581...20C}
reduces $\chi^2$ by 97.2 (for 65 points this reduction would
occur by chance only 0.6\% of the time).    
Figure \ref{croftcomp} shows the 
\cite{2002ApJ...581...20C} points,
along with the fit prediction for them.
The agreement is actually very good for 4 of the 5
redshift bins, while the $z=2.47$ points are 
obviously out of place (these 13 points 
increase $\chi^2$ by 54).  
Figure \ref{mecomp} shows the 
\cite{2000ApJ...543....1M} points,
along with the fit prediction for them 
(for this fit we removed the $z=2.47$
\cite{2002ApJ...581...20C} points).
The agreement is good, with the agreement at $z=2.41$
disfavoring the anomalous \cite{2002ApJ...581...20C} 
$z=2.47$ points; further investigation by R.\ Croft (private communication)
does not reveal any obvious error in this redshift bin that would explain the
anomaly.
Note that the agreement 
of \cite{2000ApJ...543....1M} and 
\cite{2002ApJ...581...20C} at high $k$ adds weight
to the idea that something is seriously wrong with 
\cite{2004MNRAS.347..355K}.  \cite{2004MNRAS.347..355K}
show some comparisons with past results, and claim they
agree, but these comparisons used custom redshift bins
(i.e., not the bins in their table),
and were not high precision
(for example they compare
the \cite{2002ApJ...581...20C} points at $z=2.13$ to
a bin with $z=2.04$, so evolution cancels some
of the amplitude offset, and they call the apparent
$\sim 50$\% difference at $k\sim 0.04\ikms$ 
a ``slight'' excess).

\section{Final Results Table and Directions for Use \label{directionsforuse}}

Table \ref{resultstab}  gives the primary power spectrum results.
The columns are: $z$, the redshift of the bin; $k$, the wavenumber
of the bin; \PF, our final \lyaf\ power spectrum result (along
with the square roots of the diagonal elements of the error 
covariance matrix); $P_{\rm noise}$, the noise power that was 
subtracted from each bin; and $P_{\rm background}$, the background
that was subtracted from each bin ($P_{1268,1380}$ 
adjusted according to the amount of noise in the forest, 
eqs.~\ref{expectdeltaf}, \ref{backsubeq}, and \ref{fitdepeq}).  
$P_{\rm noise}$ is just
the noise subtracted from $P_{1041,1085}$ (a roughly 
comparable amount of noise was subtracted from the background, so to some degree
these cancel in the final result).
The table and the covariance matrix of the errors are available at
{\it http://feynman.princeton.edu/$\sim$pmcdonal/LyaF/sdss.html}.
The covariance matrix must be used in any serious quantitative fitting.
When using this table to constrain models, the following allowances
should be made for residual systematic uncertainties:  
\begin{itemize}
\item Allow $\pm 5$\% rms error on the noise-power amplitude at each redshift.  
We do not have any reason to think the error is really this large, but, 
considering the complications
related to the noise, we think it is prudent to include it. 
Operationally, we suggest subtracting $f_i P_{\rm noise}(k,z_i)$ from 
$P_F(k,z_i)$, where $f_i$ are free parameters in the fit
(one for each redshift bin),
and adding $\sum_i (f_i/0.05)^2$ to $\chi^2$. 
\item Allow $\pm(7 \kms)^2$ rms overall error on the 
resolution variance (i.e.,
the square of the rms width of the Gaussian resolution kernel).  This is the
expected size of the uncertainty due to flexure in the detector, although 
Figure \ref{skypower5579} suggests that it may actually be smaller.
Specifically, multiply $P_F(k,z)$ by $\exp(\alpha k^2)$, with $\alpha$
a free parameter in the fit, and add $[\alpha/(7\kms)^2]^2$ to $\chi^2$.
\item SiIII-\lya\ cross-correlation must be accounted for.  We have suggested
a simple method -- assume the cross-correlation has the same form as the
\lya-\lya\ auto-correlation up to an amplitude that is a free parameter, and 
possibly include freedom in the correlation width and/or redshift 
evolution of the amplitude --
but others could be devised (e.g., including SiIII in the 
simulated spectra through a parameterized semi-analytic model).
\end{itemize}
\begin{deluxetable}{lcccc}
\tablecolumns{5}
\tablecaption{\PF\ Results \label{resultstab}}
\tablehead{ 
\colhead{$z$} 
& \colhead{$k$}
& \colhead{\PF} 
& \colhead{$P_{\rm noise}$} 
& \colhead{$P_{\rm background}$} 
}
\startdata
2.2 & 0.00141 & $  18.09 \pm  1.74 $ &   6.20 &   3.29 \\
2.2 & 0.00178 & $  17.55 \pm  1.34 $ &   6.07 &   2.71 \\
2.2 & 0.00224 & $  19.05 \pm  1.21 $ &   6.20 &   2.46 \\
2.2 & 0.00282 & $  18.93 \pm  1.03 $ &   6.23 &   2.45 \\
2.2 & 0.00355 & $  15.80 \pm  0.74 $ &   5.92 &   1.97 \\
2.2 & 0.00447 & $  12.68 \pm  0.60 $ &   5.57 &   1.18 \\
2.2 & 0.00562 & $  14.04 \pm  0.53 $ &   5.85 &   1.04 \\
2.2 & 0.00708 & $  11.09 \pm  0.45 $ &   5.87 &   1.17 \\
2.2 & 0.00891 & $   9.38 \pm  0.39 $ &   6.06 &   1.20 \\
2.2 & 0.01122 & $   8.09 \pm  0.38 $ &   6.87 &   1.40 \\
2.2 & 0.01413 & $   6.99 \pm  0.34 $ &   8.48 &   1.32 \\
2.2 & 0.01778 & $   4.69 \pm  0.35 $ &  10.52 &   0.87 \\
2.4 & 0.00141 & $  21.52 \pm  1.91 $ &   5.70 &   3.72 \\
2.4 & 0.00178 & $  23.66 \pm  2.09 $ &   5.65 &   2.63 \\
2.4 & 0.00224 & $  23.57 \pm  1.39 $ &   5.65 &   2.45 \\
2.4 & 0.00282 & $  22.25 \pm  1.24 $ &   5.58 &   2.47 \\
2.4 & 0.00355 & $  18.65 \pm  0.89 $ &   5.31 &   1.82 \\
2.4 & 0.00447 & $  15.74 \pm  0.65 $ &   5.15 &   1.18 \\
2.4 & 0.00562 & $  18.07 \pm  0.68 $ &   5.43 &   0.79 \\
2.4 & 0.00708 & $  13.16 \pm  0.51 $ &   5.32 &   1.07 \\
2.4 & 0.00891 & $  12.58 \pm  0.41 $ &   5.71 &   1.14 \\
2.4 & 0.01122 & $  10.42 \pm  0.41 $ &   6.27 &   1.29 \\
2.4 & 0.01413 & $   8.17 \pm  0.36 $ &   7.32 &   1.22 \\
2.4 & 0.01778 & $   6.08 \pm  0.33 $ &   9.37 &   0.91 \\
2.6 & 0.00141 & $  28.29 \pm  2.55 $ &   6.78 &   4.02 \\
2.6 & 0.00178 & $  29.04 \pm  1.85 $ &   6.68 &   3.18 \\
2.6 & 0.00224 & $  32.13 \pm  1.76 $ &   6.76 &   2.65 \\
2.6 & 0.00282 & $  27.44 \pm  1.39 $ &   6.63 &   2.41 \\
2.6 & 0.00355 & $  25.06 \pm  1.09 $ &   6.52 &   1.79 \\
2.6 & 0.00447 & $  20.67 \pm  0.85 $ &   6.40 &   1.27 \\
2.6 & 0.00562 & $  22.49 \pm  0.72 $ &   6.69 &   1.05 \\
2.6 & 0.00708 & $  17.19 \pm  0.60 $ &   6.71 &   1.17 \\
2.6 & 0.00891 & $  15.40 \pm  0.51 $ &   7.12 &   1.05 \\
2.6 & 0.01122 & $  13.25 \pm  0.48 $ &   7.91 &   1.24 \\
2.6 & 0.01413 & $  10.25 \pm  0.41 $ &   9.15 &   1.22 \\
2.6 & 0.01778 & $   8.43 \pm  0.37 $ &  11.66 &   0.95 \\
2.8 & 0.00141 & $  37.25 \pm  2.75 $ &   6.83 &   2.99 \\
2.8 & 0.00178 & $  37.52 \pm  2.20 $ &   6.75 &   2.10 \\
2.8 & 0.00224 & $  38.74 \pm  1.80 $ &   6.79 &   1.84 \\
2.8 & 0.00282 & $  37.12 \pm  1.48 $ &   6.78 &   2.29 \\
2.8 & 0.00355 & $  30.11 \pm  1.23 $ &   6.60 &   1.52 \\
2.8 & 0.00447 & $  25.67 \pm  0.92 $ &   6.52 &   1.22 \\
2.8 & 0.00562 & $  25.74 \pm  0.83 $ &   6.73 &   1.19 \\
2.8 & 0.00708 & $  22.54 \pm  0.67 $ &   6.95 &   0.98 \\
2.8 & 0.00891 & $  20.12 \pm  0.62 $ &   7.41 &   1.11 \\
2.8 & 0.01122 & $  15.89 \pm  0.48 $ &   8.06 &   1.18 \\
2.8 & 0.01413 & $  13.04 \pm  0.42 $ &   9.37 &   1.13 \\
2.8 & 0.01778 & $   9.63 \pm  0.36 $ &  11.64 &   0.90 \\
3.0 & 0.00141 & $  46.36 \pm  3.72 $ &   7.76 &   3.51 \\
3.0 & 0.00178 & $  42.53 \pm  2.87 $ &   7.63 &   2.74 \\
3.0 & 0.00224 & $  47.66 \pm  2.69 $ &   7.73 &   2.20 \\
3.0 & 0.00282 & $  42.20 \pm  2.19 $ &   7.66 &   2.84 \\
3.0 & 0.00355 & $  36.99 \pm  1.72 $ &   7.51 &   1.81 \\
3.0 & 0.00447 & $  29.47 \pm  1.20 $ &   7.34 &   1.30 \\
3.0 & 0.00562 & $  30.12 \pm  1.07 $ &   7.56 &   1.33 \\
3.0 & 0.00708 & $  24.30 \pm  0.81 $ &   7.63 &   0.99 \\
3.0 & 0.00891 & $  22.51 \pm  0.75 $ &   8.15 &   1.30 \\
3.0 & 0.01122 & $  18.75 \pm  0.66 $ &   8.83 &   1.30 \\
3.0 & 0.01413 & $  14.33 \pm  0.52 $ &   9.89 &   1.38 \\
3.0 & 0.01778 & $  11.26 \pm  0.47 $ &  11.90 &   0.91 \\
3.2 & 0.00141 & $  54.73 \pm  4.97 $ &   9.57 &   5.07 \\
3.2 & 0.00178 & $  49.72 \pm  4.12 $ &   9.44 &   3.73 \\
3.2 & 0.00224 & $  52.86 \pm  3.29 $ &   9.46 &   3.01 \\
3.2 & 0.00282 & $  48.44 \pm  2.58 $ &   9.38 &   2.59 \\
3.2 & 0.00355 & $  44.01 \pm  2.33 $ &   9.28 &   2.71 \\
3.2 & 0.00447 & $  35.12 \pm  1.54 $ &   8.95 &   1.36 \\
3.2 & 0.00562 & $  34.57 \pm  1.41 $ &   9.15 &   1.28 \\
3.2 & 0.00708 & $  31.14 \pm  1.19 $ &   9.40 &   1.29 \\
3.2 & 0.00891 & $  26.96 \pm  0.95 $ &   9.80 &   1.48 \\
3.2 & 0.01122 & $  22.21 \pm  0.83 $ &  10.49 &   1.80 \\
3.2 & 0.01413 & $  18.37 \pm  0.70 $ &  11.74 &   1.38 \\
3.2 & 0.01778 & $  15.12 \pm  0.66 $ &  14.07 &   1.18 \\
3.4 & 0.00141 & $  56.42 \pm  5.85 $ &  11.12 &   4.08 \\
3.4 & 0.00178 & $  75.75 \pm  5.33 $ &  11.32 &   2.41 \\
3.4 & 0.00224 & $  56.79 \pm  3.87 $ &  10.97 &   2.09 \\
3.4 & 0.00282 & $  58.40 \pm  3.43 $ &  11.04 &   3.16 \\
3.4 & 0.00355 & $  52.56 \pm  2.85 $ &  10.96 &   2.62 \\
3.4 & 0.00447 & $  43.43 \pm  2.21 $ &  10.76 &   2.00 \\
3.4 & 0.00562 & $  41.67 \pm  1.73 $ &  10.99 &   2.04 \\
3.4 & 0.00708 & $  37.36 \pm  1.43 $ &  11.28 &   1.55 \\
3.4 & 0.00891 & $  32.57 \pm  1.19 $ &  11.87 &   1.77 \\
3.4 & 0.01122 & $  28.51 \pm  1.15 $ &  13.06 &   2.00 \\
3.4 & 0.01413 & $  22.28 \pm  0.88 $ &  14.63 &   1.63 \\
3.4 & 0.01778 & $  18.01 \pm  0.79 $ &  17.80 &   1.23 \\
3.6 & 0.00141 & $  79.46 \pm  8.33 $ &  15.11 &   2.25 \\
3.6 & 0.00178 & $  85.12 \pm  8.28 $ &  14.90 &   2.87 \\
3.6 & 0.00224 & $  75.03 \pm  5.88 $ &  14.87 &   3.28 \\
3.6 & 0.00282 & $  66.15 \pm  4.98 $ &  14.51 &   3.30 \\
3.6 & 0.00355 & $  66.32 \pm  4.08 $ &  14.59 &   2.26 \\
3.6 & 0.00447 & $  55.66 \pm  3.36 $ &  14.22 &   1.33 \\
3.6 & 0.00562 & $  49.51 \pm  2.72 $ &  14.17 &   1.28 \\
3.6 & 0.00708 & $  43.77 \pm  2.15 $ &  14.41 &   1.62 \\
3.6 & 0.00891 & $  40.20 \pm  1.93 $ &  15.09 &   1.98 \\
3.6 & 0.01122 & $  32.04 \pm  1.63 $ &  15.72 &   1.61 \\
3.6 & 0.01413 & $  25.82 \pm  1.31 $ &  17.26 &   1.25 \\
3.6 & 0.01778 & $  21.49 \pm  1.23 $ &  21.11 &   1.51 \\
3.8 & 0.00141 & $ 118.61 \pm 15.47 $ &  22.58 &   6.89 \\
3.8 & 0.00178 & $  61.52 \pm  9.80 $ &  20.93 &   6.40 \\
3.8 & 0.00224 & $  77.29 \pm  7.09 $ &  20.91 &   5.07 \\
3.8 & 0.00282 & $  71.78 \pm  7.42 $ &  20.76 &   2.20 \\
3.8 & 0.00355 & $  77.49 \pm  5.65 $ &  21.00 &   2.51 \\
3.8 & 0.00447 & $  59.12 \pm  4.21 $ &  20.37 &   2.37 \\
3.8 & 0.00562 & $  57.53 \pm  3.72 $ &  20.78 &   2.61 \\
3.8 & 0.00708 & $  56.25 \pm  3.45 $ &  21.63 &   2.79 \\
3.8 & 0.00891 & $  42.46 \pm  2.34 $ &  21.69 &   2.57 \\
3.8 & 0.01122 & $  36.93 \pm  2.25 $ &  23.43 &   2.69 \\
3.8 & 0.01413 & $  29.52 \pm  2.25 $ &  26.14 &   2.47 \\
3.8 & 0.01778 & $  27.72 \pm  1.85 $ &  33.51 &   2.15 \\
\enddata
\tablecomments{$k$ has units $\ikms$, power spectra have 
units $\kms$.  The error covariance matrix must be 
used for any quantitative fitting.}
\end{deluxetable}
\clearpage

\section{Conclusions \label{conclusions}}

We have analyzed a sample of 3035 quasar spectra measured by SDSS 
and covering the redshift range $2<z<4$. This data set is almost 
two orders of magnitude larger than previously available data sets.
We have focused on the 
flux power spectrum in the redshift range $2.1<z<3.9$
and for modes $0.0013~\ikms<k<0.02~\ikms$. The extraordinary size of 
the data sample leads to an order of magnitude reduction in errors 
compared to previous analyses. Consequently, to do justice to this data 
set requires a much more careful analysis than was needed
in the past. To this end we have developed a new analysis pipeline 
using quadratic power spectrum estimation
with near optimal performance. We applied this analysis to 
realistic mock spectra and demonstrated (after several tweaks) that 
the method performs as expected. We emphasize that realistic mock spectra 
are essential if one is to trust the results at the level of precision 
allowed by this data set. Our error estimation is based on bootstrap 
resampling, which works well here because the individual quasars
are independent of each other.  
The errors were tested against mock
spectra and found to be accurate.
We also compared the bootstrap errors to Gaussian 
errors, finding them to be in general less 
than 20\% higher than Gaussian.

Given the small errors on the recovered flux power spectrum the 
required control of systematic effects must be improved correspondingly 
as well.  A significant part of this paper is devoted to this issue. 
We find several sources of contamination present in the data and 
develop methods to remove them. Metal absorption for metals with 
rest wavelength transitions above $\sim 1300$\AA, 
uncertainties in sky subtraction, and calibration errors can be subtracted 
essentially exactly
by measuring the power on the red side of \lyaf\, using the same 
observed wavelength range.  
We search for a contribution from metals
with transitions close to Ly-$\alpha$ 
using a correlation function analysis, assuming that they are correlated
with hydrogen. We find clear evidence of SiIII contamination
and develop a simple 
and effective scheme to remove it. This procedure improves the $\chi^2$ 
of the fit from 194 to 129 for $\sim 104$ degrees of freedom and is thus 
necessary for a satisfactory fit.
We find no evidence of any other metal line contribution to the background
subtracted flux power spectrum.

We reduce any contribution of the continuum to the flux power spectrum
by dividing each spectrum by the mean quasar continuum.  If 
contributions from quasar-to-quasar continuum differences are similar
in different regions of the spectrum, then our subtraction of power
from the red side of \lya, as described above, should remove them.
Several tests suggest that any residual contributions from continuum
fluctuations are negligible.
First, we measure the power in the mean continuum in several rest frame
regions, finding it to be always small relative to our error bars,
so power in quasar-to-quasar fluctuations has to be
larger than power in the mean continuum itself to be significant.
Second, measurements of the background in several rest frame regions 
place upper limits on the fluctuations in those regions.
Third, a split of the \lyaf\
region into two halves reveals no evidence that residual continuum 
fluctuations differ from one half to the other.
Finally, estimating the continuum quasar-by-quasar using a 
principal component analysis does not change the power spectrum
results significantly.

In section \S 4 we perform a series of tests to verify the robustness 
of the analysis against several modifications of the standard procedure
and splits of the data. This reveals an interesting correlation between 
the power in the red side and the average noise (and some other properties
of the spectra that correlate with noise, like the amplitude of the sky 
flux relative to the quasar flux).  While we do not have 
a detailed explanation for this effect, we are able to remove it
by modifying the standard procedure to include
this correlation.
From our full battery of tests, 
we conclude that systematic effects in the power spectrum measurement 
are not likely to 
significantly affect the results of cosmological fitting (i.e.,
it is likely that some effects remain
formally significant relative to the errors on \PF, but the shape of
these systematic errors does not seem to correspond to a change in the 
cosmological parameters).
This conclusion is further confirmed by the analysis of different subsets, 
which do not reveal any systematic deviations from those expected 
statistically.

In this paper we limit ourselves to the analysis of the flux 
power spectrum, without attempting to compare it to cosmological 
models. The results of this paper should thus be fairly noncontroversial 
and can be used by others who wish to perform their own cosmological 
analysis. 
Our own analysis will be presented in a separate publication, 
as will the cosmological implications that follow from it. 
We note that the expected error on the linear rms
amplitude of fluctuations is $\sim 5$\%
and on the slope is $\sim 0.024$, both at 
the pivot point $k=0.009~\ikms$. 
This should be compared to 10\% error on 
the amplitude and 0.04 on the slope from the WMAP data 
at $k=0.05{\rm Mpc}^{-1}$ \citep{2003ApJS..148..175S}. 
This data set provides very tight 
constraints on the amplitude and slope of the matter power spectrum.
Many additional 
analyses can be performed using this data set, 
among them the mean flux 
evolution, 
cross-correlations between close pairs, and a bispectrum
measurement. These will provide a wealth of additional information both 
on cosmology and on the state of the intergalactic medium at $2<z<4$, and they
will allow us to test the basic picture of the \lyaf\ that has emerged
over the last decade.
We believe 
that the unprecedented size of this data set will revolutionize our 
understanding of the high redshift universe; this work is merely a 
first step in this endeavor. 

\acknowledgements
Funding for the creation and distribution of the SDSS Archive has been 
provided by the Alfred P. Sloan Foundation, the Participating Institutions, 
the National Aeronautics and Space Administration, the National Science 
Foundation, the U.S. Department of Energy, the Japanese Monbukagakusho, and 
the Max Planck Society. The SDSS Web site is {\it http://www.sdss.org/}. 

The SDSS is managed by the Astrophysical Research Consortium (ARC) for the 
Participating Institutions. The Participating Institutions are The University 
of Chicago, Fermilab, the Institute for Advanced Study, the Japan 
Participation Group, The Johns Hopkins University, Los Alamos National 
Laboratory, the Max-Planck-Institute for Astronomy (MPIA), the 
Max-Planck-Institute for Astrophysics (MPA), New Mexico State University, 
University of Pittsburgh, Princeton University, the United States Naval 
Observatory, and the University of Washington.

Some of the computations used facilities at Princeton provided in part 
by NSF grant AST-0216105, and some computations were performed at NCSA. 

Transition wavelengths used in this paper are from
the Atomic Line List, {\it http://www.pa.uky.edu/$\sim$peter/atomic}.

JS is grateful for support from the W.M. Keck Foundation
and NSF grant PHY-0070928.  DPS is supported by NSF grant
AST03-07582.

\end{document}